\documentclass{ws-ijmpb}

\usepackage{graphicx}
\usepackage{dcolumn}
\usepackage{bm}
\usepackage{amsmath,amssymb}

\newcommand{\bq}{\begin{equation}}
\newcommand{\eq}{\end{equation}}
\newcommand{\bqa}{\begin{eqnarray}}
\newcommand{\eqa}{\end{eqnarray}}
\newcommand{\nn}{\nonumber \\}

\newcommand{\bol}[1]{\boldsymbol #1}

\def\be     {\begin{equation}}
\def\ee     {\end{equation}}
\def\bea        {\begin{eqnarray}}
\def\eea        {\end{eqnarray}}
\def\bnn    {\begin{eqnarray*}}
\def\enn    {\end{eqnarray*}}

\begin{document}

\markboth{Ki-Seok Kim and Akihiro Tanaka} {Emergent gauge fields and their nonperturbative effects in correlated electrons}

%%%%%%%%%%%%%%%%%%%%% Publisher's Area please ignore %%%%%%%%%%%%%%%
%
\catchline{}{}{}{}{}
%
%%%%%%%%%%%%%%%%%%%%%%%%%%%%%%%%%%%%%%%%%%%%%%%%%%%%%%%%%%%%%%%%%%%%

\title{Emergent gauge fields and their nonperturbative effects in correlated electrons}

\author{Ki-Seok Kim}

\address{Department of Physics, POSTECH, Pohang, Gyeongbuk 790-784, Korea \footnote{tkfkd@postech.ac.kr}}

\author{Akihiro Tanaka}

\address{Computational Materials Science Unit, National Institute for Materials Science, 1-1 Namiki, Ibaraki, Tsukuba 305-0044, Japan \footnote{TANAKA.Akihiro@nims.go.jp}}

\maketitle

\begin{history}
\received{Day Month Year}
\revised{Day Month Year}
%\accepted{(Day Month Year)}
%\comby{(xxxxxxxxxx)}
\end{history}

\begin{abstract}
%History 
The history 
of modern condensed matter physics may be regarded as 
%competition 
the competition 
and reconciliation between Stoner's and Anderson's physical pictures, where the former is based on momentum-space descriptions focusing on long wave-length fluctuations while the latter is based on real-space physics emphasizing emergent localized excitations. In particular, these two view points compete with each other in various nonperturbative phenomena, which %cover 
range 
from the problem of high T$_{c}$ superconductivity, quantum spin liquids in organic materials and frustrated spin systems, heavy-fermion quantum criticality, metal-insulator transitions in correlated electron systems such as doped silicons and two-dimensional electron systems, 
the fractional quantum Hall effect, to 
the recently discussed Fe-based superconductors. 
%An idea how 
An approach 
to reconcile these competing frameworks is to introduce topologically nontrivial excitations into the Stoner's description, which appear to be localized in either space or time and sometimes both, where scattering between itinerant electrons and topological excitations such as skyrmions, vortices, various forms of instantons, emergent magnetic monopoles, and etc. may catch nonperturbative local physics beyond the Stoner's paradigm. In this review article we discuss nonperturbative effects of topological excitations on dynamics of correlated electrons. First, we focus on the problem of scattering between itinerant fermions and topological excitations in antiferromagnetic doped Mott insulators, expected to be relevant for the pseudogap phase of high T$_{c}$ cuprates. We propose that nonperturbative effects of topological excitations can be incorporated within the perturbative framework, where an enhanced global symmetry with a topological term plays an essential role. 
%Second, we discuss rich interplays between topological excitations and topological terms %without itinerant fermions, where the nonperturbative dynamics of topological excitations is %essential in classifying interacting topological insulators protected by symmetries. We clarify %some hidden links between several effective field theories with topological terms, generalizing %one dimensional physics into higher dimensions. 
In the second part, we go on to discuss the subject of symmetry protected topological states 
in a largely similar light. While we do not introduce itinerant fermions here, the nonperturbative 
dynamics of topological excitations is again seen to be crucial in 
classifying topologically nontrivial gapped systems.  We point to some hidden links between several effective field theories with topological terms, starting with 
one dimensional physics, and subsequently finding natural generalizations to higher dimensions.

%
%Those topological excitations serve links between various field theories, which look quite different, for example, nonlinear $\sigma-$model field theories with topological terms, BF-type topological field theories, Z$_{2}$ gauge theories, and so on.
%
\end{abstract}


\section{Introduction}

Strongly coupled field theories lie at the heart of unsolved fundamental problems not only in particle physics but also in condensed matter physics, which cover from confinement in quantum chromodynamics (QCD) to mechanism of high T$_{c}$ superconductivity in doped Mott insulators. An important feature of strongly coupled field theories is that the $\beta-$function is negative in the renormalization group analysis, indicating that effective interactions between elementary excitations introduced in the UV (ultraviolet) limit are enhanced and such excitations become strongly coupled in the IR (infrared) limit. However, this does not necessarily mean that we cannot solve such strongly coupled field theories. Although it is negative the $\beta-$function of an effective interaction for superconducting instability in the Landau's Fermi-liquid state \cite{Shankar_RG}, we all know that this problem can be solved in the framework of the BCS (Bardeen, Cooper, and Schrieffer) mean-field theory \cite{BCS_Textbook}. On the other hand, even if essentially the same situation occurs in the Kondo problem \cite{Kondo_Textbook}, we do not have any mean-field types of effective theories which describe Fermi-surface instability due to a single magnetic impurity successfully except for exact methods based on Bethe ansatz \cite{Luttinger_Liquid_Textbook} and numerical renormalization group \cite{Kondo_Textbook}.

There exist other types of strongly coupled field theories, where corresponding $\beta-$functions vanish. In metals, most effective interactions between electron quasiparticles are irrelevant due to the presence of a Fermi surface while forward scattering channels remain marginal in the renormalization group sense, identified with Landau's Fermi-liquid fixed point and described by Landau's Fermi-liquid theory \cite{Shankar_RG}. This ``strongly" coupled field theory is solved within the $1/N_{\sigma}$ technique, which allows us to neglect vertex corrections, where $N_{\sigma}$ is an enhanced spin degeneracy from $\uparrow, \downarrow$ to $1, 2, ..., N_{\sigma}$. On the other hand, when the spatial dimension is reduced to be one, vertex corrections should be introduced to play a central role in the treatment of IR divergences, which lead the electron-quasiparticle peak to split into double peaks of spinons and holons with their continuum, identified with Luttinger-liquid fixed point and described by Luttinger-liquid theory \cite{Luttinger_Liquid_Textbook}. Recently, effective field theories remain to be strongly coupled in the vicinity of quantum phase transitions from Landau's Fermi-liquid state, where all planar diagrams are shown to be the same order in the $1/N_{\sigma}$ technique \cite{SSL_Large_N_Failure}, which implies that vertex corrections should be incorporated appropriately as the case of the Luttinger-liquid state.

These discussions give us an interesting question. When do vertex corrections become relevant in such strongly coupled field theories? In the above we have discussed two cases: (1) Fermi-surface instability toward the BCS superconducting state vs. Fermi-surface instability toward the local Fermi-liquid state (the Kondo effect) in the case of negative $\beta-$functions and (2) Landau's Fermi-liquid theory vs. Luttinger-liquid theory and quantum criticality from the Landau's Fermi-liquid state in the case of zero $\beta-$functions. Our speculation is that vertex corrections may encode the information of scattering between emergent localized excitations and itinerant electrons, where such localized excitations are identified with topologically nontrivial fluctuations, referred to as vortices in superconductivity, skyrmions in magnetism, and various forms of instantons localized even in time. Consistent introductions of vertex corrections in strongly coupled field theories mean that effects of topological excitations are incorporated into effective field theories appropriately. This scattering physics is expected to be responsible for Fermi-surface instabilities associated with orthogonality catastrophe \cite{Mahan_Textbook}. However, the absence of vertex corrections does not mean that the role of topological excitations is not introduced. If one considers the boson-vortex duality in the superfluid to Mott-insulator transition, the perturbative renormalization group analysis based on the charge description gives essentially the same critical physics as that based on the vortex picture \cite{Boson_Vortex_Duality_Textbook} \footnote{Unfortunately, it is not straightforward to prove explicitly that their critical physics are same, where the vortex description involves noncompact U(1) gauge fluctuations.}, implying that the information of topological excitations is introduced within the perturbative analysis.

The question is when the perturbative framework fails to incorporate physics of topological excitations. Here, the perturbative framework means that a given field theory can be solved within the self-consistent RPA (random phase approximation), equivalently the $1/N_{\sigma}$ approximation or Eliashberg theory, where only self-energy corrections are introduced self-consistently. We recall that vertex corrections are introduced self-consistently through the Ward identity in one-dimensional interacting electrons, where the resulting Green's function in the nonperturbative diagrammatic approach gives essentially the same expression as that in the bosonization framework which introduces spinons and holons explicitly, identified with topological excitations (solitons) \cite{Maslov_Review_LL}. This implies that dimensionality which controls quantum fluctuations may play an important role for nonperturbative physics. We speculate that the perturbative framework may work near the upper critical dimension while it breaks down, which requires nonperturbative approaches, in low dimensions near the lower critical dimension or much below the upper critical dimension \footnote{KS enjoyed intensive discussions with Prof. V. Dobrosavljevic when he visited POSTECH in the summer season of 2014.}. Here, the nonperturbative framework means to introduce topological excitations explicitly into the strongly coupled effective field theory and to deal with scattering physics between such localized excitations and itinerant electrons on equal footing \cite{Solitons_Instantons_Textbook}.

In this review article we discuss nonperturbative effects of topological excitations on dynamics of correlated electrons. First, we focus on the problem of scattering between itinerant fermions and topological excitations in antiferromagnetic doped Mott insulators, where dynamics of localized magnetic moments and their localized excitations are described by emergent gauge fields and their topologically nontrivial configurations. We propose that nonperturbative effects of topological excitations can be incorporated within the perturbative framework, where an enhanced global symmetry allows us to introduce effects of topological excitations into an effective field theory explicitly in the presence of a topological term. Second, we discuss rich interplays between topological excitations and topological terms without itinerant fermions, where the nonperturbative dynamics of topological excitations is essential in classifying interacting topological insulators protected by symmetries. We clarify some hidden links between several effective field theories with topological terms, generalizing one dimensional physics into higher dimensions.

%Topological excitations can be utilized to classify various phases, not characterized by condensation of local order parameters and regarded beyond the symmetry classification. In section 3 we review the role of topological excitations in the classification of symmetry protected topological phases, recently discussed in the condensed-matter physics community, where topological terms play a key role.

\section{How to simulate nonperturbative physics from topological excitations within the perturbative framework?}

\subsection{Organization of this section}

In section 2.2 we review an origin of non-Fermi liquid physics in antiferromagnetic doped Mott insulators, describing effective interactions between doped holes and hedgehog-type instanton excitations, expected to be involved with the pseudogap phase of high T$_{c}$ cuprates. In section 2.2.2 we construct an effective gauge-field theory from the t-J Hamiltonian, regarded to be one of the standard models for strongly correlated electrons, where dynamics of localized magnetic moments is described by SO(5) Wess-Zumino-Witten (WZW) theory and that of doped holes is given by QED$_{3}$ (quantum electrodynamics in one time and two spatial dimensions) with a finite chemical potential, referred to as $\mu-$QED$_{3}$ and coupled to the SO(5) WZW theory \cite{Kim_SO5_WZW_mu_QED3}. We discuss various limits of this emergent gauge theory. In section 2.2.1 we discuss the case of half filling, where hole concentration vanishes, thus reduced to the SO(5) WZW theory, which describes competing fluctuations between antiferromagnetic (three components) and valence bond (two components) order parameters. An essential aspect in this effective field theory is that space-time hedgehog fluctuations (magnetic monopoles as instantons) of the antiferromagnetic order parameter carry the quantum number of valence bond ordering near its core \cite{Haldane,Read_Sachdev}, which originates from the WZW term. Such topological excitations can be incorporated within the perturbative framework, where valence bond fluctuations are introduced explicitly and naturally through the SO(5) enhanced symmetry with the WZW term \cite{Tanaka_SO5}. We review physics of deconfined quantum criticality \cite{DQCP} based on the SO(5) WZW theory \cite{Tanaka_SO5}, which argues how deconfinement of fractionalized spin excitations referred to as spinons, regarded to be quark-like objects, can be realized near quantum criticality, where magnetic monopole excitations as instantons become suppressed to preserve the total number of skyrmions \cite{DQCP}. In section 2.2.3 we apply the $\mu-$QED$_{3}$ coupled to the SO(5) WZW theory into one dimension, where the corresponding effective field theory is given by QED$_{2}$ coupled to SO(4) WZW theory. We discuss that this effective field theory recovers the Luther-Emery phase \cite{Luttinger_Liquid_Textbook}, where spin excitations are gapped while superconducting correlations between doped holes are enhanced. In section 2.2.4 we discuss dynamics of doped holes near the deconfined quantum criticality of the SO(5) WZW theory, where the interplay between doped holes and space-time hedgehog excitations is encoded into the perturbative framework, i.e., scattering between itinerant fermions and valence bond fluctuations \cite{Kim_SO5_WZW_mu_QED3}. We propose the role of valence bond fluctuations in dynamics of doped holes for their non-Fermi liquid physics in the pseudogap phase of high T$_{c}$ cuprates.

Not only the situation of deconfinement but also that of confinement is discussed in section 2.3, based on a recently developed effective field theory for QCD at low energies in Hadron physics, referred to as Polyakov-loop extended Nambu-Jona-Lasinio (NJL) model \cite{Fukushima_PNJL,PNJL_Review}, where such quark-like objects correspond to holons and spinons, representing doped holes and fractionalized spin excitations roughly speaking. Applying the Polyakov-loop extended NJL (PNJL) model to the problem of paramagnetic doped Mott insulators, we describe non-Fermi liquid transport phenomena near optimal doping of high T$_{c}$ cuprates outside the pseudogap state, based on the confinement of spinons and holons \cite{Kim_Kim_PNJL}.

Deep inside the Mott insulating phase, spin fluctuations are only relevant degrees of freedom at half filling. However, charge fluctuations are expected to play a central role in metal-insulator transitions, which may suppress magnetic ordering to allow spin liquid states, described by emergent SU(2) gauge theories. In section 2.4 we discuss metal-insulator transitions, generalizing the t-J Hamiltonian to the Hubbard model, where charge fluctuations are introduced. Constructing an effective SU(2) gauge theory to describe interactions between spinons and holons through SU(2) gauge fluctuations \cite{Kim_SU2SR}, we discuss possible spin liquid states near the metal-insulator transition on honeycomb \cite{Kim_Tien_SU2SR} and triangular lattices. In particular, we speculate how physics of spin liquids, metal-insulator transitions, and unconventional superconductivity will emerge from such nonabelian gauge theories beyond the saddle-point analysis, where gluon condensation consistent with the lattice symmetry is suggested to play an essential role.

In section 2.5 we conclude the first part of this review article, speculating that gauge field theories can appear rather commonly than expected in strongly coupled field theories \cite{Kim_Spin_Fermion_Model_AFQCP,Kim_Spin_Fermion_Model_FMQCP}. We discuss an antiferromagnetic quantum phase transition from the Landau's Fermi-liquid state, where a critical field theory describes scattering between itinerant electrons and antiferromagnetic spin fluctuations \cite{Chubukov_Spin_Fermion_Model}. Recently, the $1/N_{\sigma}$ technique turns out to fail to describe non-Fermi liquid physics near antiferromagnetic quantum criticality \cite{Sachdev_Large_N_Failure}, where the critical field theory lies in the strongly coupled regime, meaning that vertex corrections should be incorporated consistently. We suggest that some types of instanton excitations may keep such nonperturbative physics, constructing an effective field theory with the introduction of instantons. Integrating out contributions of topological excitations, we speculate that an effective gauge-field theory emerges, regarded to generalize the scenario of the SO(5) WZW theory.

\subsection{Emergent gauge fields and their nonperturbative effects in antiferromagnetic doped Mott insulators}

\subsubsection{SO(5) Wess-Zumino-Witten theory from Heisenberg model}

\begin{figure}[htp]
\centerline{\includegraphics[scale=0.6]{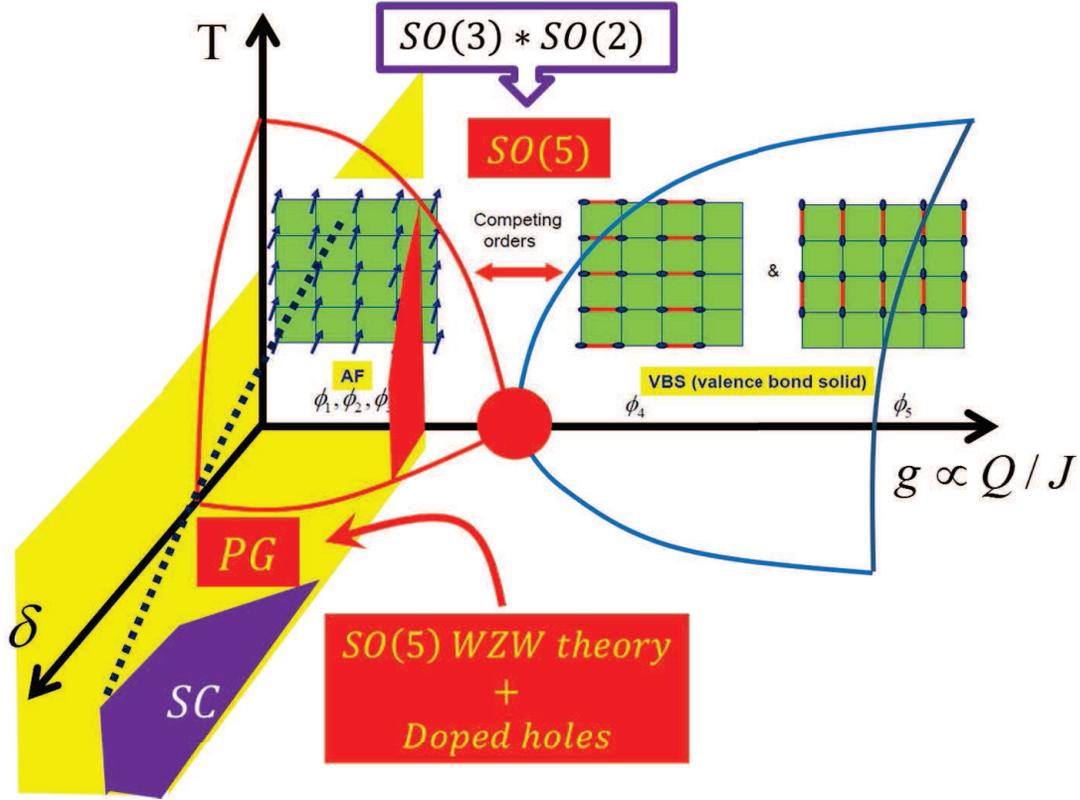}}
\caption{ (Color online) A schematic phase diagram of SO(5) WZW theory and its possible connection to the pseudogap phase of high T$_{c}$ cuprates. Since the antiferromagnetic phase breaks the SO(3) spin-rotation symmetry and the valence-bond solid state does the SO(2) lattice-translation symmetry, it is natural to propose that a critical field theory would enjoy the SO(3) $\otimes$ SO(2) symmetry within the Landau-Ginzburg-Wilson framework. However, there exists an exotic possibility that such a global symmetry becomes enhanced to SO(5), where both antiferromagnetic and valence bond fluctuations are symmetry equivalent at the quantum critical point, responsible for the emergence of fractionalized spin excitations, referred to as spinons. The high T$_{c}$ phase diagram requires an additional axis, which corresponds to hole doping concentration. If an initial point at half filling lies near the quantum critical point of the SO(5) WZW theory, we may expect that competing fluctuations between antiferromagnetic and valence bond order parameters will play an essential role in the pseudogap phase of high T$_{c}$ cuprates after the antiferromagnetic order disappears via hole doping. Proximity of deconfined quantum criticality in the SO(5) WZW theory is our view point in this review article.} \label{Phase_Diagram_SO5_QED3_HighTc}
\end{figure}

Let's start from an extended Heisenberg model on square lattice, \bqa && H = H_{J} + H_{Q} , ~~~~~ H_{J} = J \sum_{ij} \bm{S}_{i} \cdot \bm{S}_{j} , \nn && H_{Q} = - Q \sum_{\{ij kl\} \in \square} (\bm{S}_{i} \cdot \bm{S}_{j} - 1/4) (\bm{S}_{k} \cdot \bm{S}_{l} - 1/4) , \eqa where $H_{J}$ is an antiferromagnetic ($J > 0$) Heisenberg model to describe dynamics of localized magnetic moments, and $H_{Q}$ is an extended part to favor the formation of valence bond ordering ($Q > 0$) \cite{Sandvik_DQCP}. It is not difficult to speculate that an antiferromagnetic phase appears in the case of $J \gg Q$, breaking SO(3) symmetry involved with spin rotation, while a valence bond ordered state emerges in the case of $J \ll Q$, breaking Z$_{4}$ associated with lattice translation. In this respect one may expect that a critical field theory would enjoy SO(3) $\otimes$ SO(2) symmetry in terms of both antiferromagnetic and valence bond order parameters, where the Z$_{4}$ symmetry can be enhanced to SO(2) in the continuum limit. However, it has been proposed that the SO(3) $\otimes$ SO(2) symmetry may be enlarged to SO(5), where both order parameters form a superspin vector at this antiferromagnetic to valence bond quantum critical point \cite{Tanaka_SO5}. This scenario is in parallel with the well-known physics of an antiferromagnetic quantum spin chain, where an effective field theory is given by SO(4) WZW theory although its microscopic lattice model enjoys SO(3) $\otimes$ Z$_{2}$ \cite{Luttinger_Liquid_Textbook}. This effective field theory turns out to be critical due to the existence of the WZW term, allowing fractionalized spin excitations referred to as spinons \cite{Luttinger_Liquid_Textbook}. Emergence of an enhanced symmetry is suggested to play a central role in deconfined quantum criticality above one time and one space dimensions \cite{Tanaka_SO5,Tanaka_SO4}. See Fig. \ref{Phase_Diagram_SO5_QED3_HighTc}, which shows a schematic phase diagram of the SO(5) WZW theory and a possible connection to the pseudogap phase of high T$_{c}$ cuprates.

%%%%%%%%%%%%%%%%%%%%%%%%%%%%%%%%%%%%%%%%%%%%%%%%%%%%%%%%%%%%%%%%%%%%%%%%
%%%%%%%%%%%%%%%%%%%%%%%%%%%%%%%%%%%%%%%%%%%%%%%%%%%%%%%%%%%%%%%%%%%%%%%%

In order to take into account the role of Berry phase in the path-integral representation with the spin coherent basis \cite{Quantum_Spins_Textbook}, one may consider a projective representation for the spin operator as follows \bqa && \bm{S}_{i} = \frac{1}{2} f_{i\alpha}^{\dagger} \bm{\sigma}_{\alpha\beta} f_{i\beta} , \eqa backup by the single occupancy constraint $f_{i\sigma}^{\dagger} f_{i\sigma} = 1$. Here, we use the Einstein convention. Inserting this expression into the Heisenberg model and decomposing the four-fermion effective-interaction term into particle-hole and particle-particle channels within the singlet domain, we find an effective UV theory in this parton construction as follows \cite{SU2SBGT} \bqa && Z_{UV} = \int D \psi_{i \alpha} D \chi_{ij} D \eta_{ij} D a_{i\tau}^{k} e^{- \int_{0}^{\beta} d \tau L_{eff}} , \nn && L_{UV} = L_{0} + L_{s} , ~~~~~ L_{0} = J_{r} \sum_{ij} \mathbf{tr}[U_{ij}^{\dagger}U_{ij}] , \nn && L_{s} = \frac{1}{2} \sum_{i} \psi_{i\alpha}^{\dagger} (\partial_{\tau} - i a_{i\tau}^{k}\tau_{k}) \psi_{i\alpha} + J_{r} \sum_{ij} ( \psi_{i\alpha}^{\dagger}U_{ij}\psi_{j\alpha} + H.c.)  \eqa with $J_{r} = \frac{3J}{16}$. $\psi_{i\alpha} = \left(\begin{array}{c} f_{i\alpha} \\ \varepsilon_{\alpha\beta} f_{i\beta}^{\dagger} \end{array}\right)$ is a two-component Nambu-spinor, where $\varepsilon_{\alpha\beta}$ is an antisymmetric tensor. $U_{ij} = \left( \begin{array}{cc} - \chi_{ij}^{\dagger} & \eta_{ij} \\ \eta_{ij}^{\dagger} & \chi_{ij} \end{array}\right)$ is an order-parameter matrix, where $\chi_{ij}$ represents an effective hopping parameter and $\eta_{ij}$ does a pairing order parameter. $a_{i\tau}^{k}$ is a Lagrange multiplier field to impose the single occupancy constraint with $k = 1, 2, 3$, which may be identified with a time component of an SU(2) gauge field, where two constraint equations from $k = 1, 2$ are satisfied trivially by that from $k = 3$.

\begin{figure}[htp]
\centerline{\includegraphics[scale=0.6]{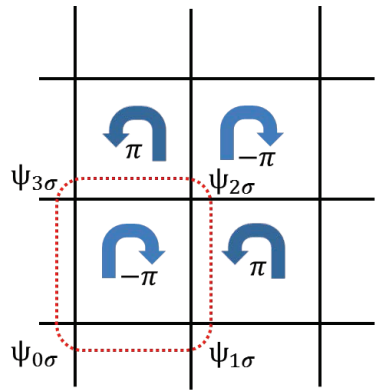}}
\caption{ (Color online) $\pi-$flux phase. When spinons hop around a plaquette, they acquire an Aharonov-Bohm phase of $\pi$, which turns out to lower the ground-state energy. A square enclosed by a red-rotted line is a unit cell in deriving a continuum field theory, which includes four components of SU(2) doublets.} \label{Pi_Flux_Phase}
\end{figure}

Performing the saddle-point analysis for the order-parameter matrix, the ground state turns out to be a $\pi-$flux phase \cite{Affleck_Marston_pi_Flux,Kotliar_pi_Flux}, where $\pi-$flux penetrates each plaquette, given by \bqa && U_{ij}^{\pi F} = - \chi \tau_{3} \exp[i(-1)^{i_{x} + i_{y}} \frac{\pi}{4} \tau_{3}] \label{Pi_Flux} \eqa with equal amplitudes between hopping and pairing order parameters as $\chi = \eta$. See Fig. \ref{Pi_Flux_Phase}. Although amplitude fluctuations of the order-parameter matrix-field are frozen, there exist low-lying transverse excitations, which can be identified with SU(2) gauge fields. Introducing such low energy fluctuations into an effective lattice field theory within the $\pi-$flux phase, we obtain \bqa && Z = \int D \psi_{i \alpha} D a_{ij}^{k} D a_{i\tau}^{k} \exp\Bigl[ - \int_{0}^{\beta} d \tau \Bigl\{ \frac{1}{2} \sum_{i} \psi_{i\alpha}^{\dagger} (\partial_{\tau} - i a_{i\tau}^{k}\tau_{k})\psi_{i\alpha} \nn && + J_{r} \sum_{ij} ( \psi_{i\alpha}^{\dagger} U_{ij}^{\pi F} e^{ia_{ij}^{k}\tau_{k}}\psi_{j\alpha} + H.c.) + J_{r} \sum_{ij} \mathbf{tr}[U_{ij}^{\pi F\dagger}U_{ij}^{\pi F}] \Bigr\} \Bigr] , \eqa where $a_{ij}^{k}$ is a spatial component of an SU(2) gauge field.

%%%%%%%%%%%%%%%%%%%%%%%%%%%%%%%%%%%%%%%%%%%%%%%%%%%%%%%%%%%%%%%%%%%%%%%%
%%%%%%%%%%%%%%%%%%%%%%%%%%%%%%%%%%%%%%%%%%%%%%%%%%%%%%%%%%%%%%%%%%%%%%%%

It is straightforward to find a continuum field theory for this lattice gauge theory. Turning off the lattice gauge field in the saddle-point approximation, performing the Fourier transformation to the momentum space, and taking the long-wavelength limit near the chemical potential, one finds an effective SU(2) gauge-field theory \cite{Wen_Symmetry} \bqa && Z_{eff} = \int D \psi D a_{\mu}^{k} e^{- \int_{0}^{\beta} d \tau \int d^{2} \bm{r} {\cal L}_{eff}} , \nn && {\cal L}_{eff} = \bar{\psi}\gamma_{\mu}(\partial_{\mu} - ia_{\mu}^{k}\tau_{k})\psi - \frac{1}{4 e^{2}} f_{\mu\nu}^{k} f_{\mu\nu}^{k} , \eqa where SU(2) gauge fluctuations have been recovered. $\psi$ is an eight-component Dirac spinor, composed of four SU(2) doublets living at each site of a plaquette (Fig. \ref{Pi_Flux_Phase}), where Dirac matrices are given by  $\gamma_{0} = \left( \begin{array}{cc} \sigma_{3} & 0 \\ 0 & - \sigma_{3} \end{array}\right)$, $\gamma_{1} = \left( \begin{array}{cc} \sigma_{1} & 0 \\ 0 & - \sigma_{1} \end{array}\right)$, and $\gamma_{2} = \left( \begin{array}{cc} \sigma_{2} & 0 \\ 0 & - \sigma_{2} \end{array}\right)$. $f_{\mu\nu}^{k} = \partial_{\mu} a_{\nu}^{k} - \partial_{\nu} a_{\mu}^{k} - i \epsilon_{klm} a_{\mu}^{l} a_{\nu}^{m}$ is an SU(2) field-strength tensor, where this Yang-Mills dynamics is expected to appear from UV fluctuations of the lattice scale. We emphasize that both the Dirac structure and SU(2) gauge field emerge from the $\pi-$flux fixed-point ansatz.

In order to discuss spontaneous chiral symmetry breaking and find an effective field theory for low energy spin fluctuations, it is necessary to consider the physical symmetry of the matter sector. It is interesting to notice that lattice symmetries such as translations, rotations, and etc. are translated into internal symmetries given by Dirac matrices, for example, where $\bm{\gamma}_{3} = \left( \begin{array}{cc} 0 & \bm{I} \\ \bm{I} & 0 \end{array}\right)$ and $\bm{\gamma}_{5} = \bm{\gamma}_{0} \bm{\gamma}_{1} \bm{\gamma}_{2} \bm{\gamma}_{3} = i \left( \begin{array}{cc} 0 & \bm{I} \\ - \bm{I} & 0 \end{array}\right)$ are associated with lattice translations along the $x-$ and $y-$ directions, respectively \cite{Wen_Symmetry} \footnote{More precisely, the translational symmetry should be backup by an appropriate gauge transformation in the projective representation \cite{Wen_Textbook}. If one sees Fig. \ref{Pi_Flux_Phase}, he realizes immediately that the translational symmetry is broken explicitly for the configuration of Eq. \ref{Pi_Flux}. However, this should be regarded to be an artifact of the mean-field ansatz, where the order parameter field is not gauge invariant, allowing us to perform an appropriate gauge transformation and to recover the translational symmetry. W$_{T_{\bm{r}}}$ $\otimes$ T$_{\bm{r}}$, where T$_{\bm{r}}$ is the lattice-translation operator and W$_{T_{\bm{r}}}$ is the corresponding gauge transformation, is an element of the projective symmetry group, suggested to classify mean-field ground states of symmetric spin liquids described by emergent gauge theories.}. In this way we have SU(2) chiral symmetry with three generators of $\bm{\gamma}_{3}$, $\bm{\gamma}_{5}$, and $i \bm{\gamma}_{3} \bm{\gamma}_{5}$ in addition to the SU(2) spin rotational one \cite{Tanaka_SO5}. This leads one to propose the SU(2) $\otimes$ SU(2) symmetry, where the former is associated with spin rotations and the latter is involved with chiral symmetry. However, an actual global symmetry turns out to be more enlarged as follows \cite{Wen_Symmetry}. It is clear that this symmetry is closely connected with both spin and Dirac spaces. Since the spin SU(2) symmetry is hidden in the present eight-component representation, one may consider the redundant representation $\Psi =  \left( \begin{array}{c} \psi \\ \hat{\psi} \end{array}\right)$ of sixteen-components with a Dirac spinor $\hat{\psi} \equiv i\tau_{2}\psi^{*} = \left( \begin{array}{c} f_{\downarrow} \\ - f_{\uparrow}^{\dagger} \end{array}\right)$, regarded to be a time-reversal partner of $\psi$. Noting that the group space is composed of $G = G_{Dirac} \otimes G_{gauge} \otimes G_{spin}$, one sees ten generators associated with SO(5) symmetry given by $\bm{I} \otimes \bm{I} \otimes \bm{\sigma}$, $\bm{\gamma}_{3} \otimes \bm{I} \otimes \bm{\sigma}$, $\bm{\gamma}_{5} \otimes \bm{I} \otimes \bm{\sigma}$, $i\bm{\gamma}_{3}\bm{\gamma}_{5} \otimes \bm{I} \otimes \bm{I}$. Since SO(8) is the largest global symmetry, considering the Majorana fermion representation for the four-component SU(2) doublet Dirac spinor, the SO(5) symmetry can be regarded to be the largest subgroup, where the emergent Lorentz symmetry forms SO(3) and the SO(8) symmetry group can be decomposed as follows SO(8) $\rightarrow$ SO(5) $\otimes$ SO(3).

The above discussion implies that symmetry equivalent operators via the SO(5) rotation have the same strength for instability in this critical spin-liquid state, i.e., the same critical exponent for each correlation function, which suggests an SO(5) superspin vector $\bm{v} = (n_{x}, n_{y}, n_{z}, v_{x}, v_{y})$ through the following fermion-mass term $- m \bar{\Psi} (\bm{v}\cdot\bm{\Gamma}) \Psi$ with $\bm{\Gamma} = (\bm{I} \otimes \bm{I} \otimes \bm{\sigma}_{x}, \bm{I} \otimes \bm{I} \otimes \bm{\sigma}_{y}, \bm{I} \otimes \bm{I} \otimes \bm{\sigma}_{z}, i \bm{\gamma}_{3} \otimes \bm{I} \otimes \bm{I}, i \bm{\gamma}_{5} \otimes \bm{I} \otimes \bm{I})$, where the former three components form Neel vectors and the latter two represent $x-$ and $y-$ valance bond fluctuations. As a result, one reaches the following Lagrangian for spontaneous chiral symmetry breaking \bqa && Z_{eff} = \int D \Psi D v_{i} D a_{\mu}^{k} \exp\Bigl[ - \int_{0}^{\beta} d \tau \Bigl\{ \bar{\Psi}\gamma_{\mu}(\partial_{\mu} - ia_{\mu}^{k}\tau_{k})\Psi - m \bar{\Psi} (\bm{v}\cdot\bm{\Gamma}) \Psi - \frac{1}{4 e^{2}} f_{\mu\nu}^{k} f_{\mu\nu}^{k} \Bigr\} \Bigr] , \nn \eqa where the mechanism of this symmetry breaking is not clarified \footnote{One may demonstrate that SU(2) gauge fluctuations are responsible for this chiral symmetry breaking.}. Integrating over massive fermion excitations and performing the gradient expansion for the superspin vector field, one finds an SO(5) WZW theory for the competing physics between antiferromagnetism and valence bond ordering as follows \cite{Tanaka_SO5} \bqa && Z_{eff} = \int D v_{i} e^{ - S_{eff}} , ~~~~ S_{eff} = S_{NLsM} + S_{WZW} , \nn && S_{NLsM} = \int {d^{3} x} \frac{1}{2g} (\partial_{\mu} v_{i})^{2} , ~~~~~ S_{WZW} = i \frac{2\pi}{Area(S^{4})} \int_{0}^{1}{dt} \int {d^{3} x} \epsilon_{abcde} v_{a} \partial_{t} v_{b} \partial_{\tau} v_{c} \partial_{x} v_{d} \partial_{y} v_{e} , \nn \eqa where $Area(S^{4}) = \frac{2\pi^{5/2}}{\Gamma(5/2)}$. Although the WZW term can be nicely derived in the absence of gauge fluctuations, an additional imaginary term may arise, a coupling term between gauge fields and Goldstone-Wilczek currents \cite{Abanov_WZW}, which correspond to skyrmion currents in the present case if we restrict ourselves only in antiferromagnetic fluctuations instead of the superspin vector. When we represent the eight-component Dirac spinor as $\psi_{n} = \left( \begin{array}{c} \chi_{n}^{+} \\ \chi_{n}^{-} \end{array}\right)$, where $\chi_{n}^{\pm}$ is a two-component SU(2) doublet with an isospin index $n = 1, 2$, we can see that each sector in the Dirac space gives rise to such a term. However, their signs are opposite, thus such terms are canceled. This is well-known to be cancelation of parity anomaly in the lattice model \cite{Anomaly_Cancelation}.

Although it is not straightforward to solve this effective field theory, it would be helpful to revisit the one-dimensional version of this field theory, referred to as SO(4) WZW theory \bqa && S_{WZW} = \int {d^{2} x} \Bigl\{ \frac{1}{2g} \sum_{k=1}^{4}(\partial_{\mu} v_{k})^{2} + i \frac{2\pi}{Area(S^{3})} \int_{0}^{1}{dt} \epsilon_{abcd} v_{a} \partial_{t} v_{b} \partial_{\tau} v_{c} \partial_{x} v_{d} \Bigr\} . \eqa Performing the renormalization group analysis in the one-loop level, one finds a conformal invariant stable fixed point, which originates from the existence of the WZW term \cite{Luttinger_Liquid_Textbook}, given by \bqa && \frac{d \ln g}{d \ln \Lambda} = 0 , \eqa where $\Lambda$ is a UV cutoff. Actually, the SO(4) WZW theory is exactly solvable, characterized by the central charge $c = 1$, where such critical boson excitations are identified with fractionalized spin fluctuations called spinons \cite{Luttinger_Liquid_Textbook}. The SO(3) nonlinear $\sigma-$model with a $\Theta-$term (Berry phase) at UV flows into the SO(4) WZW theory at IR, where valence bond fluctuations carry exactly the same conformal dimension as antiferromagnetic spin fluctuations.

\begin{figure}[htp]
\centerline{\includegraphics[scale=0.4]{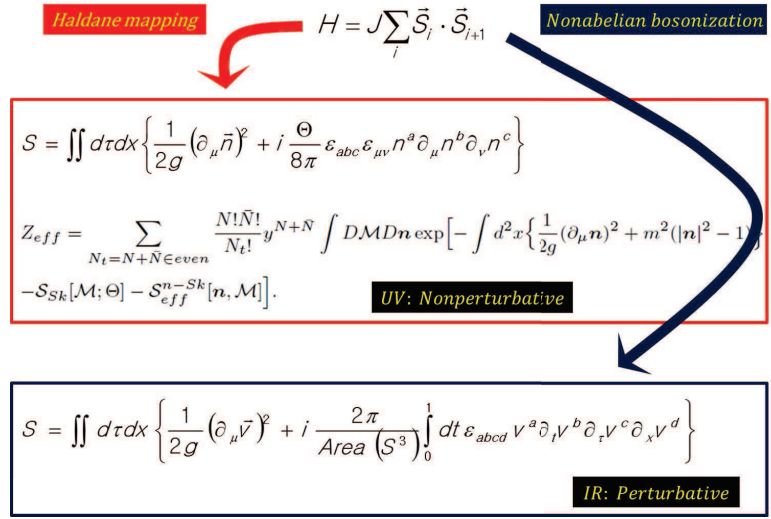}}
\caption{ (Color online) A theoretical flow chart for the Heisenberg model in one dimension. Performing the Haldane mapping for the Heisenberg spin chain, one finds an SO(3) nonlinear $\sigma-$model with a Berry phase term, which may be identified with a UV theory. In order to solve this effective field theory, one should incorporate effects of skyrmion excitations, nonperturbative as discussed in the footnote. In particular, the presence of the Berry phase term makes the role of such nonperturbative excitations more delicate. On the other hand, the other procedure based on nonabelian bosonization gives rise to the SO(4) WZW theory, regarded to be an IR theory, where the perturbative renormalization group analysis results in the quantum criticality of the spin chain. Since the UV theory is expected to flow to the IR one, which is not revealed yet as far as we know, we conclude that the SO(4) WZW theory incorporates the nonperturbative physics of instantons of the SO(3) nonlinear $\sigma-$model with the Berry-phase term. This is the main theme of the present review article that nonperturbative effects can be introduced into the perturbative framework.} \label{Flow_Chart_Theories}
\end{figure}

An important point that we would like to emphasize is as follows. One may try to solve the UV field theory directly, resorting to the CP$^{1}$ representation for the SO(3) vector field, where the SO(3) nonlinear $\sigma-$model is mapped into an emergent U(1) gauge theory with two flavors of bosonic spinons and the Berry-phase term is identified with an effective electric potential \cite{Quantum_Spins_Textbook}. Taking the easy-plane limit to map this problem into a two-flavor abelian Higgs model with an effective electric potential and performing the duality transformation to map the abelian Higgs model into an effective Sine-Gordon theory for skyrmion excitations as instantons, one may argue that such skyrmions in one-time and one-space dimensions carry the quantum number of valence bond ordering and their dynamics becomes critical, both of which originate from the Berry-phase term \cite{DQCP}. Although this argument is far from being rigorous, where the enhancement of symmetry at IR is difficult to prove, we can find how such nonperturbative physics involved with instanton excitations at UV is revealed in the perturbative analysis (renormalization group) at IR, where valence bond fluctuations are introduced explicitly into an effective field theory through the symmetry enhancement \footnote{One may write down the SO(3) nonlinear $\sigma-$model, introducing skyrmion (instanton) and anti-skyrmion (anti-instanton) fluctuations explicitly, as follows \bqa && Z_{eff} = \sum_{N_{t} = N + \bar{N} \in even} \frac{N ! \bar{N} !}{N_{t} !} y^{N+\bar{N}} \int D \mathcal{M} D \bm{n} \exp\Bigl[ - \int {d^{2} x} \Bigl\{ \frac{1}{2g} (\partial_{\mu} \bm{n})^{2} + m^{2} (|\bm{n}|^{2} - 1) \Bigr\} \nn && - \mathcal{S}_{Sk}[\mathcal{M};\Theta] - \mathcal{S}_{eff}^{n-Sk}[\bm{n}, \mathcal{M}] \Bigr] . \nonumber \eqa $N$ ($\bar{N}$) represents the number of skyrmions (anti-skyrmions), set to be equal $N = \bar{N}$ in the respect of energy cost and thus, their total number is $N_{t} = 2 N$. $y = e^{- S_{Sk}}$ is fugacity of single skyrmion excitations, where $S_{Sk}$ is an instanton action. $\mathcal{M}$ means the moduli space of skyrmions such as their sizes, center-of-mass coordinates, and so on, referred to as collective coordinates and utilized for the first quantization \cite{Solitons_Instantons_Textbook}. $m$ may be regarded to be the mass of spin fluctuations $\bm{n}$ with spin quantum number $1$, introduced to describe the unimodular constraint $|\bm{n}| = 1$. $\mathcal{S}_{Sk}[\mathcal{M};\Theta]$ describes dynamics of skyrmions, where the Berry phase term denoted by $\Theta$ should be incorporated. $\mathcal{S}_{eff}^{n-Sk}[\bm{n}, \mathcal{M}]$ describes scattering physics between smooth spin fluctuations and instanton fluctuations, regarded to be an essential part in this field theory. Unfortunately, the procedure until this field theory has not been clarified yet. This expression may be regarded to be formal. Nonperturbative physics would be encoded in this effective field theory, taking into account both topological excitations and smooth fluctuations on equal footing. An interesting point is that the Berry-phase term assigns the quantum number of valence bond ordering to the core of a skyrmion. As a result, scattering between spin fluctuations and skyrmion excitations may be translated into that between antiferromagnetic fluctuations and valence bond excitations. In this respect one may say that the SO(4) WZW theory encode the nonperturbative physics of the SO(3) nonlinear $\sigma-$model with the Berry-phase term into the perturbative framework, where the renormalization group analysis in the one-loop level reveals essential physics qualitatively.}. Figure \ref{Flow_Chart_Theories} shows one mechanism how the nonperturbative physics becomes translated into the perturbative physics.

%%%%%%%%%%%%%%%%%%%%%%%%%%%%%%%%%%%%%%%%%%%%%%%%%%%%%%%%%%%%%%%%%%%%%%%%
%%%%%%%%%%%%%%%%%%%%%%%%%%%%%%%%%%%%%%%%%%%%%%%%%%%%%%%%%%%%%%%%%%%%%%%%

One may extend the above discussion into two dimensions. Performing the duality transformation for the two-flavor abelian Higgs model in the easy-plane approximation which reduces SO(3) to SO(2) $\otimes$ Z$_{2}$, one can find another abelian Higgs model in terms of half-skyrmion (meron) excitations with two flavors, where magnetic monopole excitations give rise to meron and anti-meron pair excitations (hedgehog configurations) localized in time (instantons). However, the Berry-phase term has been proposed to make such instanton events suppressed, preserving the topological charge of meron currents and stabilizing meron excitations at the quantum critical point of this effective field theory \cite{DQCP}. Such meron fluctuations may be identified with spinon excitations in the original representation. Since the magnetic-monopole excitation carries the valence bond order near its core, assigned from the Berry-phase term, their condensation transition identifies the nature of the quantum critical point between the antiferromagnetic state and the valence bond ordered phase. One may go beyond the easy-plane limit. In this case the duality transformation is not clarified, making it difficult to describe the deconfined quantum critical point explicitly. However, it is clear that meron excitations in SO(2) $\otimes$ Z$_{2}$ should turn into skyrmions (solitons) in SO(3). As a result, the skyrmion current is conserved at the quantum critical point, where fluctuations of magnetic monopoles (instantons) become suppressed, but the conservation law breaks down in the valence-bond solid state, where the proliferation of magnetic monopoles breaks the U(1) global symmetry associated with the conservation of the skyrmion current. In this case spinon excitations can be identified with an emergent spin degree of freedom in a Z$_{4}$ vortex core, regarded to be a topological excitation in the valence-bond solid state, where the condensation of Z$_{4}$ vortices have been argued to be responsible for the quantum phase transition from the valence bond solid state to the antiferromagnetic phase \cite{Z4_Vortex_Spinon}. An essential point is that this nonperturbative physics from topological excitations at UV can be incorporated by the perturbative physics of the SO(5) WZW theory at IR, where the conformal dimension of the valence bond order parameter is the same as that of the antiferromagnetic one. The renormalization group analysis in the one-loop level is expected to allow a conformal invariant fixed point as the SO(4) WZW theory, which gives rise to deconfined critical spinon excitations. Unfortunately, we do not know an explicit result on the perturbative renormalization group analysis of the SO(5) WZW theory in two dimensions. On the other hand, the emergence of the symmetry enhancement at IR seems to be confirmed by explicit numerical simulations for the extended Heisenberg model although it is difficult to avoid the nature of weakly first ordering in the simulation \cite{Sandvik_DQCP}.

\subsubsection{$\mu-$QED$_{3}$ coupled with SO(5) WZW theory from t-J Hamiltonian}

Effects of hole doping on the SO(5) WZW theory can be investigated, based on the t-J Hamiltonian \bqa && H_{tJ} = - t \sum_{ij} (c_{i\sigma}^{\dagger} c_{j\sigma} + H.c.) + J \sum_{ij} (\bm{S}_{i}\cdot\bm{S}_{j} - \frac{1}{4}n_{i}n_{j}) , \eqa where double occupancy is prohibited. This constraint can be solved, resorting to the projective representation referred to as the SU(2) slave-boson representation for an electron operator \cite{Lee_Nagaosa_Wen_SL_Review}, \bqa && c_{i\uparrow} = \frac{1}{\sqrt{2}} h_{i}^{\dagger} \psi_{i+} = \frac{1}{\sqrt{2}} (b_{i1}^{\dagger}f_{i1} + b_{i2}^{\dagger}f_{i2}^{\dagger}) , \nn && c_{i\downarrow} = \frac{1}{\sqrt{2}} h_{i}^{\dagger} \psi_{i-} = \frac{1}{\sqrt{2}} (b_{i1}^{\dagger}f_{i2} - b_{i2}^{\dagger}f_{i1}^{\dagger}) , \eqa where spinon and holon doublets are given by $\psi_{i+} = \left(\begin{array}{c} f_{i1} \\ f_{i2}^{\dagger} \end{array}\right)$, $\psi_{i-} =  \left(\begin{array}{c} f_{i2} \\ - f_{i1}^{\dagger} \end{array}\right)$, and $h_{i} =  \left( \begin{array}{c} b_{i1} \\ b_{i2} \end{array}\right)$, respectively. Resorting to this parton construction, one may rewrite the t-J Hamiltonian as follows \cite{Lee_Nagaosa_Wen_SL_Review} \bqa && Z_{UV} = \int D \psi_{i \alpha} D h_{i} D \chi_{ij} D \eta_{ij} D a_{i\tau}^{k} e^{- \int_{0}^{\beta} d \tau L_{eff}} , \nn && L_{UV} = L_{0} + L_{s} + L_{h} , ~~~~~ L_{0} = J_{r} \sum_{ij} \mathbf{tr}[U_{ij}^{\dagger}U_{ij}] , \nn && L_{s} = \frac{1}{2} \sum_{i} \psi_{i\alpha}^{\dagger} (\partial_{\tau} - i a_{i\tau}^{k}\tau_{k})\psi_{i\alpha} + J_{r} \sum_{ij} ( \psi_{i\alpha}^{\dagger}U_{ij}\psi_{j\alpha} + H.c.) , \nn && L_{h} =  \sum_{i} h_{i}^{\dagger}(\partial_{\tau} - \mu - i a_{i\tau}^{k}\tau_{k})h_{i} + t_{r} \sum_{ij} ( h_{i}^{\dagger}U_{ij}h_{j} + H.c.) , \eqa where the Hubbard-Stratonovich transformation has been performed for particle-hole and particle-particle channels in the singlet domain, giving rise to the following order-parameter matrix field $U_{ij} = \left( \begin{array}{cc} - \chi_{ij}^{\dagger} & \eta_{ij} \\ \eta_{ij}^{\dagger} & \chi_{ij} \end{array}\right)$ with $J_{r} = \frac{3J}{16}$ and $t_{r} = \frac{t}{2}$, as discussed in the half-filled case. The time component of an SU(2) gauge field $a_{i\tau}^{k}$ is to impose the single-occupancy constraint, and $\mu$ is a chemical potential to control hole concentration.

Following the strategy of the half-filled case, the variational analysis for the order-parameter matrix field gives rise to a staggered flux state \cite{Lee_Nagaosa_Wen_SL_Review}, given by \bqa && U_{ij}^{SF} = - \sqrt{\chi^{2} + \eta^{2}} \tau_{3} \exp[i(-1)^{i_{x} + i_{y}} \Phi \tau_{3}] , \eqa where a flux through a plaquette is $4 \Phi = 4 \tan^{-1}\Bigl( \frac{\eta}{\chi} \Bigr) < \pi$ and alternating. Although the staggered flux ansatz breaks translational invariance, this formal symmetry breaking is restored via SU(2) gauge transformation between nearly degenerate U(1) mean-field states \cite{Lee_Nagaosa_Wen_SL_Review}. For example, one possible U(1) ground state, the d-wave pairing state $U_{ij}^{dSC} = - \chi \tau_{3} + (-1)^{i_{x} + j_{y}} \eta \tau_{1}$ can result from the staggered flux phase through the SU(2) rotation given by $U_{ij}^{dSC}= W_{i} U_{ij}^{SF} W_{j}^{\dagger}$, where the corresponding SU(2) matrix is $W_{i} = \exp\Bigl\{ i(-1)^{i_{x} + i_{y}}\frac{\pi}{4} \tau_{1} \Bigr\}$. Thus, this variational state should be regarded as one possible gauge choice, preserving both time reversal and translational symmetries. One can show that the staggered flux phase allows only one low-lying transverse fluctuations, identified with the third component of the SU(2) gauge field. As a result, an effective lattice field theory in the staggered flux state is given by \bqa && Z = \int D \psi_{i \alpha} D h_{i} D a_{ij}^{3} D a_{i\tau}^{3} \exp\Bigl[ - \int_{0}^{\beta} d \tau \Bigl\{ \frac{1}{2} \sum_{i} \psi_{i\alpha}^{\dagger} (\partial_{\tau} - i a_{i\tau}^{3}\tau_{3})\psi_{i\alpha} \nn && + J_{r} \sum_{ij} ( \psi_{i\alpha}^{\dagger} U_{ij}^{SF} e^{ia_{ij}^{3}\tau_{3}}\psi_{j\alpha} + H.c.) + \sum_{i} h_{i}^{\dagger}(\partial_{\tau} - \mu  - i a_{i\tau}^{3}\tau_{3})h_{i} \nn && + t_{r} \sum_{ij} ( h_{i}^{\dagger}U_{ij}^{SF} e^{ia_{ij}^{3}\tau_{3}}h_{j} + H.c.) + J_{r} \sum_{ij} \mathbf{tr}[U_{ij}^{SF\dagger}U_{ij}^{SF}] \Bigr\} \Bigr] . \eqa

An idea is to fermionize the holon sector attaching a fictitious flux to a holon field \cite{Kim_SO5_WZW_mu_QED3}, \bqa && L_{h} = \sum_{i} \eta_{i}^{\dagger}(\partial_{\tau} - \mu - i a_{i\tau}^{3}\tau_{3})\eta_{i} + t_{r} \sum_{ij} ( \eta_{i}^{\dagger}U_{ij}^{SF} e^{ia_{ij}^{3}\tau_{3}} e^{ic_{ij}\tau_{3}}\eta_{j} + H.c.) \nn && - i \sum_{i} c_{i0}\Bigl( \eta_{i}^{\dagger} \tau_{3} \eta_{i} - \frac{1}{2\Theta} (\partial_{x}c_{y} - \partial_{y}c_{x})_{i} \Bigr) , \eqa where a bosonic field variable $h_{i}$ now becomes a fermionic one $\eta_{i} = \left( \begin{array}{c} \eta_{i1} \\ \eta_{i2} \end{array}\right)$ with $\Theta = \pi$. It is important to notice that our flux attachment is performed in an opposite way for each isospin sector, confirmed by the presence of $\tau_{3}$ in $2 \Theta (\eta_{i}^{\dagger} \tau_{3} \eta_{i}) = \partial_{x}c_{y} - \partial_{y}c_{x}$. As a result, there is no net flux in the mean-field approximation of this construction, considering that the density of $b_{i1}$ bosons is the same as that of $b_{i2}$ bosons in the staggered flux phase. This observation is interesting since it suggests a connection with an SU(2) slave-fermion representation \cite{KS_MD_SC}. If $a_{ij}^{3}$ is shifted to $a_{ij}^{3} - c_{ij}$, the Chern-Simons flux is transferred to spinons, turning their statistics into bosons. Then, we have bosonic spinons with fermionic holons, nothing but the slave-fermion representation.

\begin{figure}[htp]
\centerline{\includegraphics[scale=0.6]{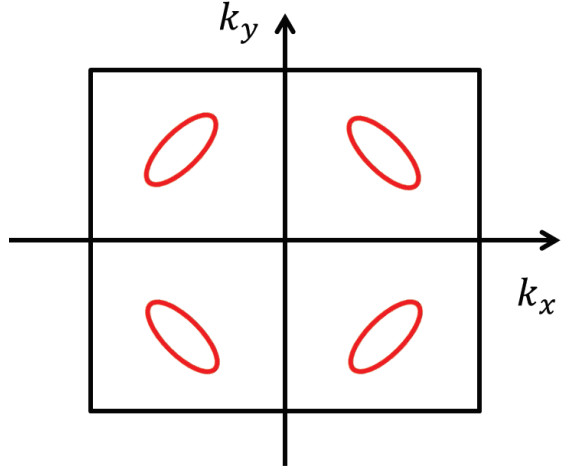}}
\caption{ (Color online) Fermi pockets near each Dirac node. This is consistent with the observed Fermi surface in the pseudogap phase of high T$_{c}$ cuprates.} \label{Fermi_Surface_Pseudogap}
\end{figure}

Following the strategy of the half-filled case, we find an effective continuum field theory \cite{Kim_SO5_WZW_mu_QED3} \bqa && Z_{eff} = \int D \psi D \eta D a_{\mu}^{3} D c_{\mu} e^{- \int_{0}^{\beta} d \tau \int d^{2} \bm{r} {\cal L}_{eff}} , \nn && {\cal L}_{eff} = \bar{\psi}\gamma_{\mu}(\partial_{\mu} - ia_{\mu}^{3}\tau_{3})\psi + \frac{1}{2e^{2}} (\epsilon_{\mu\nu\gamma}\partial_{\nu}a_{\gamma}^{3})^{2} \nn && + \bar{\eta}\gamma_{\mu}(\partial_{\mu} - ia_{\mu}^{3}\tau_{3} - i c_{\mu}\tau_{3})\eta - \mu_{h}\bar{\eta}\gamma_{0}\eta + \frac{i}{4\Theta} c_{\mu}\epsilon_{\mu\nu\lambda}\partial_{\nu}c_{\lambda} , \eqa where the Maxwell dynamics of $a_{\mu}^{3}$ is expected to appear from UV fluctuations of spinons. The Dirac structure results from the staggered flux ansatz, where both $\psi$ and $\eta$ are eight-component Dirac spinors and Dirac gamma matrices are $\gamma_{0} = \left( \begin{array}{cc} \sigma_{3} & 0 \\ 0 & - \sigma_{3} \end{array}\right)$, $\gamma_{1} = \left( \begin{array}{cc} \sigma_{1} & 0 \\ 0 & - \sigma_{1} \end{array}\right)$, and $\gamma_{2} = \left( \begin{array}{cc} \sigma_{2} & 0 \\ 0 & - \sigma_{2} \end{array}\right)$, the same as the half-filled case. It is important to understand that spinons are still at half filling even away from half filling in the SU(2) formulation \cite{Lee_Nagaosa_Wen_SL_Review}. The single-occupancy constraint in the SU(2) slave-boson representation is $f_{i1}^{\dagger}f_{i1} + f_{i2}^{\dagger}f_{i2} + b_{i1}^{\dagger}b_{i1} - b_{i2}^{\dagger}b_{i2} = 1$. Thus, if the condition of $\langle b_{i1}^{\dagger}b_{i1} \rangle = \langle b_{i2}^{\dagger}b_{i2} \rangle = \frac{\delta}{2}$ with hole concentration $\delta$ is satisfied, we see $\langle f_{i1}^{\dagger}f_{i1} + f_{i2}^{\dagger}f_{i2} \rangle = 1$, i.e., spinons are at half filling. As a result, a chemical potential term does not arise in the spinon sector. On the other hand, a chemical potential term appears in the holon sector, allowing four Fermi pockets around each Dirac node, consistent with the observed Fermi surface \cite{ARPES_Review} in the pseudogap phase of high T$_{c}$ cuprates. See Fig. \ref{Fermi_Surface_Pseudogap}.

%%%%%%%%%%%%%%%%%%%%%%%%%%%%%%%%%%%%%%%%%%%%%%%%%%%%%%%%%%%%%%%%%%%%%%%%
%%%%%%%%%%%%%%%%%%%%%%%%%%%%%%%%%%%%%%%%%%%%%%%%%%%%%%%%%%%%%%%%%%%%%%%%

Spontaneous chiral symmetry breaking in this $\mu-$QED$_{3}$ can be investigated, taking into account an emergent enhanced symmetry as the half-filled case of the $\pi-$flux state. It turns out that the group structure of $G = G_{Dirac} \otimes G_{gauge} \otimes G_{spin}$ enjoys SU(4) symmetry \cite{ASL_Mother}, allowing fifteen generators which correspond to $\bm{I} \otimes \bm{I} \otimes \bm{\sigma}$, $\bm{\gamma}_{3} \otimes \bm{I} \otimes \bm{\sigma}$, $\bm{\gamma}_{5} \otimes \bm{I} \otimes \bm{\sigma}$, $i\bm{\gamma}_{3}\bm{\gamma}_{5} \otimes \bm{I} \otimes \bm{I}$, $\bm{\gamma}_{3}\otimes \bm{\tau}_{3} \otimes \bm{I}$, $\bm{\gamma}_{5} \otimes \bm{\tau}_{3} \otimes \bm{I}$, and $i\bm{\gamma}_{3}\bm{\gamma}_{5} \otimes \bm{\tau}_{3} \otimes \bm{\sigma}$ \cite{Wen_Symmetry}. There exist additional five generators in addition to the first ten generators of the SO(5) symmetry, satisfying SU(4) algebra. A novel spin-liquid fixed point has been proposed that such an SU(4) symmetry is broken down to SO(5) \cite{Sachdev_SO5}, where most relevant spin fluctuations are Neel vector and valence bond fluctuations, giving rise to the competition between them. Such spin fluctuations are symmetry equivalent operators via chiral rotation at this emergent novel fixed point. As a result, one is allowed to construct the following effective field theory \cite{Kim_SO5_WZW_mu_QED3} \bqa && Z_{eff} = \int D \Psi D \eta D v_{i} D a_{\mu}^{3} D c_{\mu} \exp\Bigl[ - \int_{0}^{\beta} d \tau \Bigl\{ \bar{\Psi}\gamma_{\mu}(\partial_{\mu} - ia_{\mu}^{3}\tau_{3})\Psi - m \bar{\Psi} (\vec{v}\cdot\vec{\Gamma}) \Psi \nn && + \bar{\eta}\gamma_{\mu}(\partial_{\mu} - ia_{\mu}^{3}\tau_{3} - i c_{\mu}\tau_{3})\eta - \mu_{h}\bar{\eta}\gamma_{0}\eta - m_{\eta} \bar{\eta}( i \gamma_{3} v_{4} + i \gamma_{5} v_{5} ) \eta \nn && + \frac{1}{2e^{2}}(\epsilon_{\mu\nu\gamma}\partial_{\nu}a_{\gamma}^{3})^{2} + \frac{i}{4\Theta} c_{\mu}\epsilon_{\mu\nu\lambda}\partial_{\nu}c_{\lambda} \Bigr\} \Bigr] , \eqa where $\Psi =  \left( \begin{array}{c} \psi \\ \hat{\psi} \end{array}\right)$ of sixteen-components with a Dirac spinor $\hat{\psi} \equiv i\tau_{2}\psi^{*} = \left( \begin{array}{c} f_{\downarrow} \\ - f_{\uparrow}^{\dagger} \end{array}\right)$ has been introduced for the SO(5) superspin vector field. We point out that dynamics of doped holes couples to valence bond fluctuations in the form of Yukawa coupling since they do not carry spin degrees of freedom. Valence bond fluctuations may be responsible for high T$_{c}$ superconductivity in this formulation.

Integrating over massive fermion excitations and performing the gradient expansion for the superspin field, one finds an effective field theory, composed of $\mu-$QED$_{3}$ coupled to SO(5) WZW theory, \bqa && Z_{eff} = \int D v_{i} D \eta D a_{\mu}^{3} D c_{\mu} e^{ - S_{eff}} , \nn && S_{eff} = \int {d^{3} x} \Bigl\{ \frac{1}{2g} (\partial_{\mu} v_{k})^{2} - m_{\eta} \bar{\eta}( i \gamma_{3} v_{4} + i \gamma_{5} v_{5} ) \eta \Bigr\} + S_{WZW} \nn && + \int {d^{3} x} \Bigl\{ \bar{\eta}\gamma_{\mu}(\partial_{\mu} - ia_{\mu}^{3}\tau_{3} - i c_{\mu}\tau_{3} - iA_{\mu})\eta - \mu_{h}\bar{\eta}\gamma_{0}\eta + \frac{i}{4\Theta} c_{\mu}\epsilon_{\mu\nu\lambda}\partial_{\nu}c_{\lambda} + \frac{1}{2e^{2}}(\epsilon_{\mu\nu\gamma}\partial_{\nu}a_{\gamma}^{3})^{2} \Bigr\} , \nn && S_{WZW} = i \frac{2\pi}{Area(S^{4})} \int_{0}^{1}{dt} \int {d^{3} x} \epsilon_{abcde} v_{a} \partial_{t} v_{b} \partial_{\tau} v_{c} \partial_{x} v_{d} \partial_{y} v_{e} . \eqa An important observation is that the Chern-Simons contribution becomes irrelevant if the holon dynamics is in a critical phase. Shifting the slave-boson gauge field as $a_{\mu}^{3} - c_{\mu}$ and performing the integration of Chern-Simons gauge fields, we obtain $\sim (\partial\times\partial \times a^{3})\cdot(\partial\times a^{3})$. This contribution is irrelevant since it has a high scaling dimension owing to the presence of an additional derivative. Considering that the density of holons is finite to allow Fermi surfaces (pockets around Dirac points), it is natural to assume that the fermion sector lies at quantum criticality. As a result, we find an effective field theory for antiferromagnetic doped Mott insulators \bqa && S_{eff} = \int {d^{3} x} \Bigl\{ \frac{1}{2g} (\partial_{\mu} v_{k})^{2} - m_{\eta} \bar{\eta}( i \gamma_{3} v_{4} + i \gamma_{5} v_{5} ) \eta \Bigr\} + S_{WZW} \nn && + \int {d^{3} x} \Bigl\{ \bar{\eta}\gamma_{\mu}(\partial_{\mu} - ia_{\mu}^{3}\tau_{3} - iA_{\mu})\eta - \mu_{h}\bar{\eta}\gamma_{0}\eta + \frac{1}{2e^{2}} (\epsilon_{\mu\nu\gamma} \partial_{\nu} a_{\gamma}^{3})^{2} \Bigr\} , \label{Mu_QED3_SO5_WZW} \eqa which describes mutual effects on valence bond fluctuations and charge dynamics in the presence of the topological term.

We would like to emphasize that this field theoretic formulation makes effective interactions between doped holes and valence bond fluctuations explicit, allowing us to perform the perturbative analysis. If we do not take into account the valence bond order parameter explicitly, i.e., resorting to the SO(3) nonlinear $\sigma-$model description instead of the SO(5) WZW theory, we may obtain the following effective field theory for dynamics of doped holes, which scatter with magnetic monopole excitations as follows \bqa && Z_{eff} = \sum_{N_{t} = N + \bar{N} \in even} \frac{N ! \bar{N} !}{N_{t} !} y^{N+\bar{N}} \int D \mathcal{M} D \bm{n} D \chi D a_{\mu}^{3} \nn && \exp\Bigl[ - \int {d^{3} x} \Bigl\{ \frac{1}{2g} (\partial_{\mu} \bm{n})^{2} + m^{2} (|\bm{n}|^{2} - 1) + \bar{\chi} \gamma_{\mu} (\partial_{\mu} - i a_{\mu}^{3} \tau_{3} - iA_{\mu}) \chi - \mu_{h} \bar{\chi} \gamma_{0} \chi \nn && + \frac{1}{2e^{2}} (\epsilon_{\mu\nu\gamma} \partial_{\nu} a_{\gamma}^{3})^{2} \Bigr\} - \mathcal{S}_{m}[\mathcal{M};\Theta] - \mathcal{S}_{eff}^{\chi-m}[\chi, \mathcal{M}] - \mathcal{S}_{eff}^{n-m}[\bm{n}, \mathcal{M}] \Bigr] . \label{Mu_QED3_NLsM_Instanton} \eqa Since most mathematical symbols have been explained in the footnote for the SO(3) nonlinear $\sigma-$model, we do not repeat them here, where instanton excitations are identified with magnetic monopole fluctuations. $\chi$ is a Dirac spinor to represent doped holes, where such a field variable becomes modified from $\eta$ due to scattering with a pair of monopole and anti-monopole. An important point in this effective field theory is the scattering term between instanton fluctuations and doped holes, where such topological excitations carry the quantum number of valence bond ordering, given by the Berry-phase term. However, it is not straightforward to derive such effective interactions at all. On the other hand, they are incorporated by \bqa && \mathcal{S}_{int} = - \int {d^{3} x} m_{\eta} \bar{\eta}( i \gamma_{3} v_{4} + i \gamma_{5} v_{5} ) \eta \eqa explicitly in the SO(5) WZW theoretical formulation. It would be quite appealing to find any signatures of the effective field theory Eq. (\ref{Mu_QED3_SO5_WZW}) from Eq. (\ref{Mu_QED3_NLsM_Instanton}).

\subsubsection{Application to one dimension: Luther-Emery phase}

It is interesting to apply the $\mu-$QED$_{3}$ coupled to the SO(5) WZW theory to one dimension. In one dimension the spin sector is described by the SO(4) WZW theory, and the charge sector is represented by QED$_{2}$ without the chemical potential term \cite{Luttinger_Liquid_Textbook}. Accordingly, the coupling term between valence bond fluctuations and holons is adjusted. The resulting effective field theory is given by \cite{Kim_SO5_WZW_mu_QED3} \bqa && S = \int {d^{2} x} \Bigl\{ \frac{1}{2g} \sum_{k=1}^{4}(\partial_{\mu} v_{k})^{2} + i \frac{2\pi}{Area(S^{3})} \int_{0}^{1}{dt} \epsilon_{abcd} v_{a} \partial_{t} v_{b} \partial_{\tau} v_{c} \partial_{x} v_{d} \Bigr\} \nn && + \int {d^{2} x} \Bigl\{ \bar{\eta}\gamma_{\mu}(\partial_{\mu} - ia_{\mu}^{3}\tau_{3} - iA_{\mu})\eta - m_{\eta} \bar{\eta} (i \gamma_{5} v_{4} ) \eta + \frac{1}{2e^{2}}(\epsilon_{3\mu\nu}\partial_{\mu}a_{\nu}^{3})^{2} \Bigr\} , \eqa where $\bm{v} = (n_{x}, n_{y}, n_{z}, v_{x})$ is a four-component superspin vector field with three Neel components and one valence bond order parameter and $\eta$ is a four-component Dirac spinor with two-by-two Dirac matrices of $\gamma_{\mu}$ and $\gamma_{5}$. The physical origin of the SO(4) WZW term has been attributed to non-abelian chiral anomaly within the path integral formulation, where classically conserved non-abelian chiral currents with Pauli spin matrices turn out to be not preserved in a background with a nontrivial topology \cite{Tanaka_SO4}.

Performing the abelian bosonization for the fermion sector \cite{Luttinger_Liquid_Textbook}, we obtain the following expression \bqa && S = \int {d^{2} x} \Bigl\{ \frac{1}{2g} \sum_{k=1}^{4}(\partial_{\mu} v_{k})^{2} + i \frac{2\pi}{Area(S^{3})} \int_{0}^{1}{dt} \epsilon_{abcd} v_{a} \partial_{t} v_{b} \partial_{\tau} v_{c} \partial_{x} v_{d} \Bigr\} \nn && + \int {d^{2} x} \Bigl\{ \frac{1}{2}(\partial_{\mu}\phi_{+})^{2} + \frac{1}{2}(\partial_{\mu}\phi_{-})^{2} + \Bigl( \frac{\Lambda}{\pi} m_{\eta} \Bigr) v_{4} \sin \Bigl( \sqrt{4\pi}\phi_{+} \Bigr) + \Bigl( \frac{\Lambda}{\pi} m_{\eta} \Bigr) v_{4} \sin \Bigl( \sqrt{4\pi}\phi_{-} \Bigr) \nn && - ia_{\mu}^{3} \Bigl(\frac{1}{2\pi}\epsilon_{\mu\nu}\partial_{\nu}\phi_{+} - \frac{1}{2\pi}\epsilon_{\mu\nu}\partial_{\nu}\phi_{-}\Bigr) - iA_{\mu}\Bigl(\frac{1}{2\pi}\epsilon_{\mu\nu}\partial_{\nu}\phi_{+} + \frac{1}{2\pi}\epsilon_{\mu\nu}\partial_{\nu}\phi_{-}\Bigr) + \frac{1}{2e^{2}}(\epsilon_{\mu\nu}\partial_{\mu}a_{\nu}^{3})^{2} \Bigr\} , \nn \eqa where the subscript $\pm$ in the bosonic field $\phi_{\pm}$ represents the SU(2) doublet involved with $\tau_{3}$, and $\Lambda$ is a cutoff associated with band linearization. Performing integration for U(1) gauge fields, we find a mass-type term $\frac{e^{2}}{8\pi^{2}}(\phi_{+} - \phi_{-})^{2}$. This allows us to set $\phi_{+} = \phi_{-} \equiv \phi$ in the low energy limit. Shifting $\sqrt{4\pi} \phi$ with $- \frac{\pi}{2} + \sqrt{4\pi} \theta$, we are led to \bqa && S = \int {d^{2} x} \Bigl\{ \frac{1}{2g} \sum_{k=1}^{4}(\partial_{\mu} v_{k})^{2} + i \frac{2\pi}{Area(S^{3})} \int_{0}^{1}{dt} \epsilon_{abcd} v_{a} \partial_{t} v_{b} \partial_{\tau} v_{c} \partial_{x} v_{d} \Bigr\} \nn && + \int {d^{2} x} \Bigl\{ (\partial_{\mu}\theta)^{2} - \Bigl( \frac{2\Lambda}{\pi} m_{\eta} \Bigr) v_{4} \cos \Bigl( \sqrt{4\pi}\theta \Bigr) - iA_{\mu}\Bigl(\frac{1}{\pi}\epsilon_{\mu\nu}\partial_{\nu}\theta \Bigr) \Bigr\} . \eqa It is interesting to see that valence bond excitations drive charge density-wave fluctuations, consistent with our expectation. The valence bond and charge density-wave coupling term can be taken into account in the cumulant expansion, given by \bqa && \delta S = - \frac{1}{2} \Bigl( \langle S_{int}^{2} \rangle - \langle S_{int} \rangle^{2} \Bigr) \nn && = - \frac{1}{2} \Bigl( \frac{2\Lambda}{\pi} m_{\eta} \Bigr)^{2} \int d^{2}x \int d^{2}x' \Bigl\{ v_{4}(x) \langle \cos \Bigl( \sqrt{4\pi}\theta(x) \Bigr) \cos \Bigl( \sqrt{4\pi}\theta(x') \Bigr) \rangle v_{4}(x') \nn && + \cos \Bigl( \sqrt{4\pi}\theta(x) \Bigr) \langle v_{4}(x)v_{4}(x') \rangle \cos \Bigl( \sqrt{4\pi}\theta(x') \Bigr) \Bigr\} \equiv \delta S_{v_{4}} + \delta S_{\theta} , \eqa where $S_{int} = - \int d^{2} x \Bigl( \frac{2\Lambda}{\pi} m_{\eta} \Bigr) v_{4} \cos \Bigl( \sqrt{4\pi}\theta \Bigr)$ is the coupling term.

It is not difficult to evaluate the density-density correlation function since charge fluctuations are described by the noninteracting Gaussian ensemble if metallic charge dynamics is assumed. In this case we find \bqa && \langle \cos \Bigl( \sqrt{4\pi}\theta(x) \Bigr) \cos \Bigl( \sqrt{4\pi}\theta(x') \Bigr) \rangle \propto \cosh \Bigl( 4\pi \langle \theta(x) \theta(x') \rangle \Bigr) \nn && = \cosh \Bigl( 4\pi \mathcal{C}_{\theta} \ln |x - x'| \Bigr) \rightarrow |x - x'|^{4\pi\mathcal{C}_{\theta}} , \eqa where $\mathcal{C}_{\theta}$ is a positive numerical constant, and the last part is valid at large distances, i.e., $|x-x'|\rightarrow\infty$. Inserting this expression into the spin sector, we obtain an effective theory for SO(4) spin fluctuations \bqa && S_{v_{4}} = \int {d^{2} x} \Bigl[ \frac{1}{2g} \sum_{k=1}^{4}(\partial_{\mu} v_{k})^{2} + i \frac{2\pi}{Area(S^{3})} \int_{0}^{1}{dt} \epsilon_{abcd} v_{a} \partial_{t} v_{b} \partial_{\tau} v_{c} \partial_{x} v_{d} \Bigr] \nn && - \int d^{2}x \int d^{2}x' \mathcal{C}_{v_{4}} \Bigl( \frac{2\Lambda}{\pi} m_{\eta} \Bigr)^{2} v_{4}(x) |x - x'|^{4\pi\mathcal{C}_{\theta}} v_{4}(x') , \nn \eqa where $\mathcal{C}_{v_{4}}$ is a positive numerical constant. An important point is that metallic charge fluctuations give rise to confining interactions between skyrmions, suppressing such topological fluctuations. Then, the skyrmion and anti-skyrmion pair configuration may not detect the WZW term since the skyrmion pair is inert ``magnetically". If the WZW term becomes irrelevant due to the confining interaction, the SO(4) nonlinear $\sigma-$model becomes gapped, which gives rise to deconfined gapped spinon excitations due to the confinement of skyrmion excitations \cite{Witten_QED2_NLsM,Shankar_SF_tJ}. This situation differs from the half-filled case allowing critical spinon excitations, which arises from the WZW term. The emergence of a spin-gapped state seems to be consistent with our physical intuition that charge fluctuations will cut spin correlations, making their correlation length short. Then, the resulting state may be identified with the Luther-Emery phase, where spin fluctuations are gapped while charge excitations exhibit enhanced superconducting correlations \cite{Luttinger_Liquid_Textbook}.

On the other hand, if charge fluctuations are gapped, i.e., in an insulating phase, their density-density correlations will vanish at large distances as follows, $\langle \cos \Bigl( \sqrt{4\pi}\theta(x) \Bigr) \cos \Bigl( \sqrt{4\pi}\theta(x') \Bigr) \rangle \propto e^{- |x-x'|/\xi_{\eta}}$, where $\xi_{\eta}^{-1}$ is associated with their excitation gap. Then, spin dynamics will be described by the pure SO(4) WZW theory in the long wave-length limit. As a result, a critical spin-liquid Mott insulator is expected to appear in this case. We suspect that localization of doped holes may be realized in the presence of disorder \cite{Tanaka_SO4}.

\subsubsection{Non-Fermi liquid physics in antiferromagnetic doped Mott insulators}

An essential point in this review article is how to incorporate nonperturbative physics within the perturbative framework. As discussed in section 2.1.1 for the SO(5) WZW theory, an effective field theory with an enhanced symmetry may encode such nonperturbative physics in the perturbation analysis, where instanton excitations are identified with valence bond fluctuations. In this section we perform the perturbative analysis for the scattering problem between doped holes and valence bond fluctuations \cite{Kim_SO5_WZW_mu_QED3}.

\begin{figure}[htp]
\centerline{\includegraphics[scale=0.6]{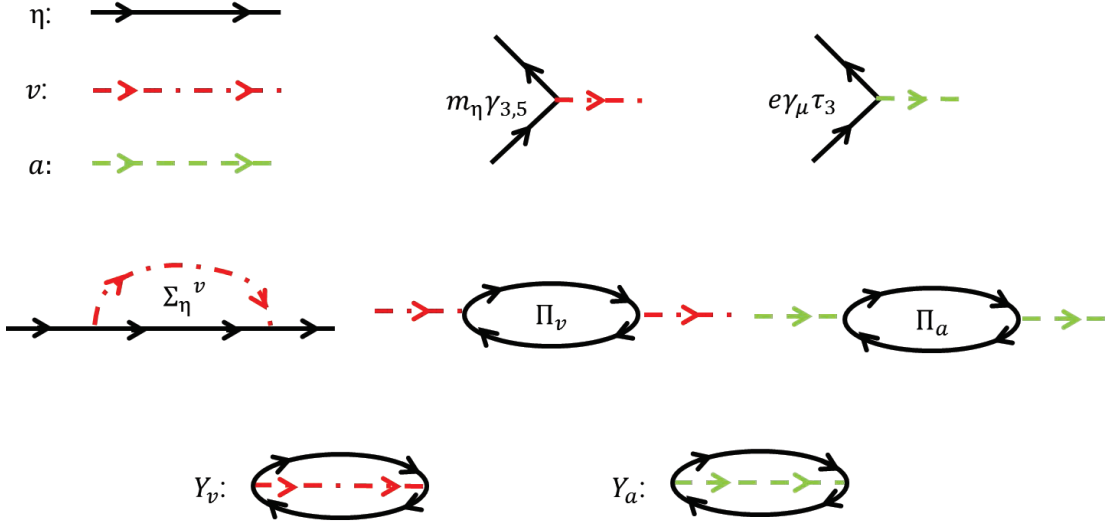}}
\caption{ (Color online) Feynman diagrams in the Eliashberg approximation. We have two kinds of interaction vertices: One represents scattering between doped holes and valence bond fluctuations and the other describes that between doped holes and U(1) gauge fluctuations. These interaction vertices give rise to self-energy corrections for holons, valence bond fluctuations, and U(1) gauge fields, respectively, introduced self-consistently via the Luttinger-Ward functional approach.} \label{Feynman_Diagrams}
\end{figure}

Performing the cumulant expansion of $\mathcal{S}_{int} = - \frac{1}{2} \Bigl(\langle S_{int}^{2} \rangle - \langle S_{int} \rangle^{2} \Bigr)$ for $S_{int} = \int{d^{3}x} \Bigl\{- m_{\eta} \bar{\eta}( i \gamma_{3} v_{4} + i \gamma_{5} v_{5} )\eta - i a_{\mu}^{3}\bar{\eta}\gamma_{\mu} \tau_{3} \eta \Bigr\}$, we derive the Luttinger-Ward functional \cite{Luttinger_Ward_Functional} in the Eliashberg framework \cite{Luttinger_Ward_Functional_Kim_Pepin}, \bqa && F_{LW} = F_{LW}^{\eta} + F_{LW}^{v} + F_{LW}^{a} + Y_{v} + Y_{a} , \nn && F_{LW}^{\eta} = - T \sum_{i\omega} \int\frac{d^{d}k}{(2\pi)^{d}} \mathbf{tr} \Bigl[\ln\Bigl\{ g_{\eta}^{-1}(k,i\omega) + \Sigma_{\eta}(i\omega) \Bigr\} - \Sigma_{\eta}(i\omega) G_{\eta}(k,i\omega) \Bigr] , \nn && F_{LW}^{v} = T \sum_{i\Omega} \int\frac{d^{d}q}{(2\pi)^{d}} \Bigl[ \mathbf{tr} \ln\Bigl\{ d_{v}^{-1}(q,i\Omega)\delta_{mn} + \Pi_{v}^{mn}(q,i\Omega) \Bigr\} - \sum_{m,n=1}^{5} \Pi_{v}^{mn}(q,i\Omega)D_{v}^{mn}(q,i\Omega)\Bigr] - m_{v}^{2} , \nn && F_{LW}^{a} = T \sum_{i\Omega} \int\frac{d^{d}q}{(2\pi)^{d}} \Bigl[\ln\Bigl\{ d_{a}^{-1}(q,i\Omega) + \Pi_{a}(q,i\Omega) \Bigr\} - \Pi_{a}(q,i\Omega) D_{a}(q,i\Omega) \Bigr] , \nn && Y_{v} = - \frac{m_{\eta}^{2}}{2} T \sum_{i\omega} \int\frac{d^{d}k}{(2\pi)^{d}} T \sum_{i\Omega} \int\frac{d^{d}q}{(2\pi)^{d}} \sum_{m,n=4}^{5} \mathbf{tr}[ D_{v}^{mn}(q,i\Omega) \gamma_{m} G_{\eta}(k+q,i\omega+i\Omega) \gamma_{n}G_{\eta}(k,i\omega)] , \nn && Y_{a} = - \frac{1}{2} T \sum_{i\omega} \int\frac{d^{d}k}{(2\pi)^{d}} T \sum_{i\Omega} \int\frac{d^{d}q}{(2\pi)^{d}} \mathbf{tr} [ D_{a}^{\mu\nu}(q,i\Omega) \gamma_{\mu}\tau_{3} G_{\eta}(k+q,i\omega+i\Omega) \gamma_{\nu}\tau_{3}G_{\eta}(k,i\omega)] , \eqa where self-energy corrections are incorporated self-consistently but vertex corrections are not introduced. $G_{\eta}(k,i\omega)$ in $F_{LW}^{\eta}$ is a renormalized propagator for holons, given by $G_{\eta}(k,i\omega) = \Bigl\{g_{\eta}^{-1}(k,i\omega) + \Sigma_{\eta}(i\omega) \Bigr\}^{-1}$, where $g_{\eta}(k,i\omega) = \Bigl( i \gamma_{0} \omega + i \gamma_{i} k_{i} + \mu_{h} \gamma_{0} \Bigr)^{-1}$ is its bare propagator, and $\Sigma_{\eta}(i\omega)$ is its momentum-independent self-energy. $D_{v}^{mn}(q,i\Omega)$ in $F_{LW}^{v}$ is a renormalized propagator for superspin vector fields, given by $D_{v}^{mn}(q,i\Omega) = \Bigl\{d_{v}^{-1}(q,i\Omega)\delta_{mn} + \Pi_{v}^{mn}(q,i\Omega)\Bigr\}^{-1}$, where $d_{v}(q,i\Omega) = \Bigl( \frac{q^{2} + \Omega^{2}}{2g} + m_{v}^{2} \Bigr)^{-1}$ is its bare propagator, and $\Pi_{v}^{mn}(q,i\Omega)$ is its self-energy. $D_{a}(q,i\Omega)$ in $F_{LW}^{a}$ is a renormalized kernel for the gauge propagator $D_{a}^{\mu\nu}(q,i\Omega) = D_{a}(q,i\Omega)\Bigl( \delta_{\mu\nu} - \frac{q_{\mu}q_{\nu}}{q^{2}} \Bigr)$, given by $D_{a} (q,i\Omega) = \Bigl\{d_{a}^{-1}(q,i\Omega) + \Pi_{a}(q,i\Omega) \Bigr\}^{-1}$, where $d_{a}(q,i\Omega) = \Bigl( \frac{q^{2}+\Omega^{2}}{2e^{2}} \Bigr)^{-1}$ is its bare kernel, and $\Pi_{a}(q,i\Omega)$ is its self-energy in $\Pi_{a}^{\mu\nu}(q,i\Omega) = \Pi_{a}(q,i\Omega)\Bigl( \delta_{\mu\nu} - \frac{q_{\mu}q_{\nu}}{q^{2}} \Bigr)$. $Y_{v}$ is introduced for self-energy corrections resulting from effective interactions between holons and valence bond fluctuations while $Y_{a}$, arising from those between holons and U(1) gauge fluctuations. See Fig. \ref{Feynman_Diagrams}, which shows Feynman diagrams for self-energy corrections.

%%%%%%%%%%%%%%%%%%%%%%%%%%%%%%%%%%%%%%%%%%%%%%%%%%%%%%%%%%%%%%%%%%%%%%%%
%%%%%%%%%%%%%%%%%%%%%%%%%%%%%%%%%%%%%%%%%%%%%%%%%%%%%%%%%%%%%%%%%%%%%%%%

Performing variation for the Luttinger-Ward functional with respect to each self-energy, i.e., $\frac{\delta F_{LW}}{\delta \Sigma_{\eta}(i\omega)} = 0$, $\frac{\delta F_{LW}}{\delta \Pi_{v}^{mn}(q,i\Omega)} = 0$, and $\frac{\delta F_{LW}}{\delta \Pi_{a}^{\mu\nu}(q,i\Omega)} = 0$, we find self-consistent Eliashberg equations \bqa \Sigma_{\eta}(i\omega) &=& m_{\eta}^{2} T \sum_{i\Omega} \int \frac{d^{d}q}{(2\pi)^{d}} \sum_{m,n=4}^{5} D_{v}^{mn}(q,i\Omega) \gamma_{m} G_{\eta}(k_{F}+q,i\omega+i\Omega) \gamma_{n} \nn &+& T \sum_{i\Omega} \int \frac{d^{d}q}{(2\pi)^{d}} D_{a}^{\mu\nu}(q,i\Omega) \gamma_{\mu}\tau_{3} G_{\eta}(k_{F}+q,i\omega+i\Omega) \gamma_{\nu}\tau_{3} , \nn \Pi_{v}^{mn}(q,i\Omega) &=& T \sum_{i\omega}\int \frac{d^{d}k}{(2\pi)^{d}} \sum_{m,n=4}^{5} \Bigl( - \frac{m_{\eta}^{2}}{2} \mathbf{tr} [\gamma_{m}G_{\eta}(k+q,i\omega+i\Omega)\gamma_{n}G_{\eta}(k,i\omega)] \Bigr) , \nn \Pi_{a}^{\mu\nu}(q,i\Omega) &=& T \sum_{i\omega} \int\frac{d^{d}k}{(2\pi)^{d}} \Bigl( - \frac{1}{2} \mathbf{tr} [ \gamma_{\mu}\tau_{3} G_{\eta}(k+q,i\omega+i\Omega) \gamma_{\nu}\tau_{3}G_{\eta}(k,i\omega)] \Bigr) . \eqa The holon
self-energy results from both valance bond and gauge fluctuations, where $\gamma_{4} \rightarrow \gamma_{3}$ is performed in our convention. The superspin vector self-energy arises from holon fluctuations, where $\Pi_{v}^{mn}(q,i\Omega) = 0$ for $m, n = 1, 2, 3$. The gauge-field self-energy appears from holon current fluctuations. These self-consistent equations simplify the Luttinger-Ward functional as follows \bqa && F_{LW} = - T \sum_{i\omega} \int\frac{d^{d}k}{(2\pi)^{d}} \mathbf{tr} \Bigl[\ln\Bigl\{ g_{\eta}^{-1}(k,i\omega) + \Sigma_{\eta}(i\omega) \Bigr\} - \Sigma_{\eta}(i\omega) G_{\eta}(k,i\omega) \Bigr] \nn && + T \sum_{i\Omega} \int\frac{d^{d}q}{(2\pi)^{d}} \Bigl[ \sum_{m,n=1}^{5} \ln\Bigl\{ d_{v}^{-1}(q,i\Omega)\delta_{mn} + \Pi_{v}^{mn}(q,i\Omega) \Bigr\} \Bigr] - m_{v}^{2} \nn && + T \sum_{i\Omega} \int\frac{d^{d}q}{(2\pi)^{d}} \ln\Bigl\{ d_{a}^{-1}(q,i\Omega) + \Pi_{a}(q,i\Omega) \Bigr\} , \eqa where both self-energy parts for superspin vector and U(1) gauge fields are canceled. We see that the holon free energy is nothing but the free energy of Fermi liquid as follows \bqa && F_{LW}^{\eta} \approx - N_{\eta} T \sum_{i\omega} \int\frac{d^{d}k}{(2\pi)^{d}} \ln\Bigl\{ 2 \mu_{h}(i\omega) + \mu_{h}^{2} - \omega^{2} - k^{2} \Bigr\} \approx - \frac{\pi N_{\eta} \rho_{\eta}}{6} T^{2} = F_{FL}^{\eta} , \nonumber \eqa where $\rho_{\eta}$ is the density of states around the Dirac node, and $N_{\eta}$ is the number of Dirac nodes. As a result, we find the Eliashberg free energy of the effective field theory \bqa && F_{LW} = T \sum_{i\Omega} \int\frac{d^{d}q}{(2\pi)^{d}} \Bigl[ \sum_{m,n=1}^{5} \ln\Bigl\{ d_{v}^{-1}(q,i\Omega)\delta_{mn} + \Pi_{v}^{mn}(q,i\Omega) \Bigr\} \Bigr] \nn && + T \sum_{i\Omega} \int\frac{d^{d}q}{(2\pi)^{d}} \ln\Bigl\{ d_{a}^{-1}(q,i\Omega) + \Pi_{a}(q,i\Omega) \Bigr\} - \frac{\pi N_{\eta} \rho_{\eta}}{6} T^{2} - m_{v}^{2} . \eqa

It is straightforward to evaluate both self-energies of valence bond and gauge fluctuations, given by \bqa && \Pi_{v}(q,i\Omega) \approx \frac{\pi N_{\eta} m_{\eta}^{2}\rho_{\eta}}{4} \frac{|\Omega|}{q} , ~~~~~ \Pi_{a}(q,i\Omega) = \frac{\pi N_{\eta} \rho_{\eta}}{4} \frac{|\Omega|}{q} , \eqa respectively, where Landau damping occurs from fermion excitations near the Fermi surface due to hole doping \cite{Landau_Damping_QCP_Review}. Then, the renormalized propagator for superspin fluctuations and that for emergent U(1) gauge fields are given by \bqa && D_{v}^{mn}(q,i\Omega) = \frac{\delta_{mn}}{ \frac{q^{2} + \Omega^{2}}{2g} + m_{v}^{2} } , ~~~~~ \mbox{for} ~~~ m, n = 1, 2, 3 , \nn && D_{v}^{mn}(q,i\Omega) = \frac{\delta_{mn}}{ \frac{q^{2} + \Omega^{2}}{2g} + m_{v}^{2} + \frac{\pi N_{\eta} m_{\eta}^{2}\rho_{\eta}}{4} \frac{|\Omega|}{q} } \approx \frac{\delta_{mn}}{ \frac{q^{2}}{2g} + m_{v}^{2} + \frac{\pi N_{\eta} m_{\eta}^{2}\rho_{\eta}}{4} \frac{|\Omega|}{q} }  , ~~~~~ \mbox{for} ~~~ m, n = 4, 5 \nn \eqa and \bqa && D_{a}(q,i\Omega) \approx \frac{1}{\frac{q^{2}}{2e^{2}} + \frac{\pi N_{\eta} \rho_{\eta}}{4} \frac{|\Omega|}{q}} , \eqa respectively. Antiferromagnetic spin fluctuations are described by $z = 1$ theory while valence bond fluctuations are expressed by $z = 3$ theory, where $z$ is the dynamical exponent. Gauge fluctuations are described by $z = 3$ critical theory. Inserting these bosonic self-energies into the free energy, we reach the final expression for the Eliashberg free energy \bqa && F_{LW} = T \sum_{i\Omega} \int\frac{d^{d}q}{(2\pi)^{d}} \Bigl\{ 3 \ln\Bigl(q^{2} + \Omega^{2} + \xi^{-2}\Bigr) + 2 \ln \Bigl( q^{2} + \xi^{-2} + \gamma_{v} \frac{|\Omega|}{q} \Bigr) \Bigr\} \nn && + T \sum_{i\Omega} \int\frac{d^{d}q}{(2\pi)^{d}} \ln\Bigl( q^{2} + \gamma_{a} \frac{|\Omega|}{q} \Bigr) - \frac{\pi N_{\eta} \rho_{\eta}}{6} T^{2} - \frac{\xi^{-2}}{2g} , \eqa where \bqa && \xi^{-2} = 2g m_{v}^{2} , ~~~ \gamma_{v} = \frac{ \pi g N_{\eta} m_{\eta}^{2}\rho_{\eta}}{2} , ~~~ \gamma_{a} = \frac{ \pi e^{2} N_{\eta} \rho_{\eta}}{2} \nonumber \eqa represent the correlation length for superspin fluctuations, Landau damping coefficient for superspin fields, and that for gauge fields, respectively.

\begin{figure}[htp]
\centerline{\includegraphics[scale=0.6]{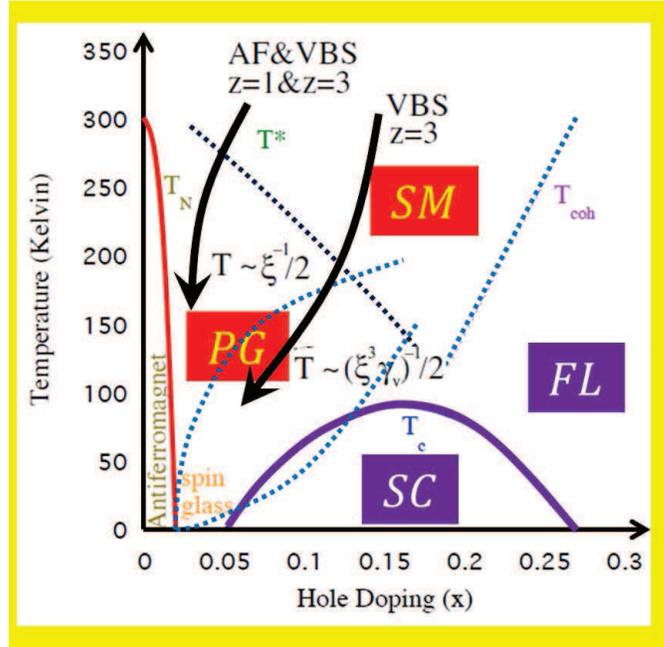}}
\caption{ (Color online) A schematic phase diagram of $\mu-$QED$_{3}$ with SO(5) WZW theory [Eq. (\ref{Mu_QED3_SO5_WZW})]. PG, SC, FL, and SM represent phases of pseudogap, superconductivity, Fermi-liquid, and strange metal, respectively. T$^{*}$ shows a pseudogap crossover temperature, and T$_{coh}$ identifies a Fermi-liquid temperature, also crossover. This phase diagram should be regarded as a part of Fig. \ref{Phase_Diagram_SO5_QED3_HighTc}, which lies on the yellow plane. When the antiferromagnetic order becomes destroyed via hole doping, there emerge competing fluctuations with antiferromagnetic spin excitations, identified with valence bond fluctuations associated with the SO(5) symmetry. Since only valence bond fluctuations are coupled to dynamics of doped holes, their dynamics is rather dissipative, characterized by the dynamical critical exponent $z = 3$, while dynamics of antiferromagnetic spin fluctuations remains to be $z = 1$. The coexistence of $z = 3$ and $z = 1$ collective excitations gives rise to three regimes inside the PG phase near the antiferromagnetic quantum critical point. It is quite interesting to observe that a large portion of the PG phase just above the SC state is governed by $z = 3$ critical valence bond fluctuations, regarded to be a characteristic feature of the $\mu-$QED$_{3}$ with SO(5) WZW theory.} \label{Phase_Diagram_QED_SO5}
\end{figure}

Performing the variation for this free energy with respect to the correlation length, i.e., $\frac{\partial F_{LW}}{\partial \xi^{-2}} = 0$, we obtain the self-consistent equation for the correlation length in the Eliashberg framework \bqa && 1 = 2g T \sum_{i\Omega} \int\frac{d^{d}q}{(2\pi)^{d}} \Bigl( \frac{3 }{ q^{2} + \Omega^{2} + \xi^{-2} } + \frac{2}{ q^{2} + \xi^{-2} + \gamma_{v} \frac{|\Omega|}{q} }\Bigr) . \eqa Notice that interactions between valence bond fluctuations and holons result in the $z=3$ part. Analyzing this self-consistent equation, we find that there exist three regimes, (A) $T < \frac{(\xi^{3}\gamma_{v})^{-1}}{2} < \frac{\xi^{-1}}{2}$, (B) $\frac{(\xi^{3}\gamma_{v})^{-1}}{2} < T < \frac{\xi^{-1}}{2}$, and (C) $\frac{(\xi^{3}\gamma_{v})^{-1}}{2} < \frac{\xi^{-1}}{2} < T$, emerging from the coexistence of $z = 1$ (antiferromagnetic) and $z = 3$ (valence bond) fluctuations. In regime (A) both $z = 1$ and $z = 3$ fluctuations are gapped while in regime (C) both spin fluctuations are critical, that is, in the quantum critical regime. In regime (B) only valence bond fluctuations ($z = 3$) are critical, and $z = 1$ antiferromagnetic ones are gapped. The phase diagram is shown in Fig. \ref{Phase_Diagram_QED_SO5}, lying in the yellow plane of Fig. \ref{Phase_Diagram_SO5_QED3_HighTc}.

%%%%%%%%%%%%%%%%%%%%%%%%%%%%%%%%%%%%%%%%%%%%%%%%%%%%%%%%%%%%%%%%%%%%%%%%
%%%%%%%%%%%%%%%%%%%%%%%%%%%%%%%%%%%%%%%%%%%%%%%%%%%%%%%%%%%%%%%%%%%%%%%%

Considering $(\xi^{3}\gamma_{v})^{-1} \ll \xi^{-1}$ near the quantum critical point, we find that antiferromagnetic fluctuations play an important role in determining the SO(5) superspin correlation length near the quantum critical point. As a result, we obtain the following expression for the correlation length \bqa \xi^{-1}\Bigl(g \sim g_{c}; T < \frac{\xi^{-1}}{2}\Bigr) && \approx \Bigl\{ \Bigl( 1 + \frac{2\pi}{9} \Bigr) \Lambda - \frac{2\pi}{3g} \Bigr\} \equiv \frac{2\pi}{3} \Bigl( \frac{1}{g_{c}} - \frac{1}{g} \Bigr) , \nn \xi^{-1} \Bigl(g \sim g_{c}; T > \frac{\xi^{-1}}{2}\Bigr) && = 2 T e^{\frac{\pi}{3} \frac{\Bigl( \frac{1}{g_{c}} - \frac{1}{g} \Bigr)}{T}} , \label{Correlation_Length} \eqa where $g_{c} = \Bigl( \frac{3}{2\pi} + \frac{1}{3} \Bigr) \Lambda$ is the quantum critical point. Recalling $g(\delta) \propto m^{2}(\delta)$ in the gradient expansion and $m^{2}(\delta) \propto |\delta - \delta_{0}|$ with hole concentration $\delta < \delta_{0}$ near the quantum critical point, we find the critical hole concentration in the Eliashberg framework \bqa && \delta_{c} = \delta_{0} - \frac{1}{c} \Bigl\{\Bigl( \frac{3}{2\pi} + \frac{1}{3} \Bigr) \Lambda\Bigr\} , \eqa where $c$ is a positive numerical constant. Thus, an antiferromagnetically ordered phase appears in $\delta < \delta_{c}$, and a quantum disordered state arises in $\delta > \delta_{c}$, where the nature of the disordered state is determined by an SO(5) symmetry-breaking effective potential. These regimes given by Eq. (\ref{Correlation_Length}) are shown in Fig. \ref{Phase_Diagram_QED_SO5} by blue dotted lines inside the pseudogap phase.

The specific heat coefficient $\gamma(T) = \frac{C(T)}{T} = - \frac{\partial^{2} F_{LW}(T)}{\partial T^{2}}$ is given by \bqa && \gamma\Bigl(T > \frac{(\gamma_{v}\xi^{3})^{-1}}{2}\Bigr) \approx 2^{8/3} \Bigl( \frac{1}{2\pi^{2}} + \frac{1}{8\pi}\Bigr) \Bigl( 2 \gamma_{v}^{2/3} + \gamma_{a}^{2/3} \Bigr) \Bigl[ \int_{0}^{\infty} {d y} \Bigl\{ - \frac{y^{5/3}}{\sinh^{2}y} + \frac{y^{8/3} \coth y}{\sinh^{2} y} \Bigr\} \Bigr] T^{-1/3} , \nn && \gamma\Bigl(T < \frac{(\gamma_{v}\xi^{3})^{-1}}{2}\Bigr) \approx 2^{8/3} \Bigl( \frac{1}{2\pi^{2}} + \frac{1}{8\pi}\Bigr) \gamma_{a}^{2/3} \Bigl[ \int_{0}^{\infty} {d y} \Bigl\{ - \frac{y^{5/3}}{\sinh^{2}y} + \frac{y^{8/3} \coth y}{\sinh^{2} y} \Bigr\} \Bigr] T^{-1/3} , \nn \eqa where only dominant contributions are shown. As discussed previously, we have three regimes, (A) $T < \frac{(\xi^{3}\gamma_{v})^{-1}}{2}$ where both antiferromagnetic and valence bond fluctuations are gapped, (B) $\frac{(\xi^{3}\gamma_{v})^{-1}}{2} < T < \frac{\xi^{-1}}{2}$ where only valence bond fluctuations are critical, and (C) $\frac{\xi^{-1}}{2} < T$ where both antiferromagnetic and valence bond fluctuations are critical. In the regime (A) contributions from superspin fluctuations exhibit an exponential dependence in temperature, thus ignored in the low energy limit. Dominant contributions are driven by $z = 3$ critical gauge fluctuations, resulting in $\gamma(T) \sim T^{-1/3}$ \cite{Z3_Gauge_Fields_Perturbation}. In the regime (B) antiferromagnetic fluctuations cause an exponential dependence in temperature while both valence bond and gauge fluctuations give rise to $\gamma(T) \sim T^{-1/3}$ due to their $z = 3$ criticality. In the regime (C) $z = 3$ critical valence bond excitations and gauge fluctuations allow $\gamma(T) \sim T^{-1/3}$ while $z = 1$ critical antiferromagnetic fluctuations result in $\gamma_{AF}(T>\frac{\xi^{-1}}{2}) = \frac{6}{\pi} \Bigl( \int_{0}^{\infty} d x \frac{x^{3}}{\sinh^{2} x}\Bigr) T$, sub-leading and ignored in the low energy limit.

The holon self-energy due to valence bond fluctuations is \bqa && \Sigma_{\eta}^{v}(i\omega) \approx - \frac{2 g m_{\eta}^{2} }{\mu_{h}} T \sum_{i\Omega} \int \frac{d^{d}q}{(2\pi)^{d}} \frac{1}{ q^{2} + \xi^{-2} + \gamma_{v} \frac{|\Omega|}{q} } \frac{i \gamma_{0} \omega + i \gamma_{i} k_{i}^{F} + \mu_{h} \gamma_{0}} { i \omega + i \Omega - q\cos\theta} , \eqa where $k_{F} = \sqrt{k_{x}^{F2} + k_{y}^{F2}} = \mu_{h}$ is the holon Fermi momentum. Then, the imaginary part of the self-energy is \bqa && \Im \Sigma_{\eta}^{v}(\omega+i\delta) = \frac{g m_{\eta}^{2} }{2\pi^{3} \mu_{h}} [\gamma_{0} \omega + i \gamma_{i} k_{i}^{F} + \mu_{h} \gamma_{0}] \int_{0}^{|\omega|} {d\Omega_{1}} \int_{\xi^{-1}}^{\infty} d q \frac{q}{\sqrt{q^{2} - (\omega + \Omega_{1})^{2}}}\frac{\gamma_{v} \Omega_{1}q}{ q^{6} + \gamma_{v}^{2}\Omega_{1}^{2} } , \nonumber \eqa where Wick rotation has been performed at zero temperature in order to see the frequency dependence of the self-energy. Performing momentum and frequency integrals, we find \bqa && \Im \Sigma_{\eta}^{v}\Bigl(\omega > \frac{(\gamma_{v}\xi^{3})^{-1}}{2}\Bigr) \approx \frac{g m_{\eta}^{2} }{4\sqrt{3}\pi^{2} \gamma_{v}^{1/3}} \gamma_{0} |\omega|^{2/3} , \nn && \Im \Sigma_{\eta}^{v}\Bigl(\omega < \frac{(\gamma_{v}\xi^{3})^{-1}}{2}\Bigr) \approx \frac{g m_{\eta}^{2} \xi}{\sqrt{3} \pi^{2} } \gamma_{0} \omega + \gamma_{0} \mathcal{O}(\omega^{2}) . \eqa Note that the $|\omega|^{2/3}$ behavior is the hallmark of $z = 3$ criticality in two dimensions \cite{Landau_Damping_QCP_Review}. The self-energy correction due to gauge fluctuations also gives rise to $\Im \Sigma_{\eta}^{a}(\omega) \propto |\omega|^{2/3}$ \cite{Nagaosa_Lee_SM}.

At finite temperatures the zero-frequency self-energy corrections turn out to diverge in the one-loop approximation \cite{Lee_Nagaosa_Wen_SL_Review}. However, such divergences due to both gauge and valence bond fluctuations need not be given much attention because such self-energies are not gauge-invariant, thus they do not have any physical meaning. These divergences should be considered as an artifact of gauge non-invariance. Gauge invariance can be incorporated via vertex corrections, which cancel the divergent parts in the self-energies, giving rise to gauge invariant finite contributions \cite{YB_Wen_Lee}. This corresponds to the transport time, given by $q^{2} \sim T^{\frac{2}{z}}$ multiplication in the quasiparticle life time. As a result, we find the following expression for the electrical resistivity \bqa && \Delta \rho(T) \propto T^{4/3} , \eqa consistent with the previous results \cite{Lee_Nagaosa_Wen_SL_Review}. We would like to emphasize that this non-Fermi liquid physics is robust in the pseudogap phase even if U(1) gauge fluctuations become massive, which can result from pairing correlations of doped holes, where $z = 3$ critical valence bond fluctuations are responsible, protected by the SO(5) symmetry.

We conclude this section, discussing how effective interactions between valence bond fluctuations and holons affect the deconfined quantum critical point of the SO(5) WZW theory. Introducing $\Psi = \frac{1}{\sqrt{2}} (v_{4} + i v_{5})$ and considering SO(5) symmetry breaking for the WZW term \cite{Tanaka_SO5}, we find an effective field theory for valence bond fluctuations \bqa && S_{VB} = T \sum_{i\Omega} \int \frac{d^{2}q}{(2\pi)^{2}} \Psi^{\dagger}(q,i\Omega) \Bigl( \frac{q^{2}}{2g} + \frac{\pi N_{\eta} m_{\eta}^{2}\rho_{\eta}}{4} \frac{|\Omega|}{q} \Bigr) \Psi(q,i\Omega) - \int_{0}^{\beta}{d\tau} \int d^{2} r y_{m} (\Psi^{4} + \Psi^{\dagger 4}) . \nn \eqa Here, the cubic power in the last term results from the WZW term with SO(5) symmetry breaking \cite{DQCP}, where $y_{m}$ is the monopole fugacity. If the topological term is not taken into account, the condensation-induced term will be given by $- \int_{0}^{\beta}{d\tau} \int d^{2} r y_{m} (\Psi + \Psi^{\dagger})$ \cite{DQCP}. An important point is that dynamics of valence bond excitations is described by $z = 3$ critical theory at the quantum critical point. Recall that their dynamics is described by $z = 1$ criticality in the absence of doped holes. As a result, two spacial dimensions are already above the upper critical dimension, thus higher order interactions beyond the Gaussian term are irrelevant, more precisely, dangerously irrelevant, which means that the WZW-induced cubic term can be neglected safely at zero temperature while scaling properties in thermodynamics are governed by such irrelevant operators at finite temperatures \cite{Landau_Damping_QCP_Review}. Equivalently, the monopole fugacity vanishes at the quantum critical point, indicating deconfinement of bosonic spinons. We point out that the topological term plays an important role in deconfinement even away from half filling. If the topological term is ignored, the monopole-fugacity term of the linear in $\Psi$ is relevant, expected to result in confinement.

It is quite interesting to investigate possible feedback effects on holon dynamics beyond their non-Fermi liquid physics. Integrating over such $z = 3$ critical valence bond fluctuations, we observe that doped holes feel effective long-ranged attractive interactions, which may enhance superconducting correlations, regarded to be a two-dimensional analogue of the Luther-Emery phase. This would serve a novel mechanism for superconductivity in antiferromagnetic doped Mott insulators.

\subsection{Paramagnetic doped Mott insulators: A Polyakov-loop extended Nambu-Jona-Lasinio model}

\begin{figure}[htp]
\centerline{\includegraphics[scale=0.6]{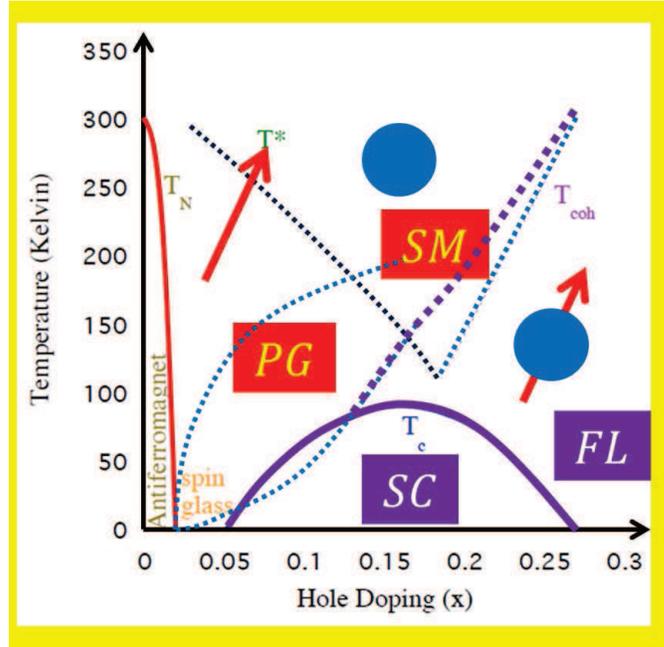}}
\caption{ (Color online) A schematic phase diagram for high T$_{c}$ cuprates. The underdoped pseudogap phase is governed by the SO(5) competing physics between antiferromagnetic and valance bond fluctuations, where $z = 3$ critical valence bond fluctuations are responsible for non-Fermi liquid physics of doped holes and possible high T$_{c}$ superconductivity. On the other hand, such spinons and holons become confined to form coherent electron excitations in the overdoped region, where the SO(5) competing physics disappears. Below the purple-dotted line from the optimal doping region, spinons and holons are confined, allowing only electron excitations.} \label{Phase_Diagram_QED_SO5_PNJL}
\end{figure}

Competing physics between antiferromagnetic and valance bond fluctuations, responsible for fractionalized critical spin excitations, would disappear when hole concentration increases further. Valence bond fluctuations are expected to remain relatively stronger than antiferromagnetic spin excitations inside but near the border of the pseudogap phase, which makes the skyrmion current not preserved any more and causes confinement between spin and charge degrees of freedom. Further hole doping makes the pseudogap state gone, where even valence bond excitations are difficult to survive. Then, it is natural to expect the confinement of spinons and holons, forming coherent electrons and describing physics of the overdoped region in high T$_{c}$ cuprates. Disappearance of the SO(5) competing physics is proposed to be responsible for the confinement of spinons and holons in the overdoped region, where an SU(2) gauge theory emerges to govern the low energy physics of the overdoped state. See Fig. \ref{Phase_Diagram_QED_SO5_PNJL}. This confinement physics is discussed within a phenomenological framework, referred to as Polyakov-loop extended Nambu-Jona-Lasinio (PNJL) model \cite{Fukushima_PNJL,PNJL_Review}.

%%%%%%%%%%%%%%%%%%%%%%%%%%%%%%%%%%%%%%%%%%%%%%%%%%%%%%%%%%%%%%%%%%%%%%%%
%%%%%%%%%%%%%%%%%%%%%%%%%%%%%%%%%%%%%%%%%%%%%%%%%%%%%%%%%%%%%%%%%%%%%%%%

We start from the SU(2) slave-boson representation of the t-J model \cite{Lee_Nagaosa_Wen_SL_Review} \begin{eqnarray} Z &=& \int D \psi_{i\alpha} D h_{i} D U_{ij} D a_{i\tau}^{k} e^{- \int_{0}^{\beta} d \tau L} , \nn L &=& \frac{1}{2} \sum_{i} \psi_{i\alpha}^{\dagger} (\partial_{\tau} - i a_{i\tau}^{k}\tau_{k})\psi_{i\alpha} + J_{r} \sum_{ij} ( \psi_{i\alpha}^{\dagger}U_{ij}\psi_{j\alpha} + H.c.) \nn &+& \sum_{i} h_{i}^{\dagger}(\partial_{\tau} - \mu  - i a_{i\tau}^{k}\tau_{k}) h_{i} + t_{r} \sum_{ij} ( h_{i}^{\dagger}U_{ij}h_{j} + H.c.) + J_{r} \sum_{ij} \mathbf{tr}[U_{ij}^{\dagger}U_{ij}] , \end{eqnarray} which has been discussed before. Further hole doping to the pseudogap phase makes the staggered flux state unstable, resulting in a uniform state given by $U_{ij}^{SM} = - i \chi I$, where the superscript ``SM" indicates ``strange metal". Incorporating low-lying transverse fluctuations around this mean-field ground state as follows $U_{ij}^{SM} = - i \chi e^{i a_{ij}^{k} \tau_{k} }$ and performing the continuum limit, we obtain an effective field theory \cite{Kim_Kim_PNJL} \begin{eqnarray} {\cal L}_{eff} &=& \psi_{\alpha}^{\dagger} (\partial_{\tau} - \mu_{s}\tau^{3} - ia_{\tau}^{k} \tau_{k} ) \psi_{\alpha} + \frac{1}{2m_{\psi}} |(\partial_{i} - i a_{i}^{k} \tau_{k} )\psi_{\alpha}|^{2} \nn &+& h^{\dagger}(\partial_{\tau} - \mu - ia_{\tau}^{k}\tau_{k})h + \frac{1}{2m_{h}} |(\partial_{i} - ia_{i}^{k}\tau_{k})h|^{2} - \frac{1}{4 e^{2}} f_{\mu\nu}^{k} f_{\mu\nu}^{k} , \end{eqnarray} where time and space components of SU(2) gauge fields arise from the Lagrange multiplier field and phase of the order-parameter matrix, respectively. $\mu_{s}$ represents a spinon chemical potential, given by the time component of the SU(2) gauge field with $k = 3$ and determined self-consistently. In this effective field theory spinons interact with holons via SU(2) gauge fluctuations.

An idea is to introduce confinement between spinons and holons phenomenologically, introducing the Polyakov-loop parameter \cite{Ployakov_Loop}. Defining the covariant derivative as \begin{equation} D_{\mu} = \partial_{\mu} - i \phi \tau_{3} \delta_{\mu \tau} - ia_{\mu}^{k} \tau_{k}, \nonumber \end{equation} where $\phi$ is the mean-field part of the gauge field associated with the Polyakov-loop parameter, and incorporating quantum fluctuations $a_{\mu}^{k}$, we write down an effective PNJL model for the matter sector \begin{eqnarray} {\cal L}_{PNJL}^{M} &=& \psi_{\alpha}^{\dagger} (\partial_{\tau} - i \phi \tau_{3} - \mu_{s}\tau_{3}) \psi_{\alpha} + \frac{1}{2m_{\psi}} | \partial_{i} \psi_{\alpha}|^{2} \cr & + & h^{\dagger}(\partial_{\tau} - i \phi \tau_{3} - \mu )h + \frac{1}{2m_{h}} | \partial_{i} h|^{2} \cr &+& g_{\psi} \psi_{\alpha n}^{\dagger} \psi_{\alpha p} \psi_{\beta p}^{\dagger} \psi_{\beta n} + g_{c} \psi_{\alpha n}^{\dagger} \psi_{\alpha p} h^{\dagger}_{p} h_{n}, \end{eqnarray} where interactions between spinons and holons are assumed to be local. This local approximation is well-utilized in the QCD context, realizing spontaneous chiral symmetry breaking successfully \cite{PNJL_Review}. Local current-current interactions are expected to be irrelevant in the renormalization group sense, thus neglected for simplicity. Here, the spinon-exchange interaction can be ignored phenomenologically in the SM phase while the electron resonance term will be allowed as quantum corrections. Then, it is straightforward to find an effective free energy from the "non-interacting" theory with the Polyakov-loop parameter $\Phi = \cos \beta \phi$. Minimizing the free energy with respect to $\Phi$, one always finds $\Phi = 1$, giving rise to deconfinement of spinons and holons. Matter fluctuations favor deconfinement as expected.

Confinement of spinons and holons can be realized by an effective Polyakov-loop action from gauge dynamics. One can derive an effective theory of the Polyakov-loop order parameter from pure Yang-Mills theory, integrating over quantum fluctuations. Unfortunately, the gauge free energy from one-loop approximation always gives rise to $\Phi = 1$ that corresponds to the deconfinement \cite{Weiss_PNJL}. It is necessary to take quantum fluctuations into account in a non-perturbative way. Such a procedure is not theoretically known yet, and we construct an effective free energy as follows \begin{eqnarray} F_{G}[\Phi;T] = A_{4} T^{3} \Bigl\{ \frac{A_{2} T_{0}}{A_{4}} \Bigl( 1 - \frac{T_{0}}{T} \Bigr) \Phi^{2} - \frac{A_{3}}{A_{4}} \Phi^{3} + \Phi^{4} \Bigr\} , \end{eqnarray} where the constants $A_{i=2,3,4}$ are positive definite, and $T_{0}$ is identified with the critical temperature for the confinement-deconfinement transition. Since the confinement-deconfinement transition is known as the first order from the lattice simulation \cite{PNJL_Review}, the cubic-power term with a negative constant is introduced such that $\Phi = 0$ in $T < T_{0}$ while $\Phi = 1$ in $T > T_{0}$, corresponding to the center symmetry ($Z_{2}$) breaking \cite{Fukushima_PNJL}.

\begin{figure}[htp]
\vspace{0.5cm}
\centerline{\includegraphics[scale=1.2]{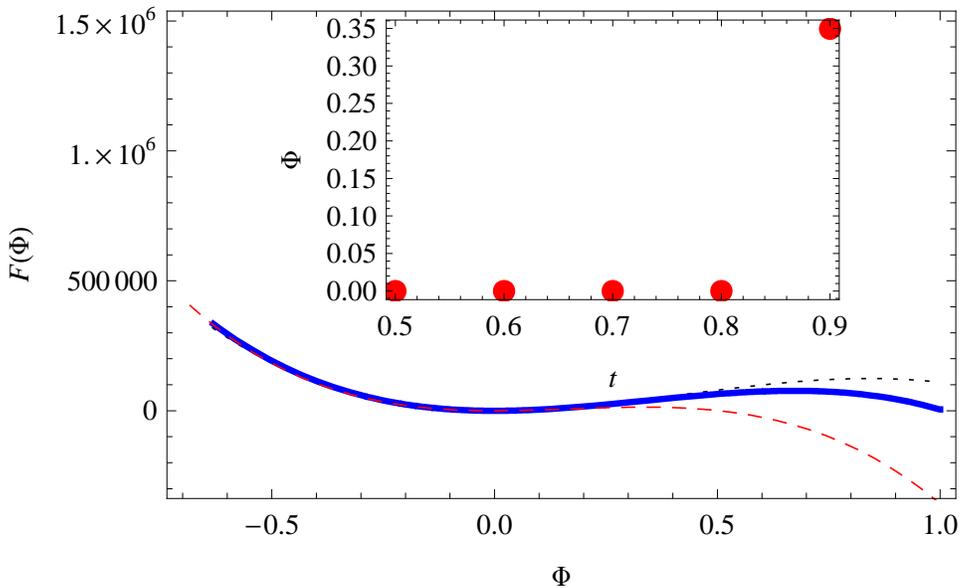} } \caption{
(Color online) The effective PNJL free energy as a function of the
Polyakov-loop parameter with $T < T_{CD}$ (Black-Dotted), $T =
T_{CD}$ (Blue-Thick), and $T > T_{CD}$ (Red-Dashed). Inset: The
Polyakov-loop parameter as a function of temperature scaled with
$T_{0}$. The Polyakov-loop parameter vanishes in $T < T_{CD}$,
causing confinement, while it becomes condensed in $T > T_{CD}$,
resulting in deconfinement. (Figure from Ki-Seok Kim and Hyun-Chul Kim [17])}
\label{FEPhiyT}
\end{figure}

The PNJL free energy is then obtained as \bqa F_{PNJL}[\Phi,\mu;\delta,T] &=& F_{M}[\Phi,\mu;\delta,T] + F_{G}[\Phi;T] \cr &=& -\frac{N_{s}}{\beta} \sum_{k} \ln \Bigl( 1 + 2 \bigl[ \Phi \cosh \beta \mu_{s} - \sqrt{1-\Phi^{2}} \sinh \beta \mu_{s} \bigr] e^{- \beta \frac{k^{2}}{2m_{\psi}} } + e^{- 2\beta \frac{k^{2}}{2m_{\psi}} } \Bigr) \cr && +\; \frac{1}{\beta} \sum_{q} \ln \Bigl( 1 - 2 \Phi e^{- \beta (\frac{q^{2}}{2m_{h}} - \mu )} + e^{- 2 \beta (\frac{q^{2}}{2m_{h}} - \mu )} \Bigr) + \mu \delta + \mu_{s} \cr && +\; A_{4} T^{3} \Bigl\{ \frac{A_{2} T_{0}}{A_{4}} \Bigl( 1 - \frac{T_{0}}{T} \Bigr) \Phi^{2} - \frac{A_{3}}{A_{4}} \Phi^{3} + \Phi^{4} \Bigr\} , \eqa where $F_{M}^{SM}[\Phi,\mu,\mu_{s};\delta,T]$ comes from matter dynamics. The confinement-deconfinement transition is driven by the gauge sector while the matter fluctuations turn the first order transition into the confinement-deconfinement crossover because the $Z_{2}$ center symmetry is explicitly broken in the presence of matters, so that the Polyakov-loop does not become an order parameter in a rigorous sense \cite{Fukushima_PNJL}. One may regard this PNJL construction as our point of view for the present problem, motivated from the crossover without the Higgs mechanism in the overdoped regime. Actually, one can construct the PNJL free energy, precisely speaking, the gauge sector to result in $\Phi = 0$ in $T < T_{CD}$ and $\Phi = 1$ in $T > T_{CD}$, where $T_{CD}$ is the confinement-deconfinement crossover temperature in the presence of matters, smaller than $T_{0}$ because matters favor the deconfinement. It is also consistent with confinement that the holon chemical potential of a negative value becomes much larger in $T < T_{CD}$ than in $T > T_{CD}$. See Fig. \ref{FEPhiyT}.

%%%%%%%%%%%%%%%%%%%%%%%%%%%%%%%%%%%%%%%%%%%%%%%%%%%%%%%%%%%%%%%%%%%%%%%%
%%%%%%%%%%%%%%%%%%%%%%%%%%%%%%%%%%%%%%%%%%%%%%%%%%%%%%%%%%%%%%%%%%%%%%%%

An interesting result in the mean-field approach of the PNJL model is that the condensation of holons is not allowed, since $D[\Phi,\mu] = 1 - 2 \Phi e^{- \beta (\frac{q^{2}}{2m_{h}} - \mu )} + e^{- 2 \beta (\frac{q^{2}}{2m_{h}} - \mu )}$ in $\ln D[\Phi,\mu]$ of the holon sector cannot reach the zero value because of $0 \leq \Phi < 1$ except for $\Phi = 1$. In other words, Higgs phenomena are not compatible with the confinement in this description. It should be noted that the mean-field approximation does not take into account feedback effects from matters to gauge fluctuations. In fact, Fermi surface fluctuations are not introduced, thus Landau-damped dynamics for gauge fluctuations is still missing. It is desirable to introduce quantum corrections beyond the PNJL mean-field theory.

The central question of this section is on the fate of the spinons and holons when the Polyakov-loop parameter vanishes \cite{Kim_Kim_PNJL}. The spinon-holon coupling term can be expressed as follows \begin{equation} \mathcal{S}_{el} \;=\; \int_{0}^{\beta} d \tau \int d^{2} r \left( \psi_{\sigma n}^{\dagger} h_{n} c_{\sigma} + c_{\sigma}^{\dagger} h^{\dagger}_{p} \psi_{\sigma p} - \frac{1}{g_{c}} c_{\sigma}^{\dagger} c_{\sigma} \right)  ,\nonumber \end{equation} where $\sigma$ and $n(p)$ represent spin and SU(2) indices, respectively. Since the Grassmann variable $c_{\sigma}$ carries exactly the same quantum numbers with the electron, one may identify it as the Hubbard-Stratonovich field $c_{\sigma}$. The effective coupling constant $g_{c}$ plays a role of the chemical potential for electrons. Note that the Fermi surface of the electrons differs from that of the spinons in principle. One can introduce the quantum corrections self-consistently in the Luttinger-Ward functional approach, as discussed in the previous section. We arrive at the self-consistent equations for self-energies \begin{eqnarray} \label{eq:8} \Sigma^{c}_{\sigma\sigma}(k,i\omega) = - \frac{1}{\beta} \sum_{i\Omega} \sum_{q} G^{h}_{p'p}(q,i\Omega) G^{\psi}_{\sigma\sigma,pp'} (k-q,i\omega-i\Omega) , \cr \Sigma^{\psi}_{\sigma\sigma,pp'} (k,i\omega) = - \frac{1}{\beta} \sum_{i\Omega} \sum_{q} G^{c}_{\sigma\sigma} (k+q,i\omega+i\Omega) G^{h}_{p'p}(q,i\Omega) , \cr \Sigma^{h}_{pp'}(q,i\Omega) = \frac{1}{\beta} \sum_{i\omega} \sum_{k} G^{c}_{\sigma\sigma}
(k+q,i\omega+i\Omega) G^{\psi}_{\sigma\sigma,pp'} (k,i\omega) , \end{eqnarray} where the Green's functions for the electron, the spinon, and the holon are given as \begin{eqnarray} G^{c -1}_{\sigma\sigma} (k,i\omega) &=& g_{c}^{-1} - \Sigma^{c}_{\sigma\sigma}(k,i\omega) , \cr G^{\psi -1}_{\sigma\sigma,pp'} (k,i\omega) &=& i [\omega + p (\phi - i \mu_{s})] \delta_{pp'} - \frac{k^{2}}{2m_{\psi}} \delta_{pp'} - \Sigma^{\psi}_{\sigma\sigma,pp'} (k,i\omega) , \cr G^{h -1}_{pp'}(q,i\Omega) &=& i [ ( \Omega + p \phi) - i \mu] \delta_{pp'} - \frac{q^{2}}{2m_{h}} \delta_{pp'} - \Sigma^{h}_{pp'}(q,i\Omega) , \end{eqnarray} respectively.

\begin{figure}[htp]
\centerline{\includegraphics[scale=0.5]{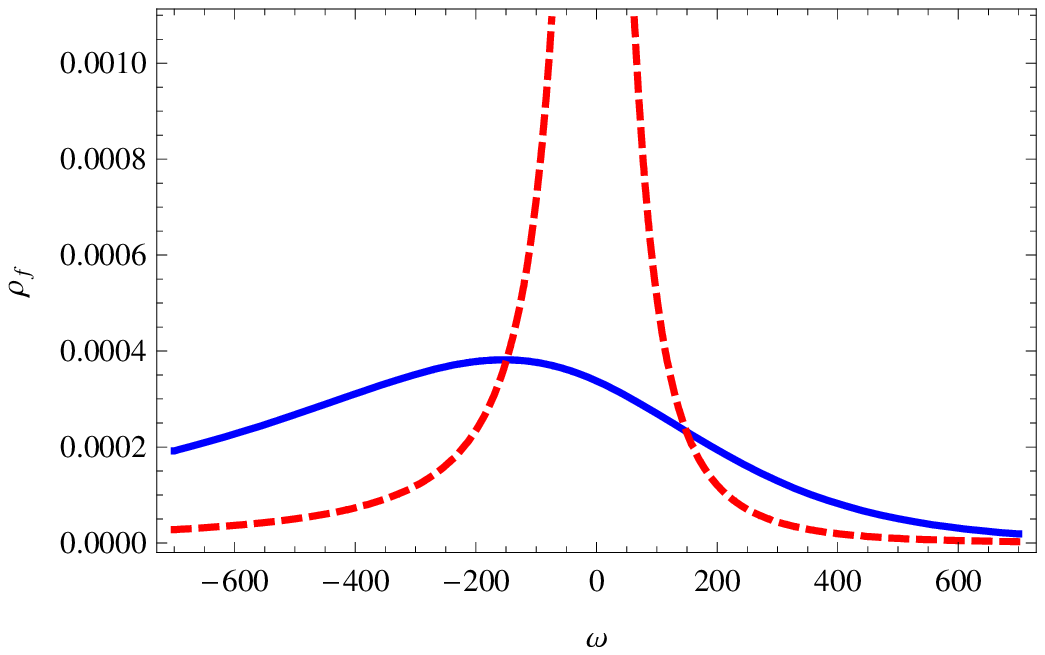} \includegraphics[scale=0.5]{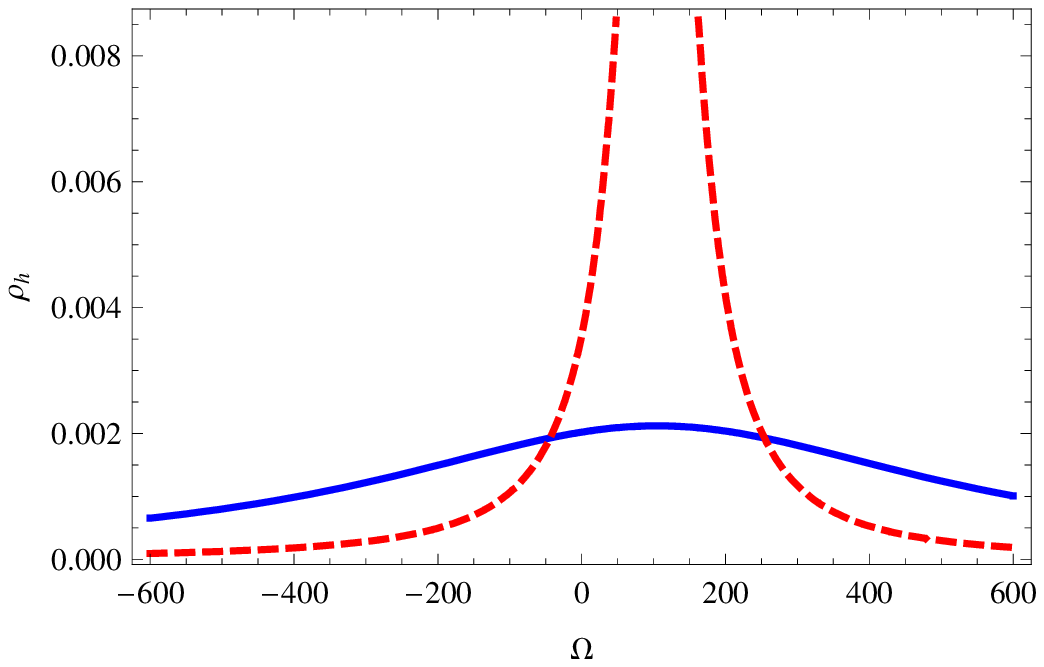}} \centerline{\includegraphics[scale=0.6]{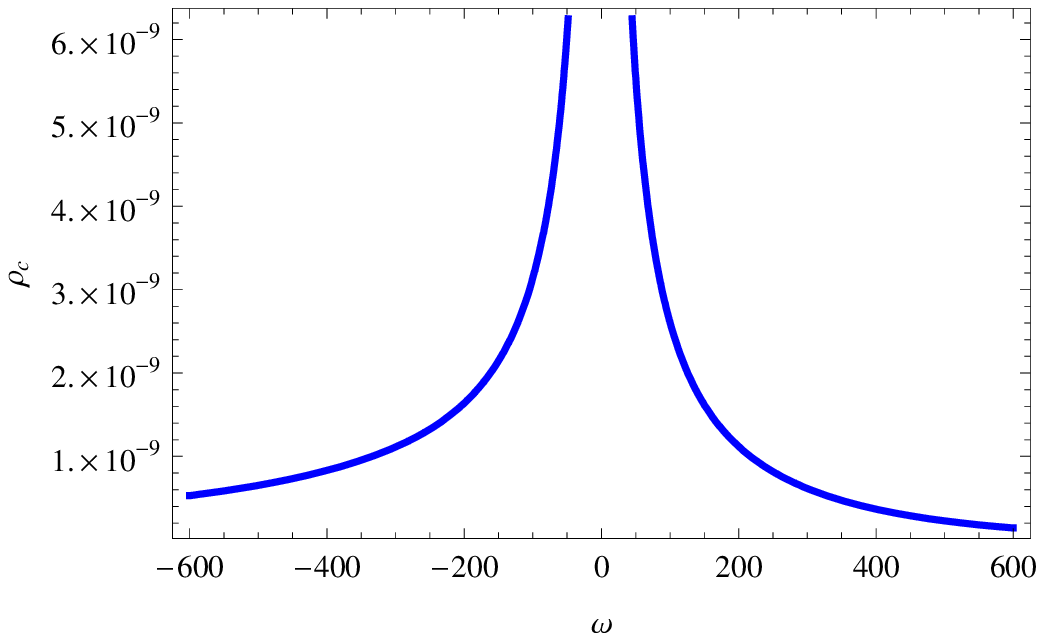}}
\caption{ (Color online) Spinon, holon, and electron spectra: Confinement (blue-thick line) makes spinon and holon spectra, sharply defined in the deconfined state of $\phi = 0$ (red-dashed line),
extremely broader, but allowing electron excitations. (Figure from Ki-Seok Kim and Hyun-Chul Kim [17])} \label{Electron_Spectrum}
\end{figure}

In the confinement phase the spectral function of the spinon should not be reduced to the delta function owing to the presence of the background potential $\phi$ even if the self-energy correction is ignored. Actually, the Polyakov-loop parameter plays a role of the imaginary part of the self-energy, which makes the spinon resonance disappear. The holon spectrum also features a broad structure. It indicates that both the spinon and the holon are not well-defined excitations in the confinement phase. On the other hand, the electron as a spinon-holon composite exhibits a rather sharp peak, since the imaginary part of their self-energy vanishes at the Fermi surface in spite of no pole structure in the Green's function. See Fig. \ref{Electron_Spectrum}.

%%%%%%%%%%%%%%%%%%%%%%%%%%%%%%%%%%%%%%%%%%%%%%%%%%%%%%%%%%%%%%%%%%%%%%%%
%%%%%%%%%%%%%%%%%%%%%%%%%%%%%%%%%%%%%%%%%%%%%%%%%%%%%%%%%%%%%%%%%%%%%%%%

The holon self-energy is found to be of the standard form in two dimensions \begin{eqnarray} && \Sigma_{p}^{b}(q,i\Omega) - \Sigma_{p}^{b}(q,0) \cr &&= - \frac{\rho_{c} }{i(\alpha-1)} \Bigl\{\tan^{-1}\Bigl(\frac{i\Omega+ip\phi- v_{F}^{c} q^{*} + v_{F}^{c} q}{- i\Omega - i p\phi + v_{F}^{c} q^{*} + v_{F}^{c} q}\Bigr) - \tan^{-1}\Bigl(\frac{i\Omega+ip\phi-\alpha v_{F}^{c} q^{*} + \alpha v_{F}^{c} q}{- i\Omega - i p\phi + \alpha v_{F}^{c} q^{*} + \alpha v_{F}^{c} q}\Bigr) \Bigr\} \nn \end{eqnarray} except for $i\Omega \rightarrow i\Omega+ip\phi$. $\rho_{c}$ is the density of states for the confined electron, and $v_{F}^{c}$ stands for the corresponding Fermi velocity. $\alpha$ denotes the ratio of the electron band mass to the spinon one, given as almost unity. $q^{*}$ designates the Fermi-momentum mismatch between the confined electron and the spinon.

An important energy scale is given by the holon chemical potential $\mu$. In $T > |\mu|$ holon dynamics is described by the dynamical exponent $z = 3$, resulting from the Landau damping of the electron and spinon. The imaginary part of the self-energy turns out to be proportional to $T^{2/3}$, since the confined electrons are scattered with such $z = 3$ dissipative modes. On the other hand, the holon excitations have gaps in $T < |\mu|$, and scattering with confined electrons becomes suppressed, recovering the Fermi liquid. Thus, the Fermi liquid appears as the coherence effect in the confinement phase rather than the Higgs in the deconfinement state. This mechanism resolves the artificial transition at finite temperatures \cite{Kim_Kim_PNJL}, which occurs when the Fermi-liquid state is assumed to result from condensation of holons.

%%%%%%%%%%%%%%%%%%%%%%%%%%%%%%%%%%%%%%%%%%%%%%%%%%%%%%%%%%%%%%%%%%%%%%%%
\begin{figure}[htp]
\centerline{\includegraphics[scale=0.6]{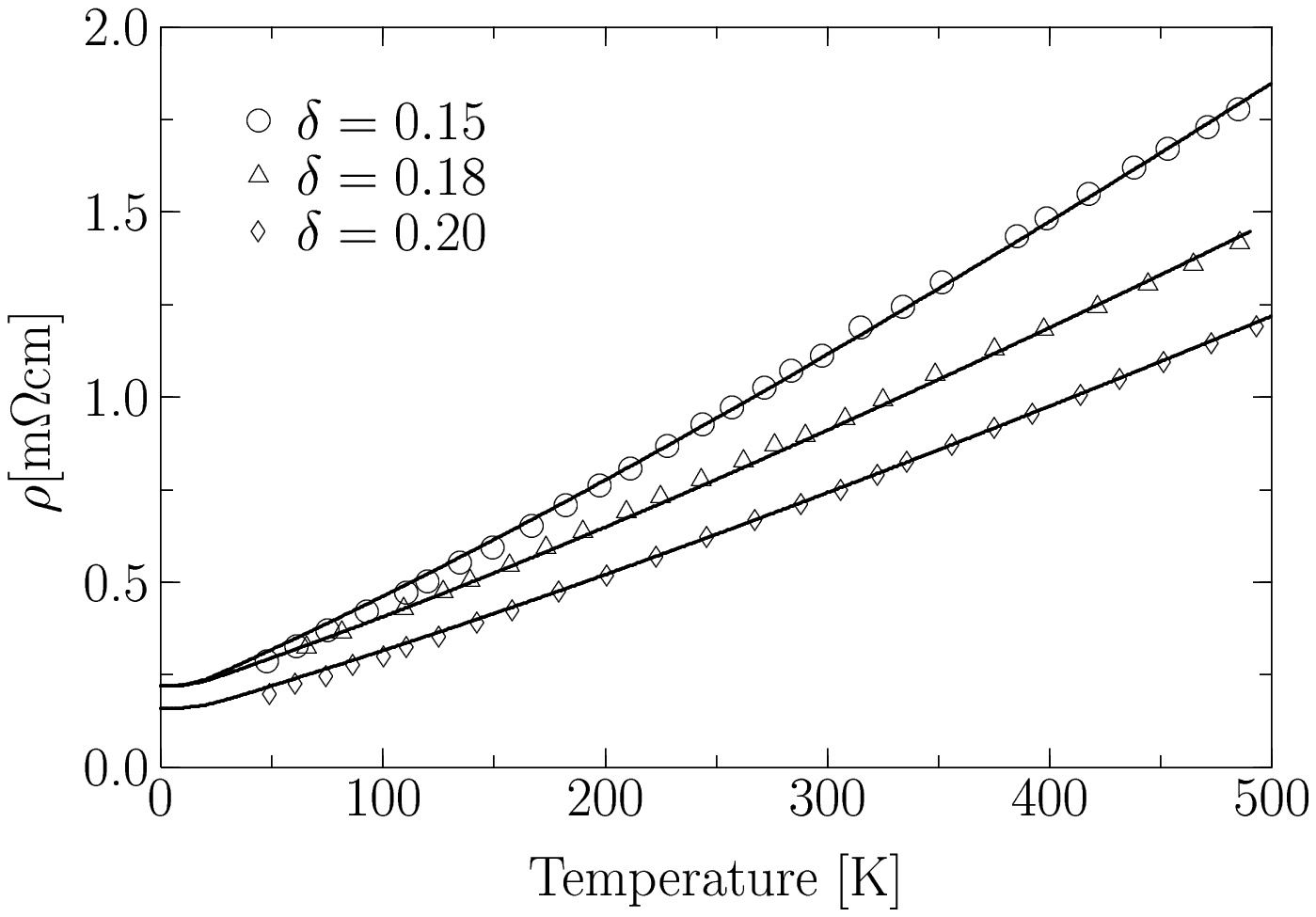}}
\caption{ (Color online) The electrical resistivity with
parameter $\mathcal{C}$ fitted. (Figure from Ki-Seok Kim and Hyun-Chul Kim [17])} \label{Electrical_Resistivity}
\end{figure}
%%%%%%%%%%%%%%%%%%%%%%%%%%%%%%%%%%%%%%%%%%%%%%%%%%%%%%%%%%%%%%%%%%%%%%%%

The coherence crossover is reflected in the electrical transport. It should be realized that the Ioffe-Larkin composition rule \cite{IoffeLarkin} for transport does not apply to the confinement phase. Instead, electrical currents would be carried by confined electrons dominantly. The relaxation time differs from the transport time, and the back scattering contribution is factored out by vertex corrections, corresponding to $T^{2/3}$ for two dimensional $z = 3$ fluctuations \cite{Kim_TR_Boltzmann}. Introducing both self-energy and vertex corrections, we reach the final expression of the electrical resistivity \cite{Kim_Kim_PNJL} \begin{equation} \rho_{el}(T) \;=\; \rho_{0} + \mathcal{C} \Bigl( N_{s} \rho_{c} \frac{v_{F}^{c2}}{3} \Bigr)^{-1} T^{2/3} \Im \Sigma_{c}(T) , \end{equation} where $\rho_{0}$, $\mathcal{C}$, and $N_{s}$ denote, respectively, the residual resistivity due to disorder, the strength for vertex corrections, and the spin degeneracy, among which $\rho_0$ and $\mathcal{C}$ are free parameters. See Fig. \ref{Electrical_Resistivity}. The results are in good agreement with the data \cite{Data}, which supports our confinement scenario. In addition, the $\mathrm{T}^{2}$ behavior is clearly observed at low temperatures, confirming our statement that the crossover from the SM phase to the Fermi liquid state is described by the coherence effect with the confinement.

\subsection{Emergent nonabelian gauge theory near Mott quantum criticality}

Deep inside Mott insulating phases, spin fluctuations are only relevant degrees of freedom at half filling, where charge fluctuations are frozen completely. Actually, the Heisenberg Hamiltonian does not allow charge fluctuations. However, charge fluctuations play an essential role in metal-insulator transitions at half filling, regarded to be one of the central problems in condensed matter physics. In this section we discuss how gauge fields emerge near metal-insulator transitions.

We start from the Hubbard model \begin{equation} H=-t\sum\limits_{ij}c_{i\sigma}^{\dagger}c_{j\sigma}+\text{H.c.} +U\sum\limits_{i}n_{i\uparrow}n_{i\downarrow}, \end{equation} where $c_{i\sigma}$ ($c_{i\sigma}^{\dagger}$) is the annihilation (creation) operator for an electron at site $i$ with spin $\sigma$. $t$ is the hopping integral, and $U$ is the on-site Coulomb interaction, where $n_{i\sigma}=c_{i\sigma}^{\dagger}c_{i\sigma}$ represents the density of electrons with spin $\sigma$. This model is reduced to the t-J Hamiltonian of the previous section in the $U/t \rightarrow \infty$ limit, where charge fluctuations are frozen completely at half filling.

Introducing the Nambu-spinor representation $\psi_{i}=\left( \begin{array} [c]{c} c_{i\uparrow}\\ c_{i\downarrow}^{\dagger}\end{array}\right)$ and performing the Hubbard-Stratonovich transformation for the pairing and density channels within the singlet domain, we obtain an effective Lagrangian \begin{eqnarray} L &=& \sum\limits_{i}\psi_{i}^{\dagger} (\partial_{\tau}1-\mu\tau_{z})\psi_{i}-t\sum\limits_{ij} \psi_{i}^{\dagger}\tau_{z}\psi_{j}+ {\rm H.c.} \nn &-& i\sum\limits_{i}[\Phi_{i}^{R}(\psi_{i}^{\dagger}\tau_{x}\psi_{i})+\Phi _{i}^{I}(\psi_{i}^{\dagger}\tau_{y}\psi_{i})+\varphi_{i}(\psi_{i}^{\dagger }\tau_{z}\psi_{i})] \nn &+& \frac{3}{2 U}\sum\limits_{i}[(\Phi_{i}^{R})^{2}+(\Phi _{i}^{I})^{2}+(\varphi_{i})^{2}] , \label{HS_transformed_Hubbard} \end{eqnarray} where $\Phi_{i}^{R(I)}$ and $\varphi_{i}$ are associated with pairing-fluctuation and density-excitation potentials, respectively, introduced to decouple the charge channel.

The SU(2) slave-rotor representation \cite{Kim_SU2SR} means to write down an electron field as a composite field in terms of a charge-neutral spinon field and a spinless holon field \begin{equation} \psi_{i}=Z_{i}^{\dagger}F_{i}, \end{equation} where $F_{i}=\left( \begin{array} [c]{c} f_{i\uparrow}\\ f_{i\downarrow}^{\dagger}\end{array} \right)$ is a fermion operator in the Nambu representation, and $Z_{i}$ is an SU(2) matrix \begin{eqnarray} Z_{i}=\left( \begin{array}[c]{cc} z_{i\uparrow} & -z_{i\downarrow}^{\dagger}\\ z_{i\downarrow} & z_{i\uparrow}^{\dagger} \end{array} \right) . \end{eqnarray} Here, $z_{i\sigma}$ is a boson operator, satisfying the unimodular (rotor) constraint, $z_{i\uparrow }^{\dagger} z_{i\uparrow} + z_{i\downarrow}^{\dagger} z_{i\downarrow} = 1$.

A key point of the slave-rotor representation \cite{Florens_Georges} is to extract out collective charge dynamics explicitly from correlated electrons. Such charge fluctuations are identified with zero sound modes in the case of short range interactions while plasmon modes in the case of long range interactions. Actually, one can check that the dispersion of the rotor variable ($z_{i\uparrow}$) is exactly the same as that of such collective charge excitations. In the slave-rotor theory the Mott transition is described by gapping of such rotor excitations. Until now, the Mott transition has not been achieved successfully, based on the diagrammatic (perturbative) approach starting from the Fermi-liquid theory in the absence of symmetry breaking.

Resorting to the SU(2) slave-rotor representation, we rewrite the effective Lagrangian as follows \cite{Kim_SU2SR} \begin{eqnarray} && L_{eff} = L_0 + L_F + L_Z , ~~~~~ L_{0} = t \sum \limits_{ij} \bm{tr} (X_{ij} Y_{ij}^{\dagger} + Y_{ij} X_{ij}^{\dagger}) , \nn && L_{F} = \sum \limits_{i} F_{i}^{\dagger} (\partial_{\tau} 1 - i \mathbf{\Omega }_{i} \cdot \bm{\tau})F_{i} - t \sum \limits_{ij} (F_{i}^{\dagger} X_{ij} F_{j} + \text{H.c.}) , \nn && L_{Z} = \frac{3}{4 U} \sum \limits_{i} \bm{tr} (\mathbf{\Omega}_{i} \cdot \bm{\tau} - i Z_{i} \partial_{\tau} Z_{i}^{\dagger} + i \mu Z_{i} \tau_{z} Z_{i}^{\dagger})^{2} - t \sum \limits_{ij} \bm{tr} (Z_{i} \tau^{z} Z_{j}^{\dagger} Y_{ij}^{\dagger} + \text{H.c.}) . ~~~~~ \label{SU2SR} \end{eqnarray} It is not difficult to see the equivalence between the SU(2) slave-rotor effective Lagrangian [Eq. (\ref{SU2SR})] and the Hubbard-Stratonovich transformed Hubbard model [Eq. (\ref{HS_transformed_Hubbard})]. Integrating over field variables of $X_{ij}$ and $Y_{ij}$, and shifting $\mathbf{\Omega}_{i} \cdot \bm{\tau}$ as \begin{eqnarray} \mathbf{\Omega}_{i} \cdot \bm{\tau} + i Z_{i} \partial_{\tau} Z_{i}^{\dagger} - i \mu Z_{i} \tau_{z} Z_{i}^{\dagger} \nonumber , \end{eqnarray} where $\mathbf{\Omega}_{i}=(\Phi_{i}^{R}, \Phi_{i}^{I}, \varphi_{i})$ is the pseudospin potential field, we recover the Hubbard-Stratonovich transformed Hubbard model exactly with an introduction of an electron field $Z_{i}^{\dagger} F_{i} \rightarrow \psi_{i}$. An important feature in the SU(2) slave-rotor description is appearance of pairing correlations between nearest neighbor electrons, given by off diagonal hopping in $X_{ij}$ which results from on-site pairing (virtual) fluctuations, captured by the off diagonal variable $z_{i\downarrow}$ of the SU(2) matrix field $Z_{i}$. We recall that the diagonal rotor field $z_{i\uparrow}$ corresponds to the zero sound mode, giving rise to the Mott transition via gapping of their fluctuations. The additional boson rotor variable $z_{i\downarrow}$ allows us to catch super-exchange correlations in the Mott transition, responsible for superconducting fluctuations. In this respect the SU(2) slave-rotor representation of the Hubbard model may be regarded as a generalized version of the SU(2) slave-boson representation of the t-J model, where the former works near Mott transitions while the latter does deep inside Mott insulating phases.

\begin{figure}[htp]
\includegraphics[width=0.6\textwidth]{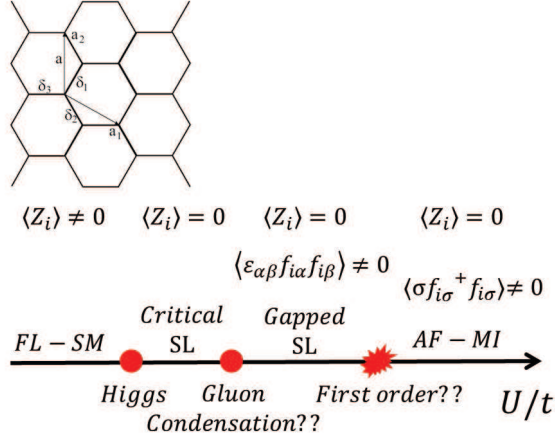}
\caption{A phase diagram of QCD$_{3}$ plus nonabelian Higgs model [Eq. (\ref{QCD3_NAHM})] based on the mean-field analysis. In honeycomb lattice, $\mathbf{a}_1$ and $\mathbf{a}_2$ are primitive translation vectors and $\delta_1$,
$\delta_2$, and $\delta_3$ are three nearest neighbor bonds. Abbreviations: FL-SM is the Fermi-liquid semi-metal phase, SL is the spin liquid state, and AF-MI is the antiferromagnetic Mott insulating phase.
gapped spin liquid, and AFM is antiferromagnetism. The FL-SM to critical SL and critical SL to gapped SL quantum phase transitions belong to the second order while the gapped SL to AF-MI quantum phase transition is the first order, where the last transition is not discussed here. In particular, we propose that the critical SL to gapped SL quantum phase transition is driven by gluon condensation. See the discussion below Eq. (\ref{SU2GT_Higgs})}
\label{Phase_Diagram_Honeycomb_Lattice}
\end{figure}

If we apply this effective field theory to the case of honeycomb lattice (graphene) and take the continuum limit as discussed in the previous section, we obtain an effective SU(2) gauge theory at half filling \cite{Kim_Tien_SU2SR} \bqa && Z_{eff}(m^{2};v^{2}) = \int D F D Z D a_{\mu}^{k} e^{- \int_{0}^{\beta} d \tau \int d^{2} \bm{r} {\cal L}_{eff}(m^{2};v^{2})} , \nn && {\cal L}_{eff}(m^{2};v^{2}) = \bar{F}\gamma_{\mu}(\partial_{\mu} - ia_{\mu}^{k}\tau_{k})F - \frac{1}{4 e^{2}} f_{\mu\nu}^{k} f_{\mu\nu}^{k} \nn && + \bm{tr} \Bigl(- i Z \partial_{\tau} Z^{\dagger} + a_{\tau}^{k} \tau_{k}\Bigr)^{2} + v^{2} \bm{tr} \Bigl(- i Z \bm{\nabla} Z^{\dagger} + \bm{a}^{k} \tau_{k}\Bigr)^{2} + m^{2} \Bigl( \bm{tr} Z Z^{\dagger} - 1 \Bigr) , \label{QCD3_NAHM} \eqa where $F$ is a Dirac spinon field and $Z$ is a matrix holon field. $m^{2}$ is introduced to enforce the unimodular constraint for the rotor variable, where the mean-field value assigns a mass gap to charge fluctuations. The holon velocity $v$ is given by a function of the Hubbard interaction $U$. Decreasing $U/t$ results in increasing $v^{2}$, favoring holon condensation (Higgs phase). We would like to emphasize that the spinon sector is quite analogous to the effective field theory in the $\pi-$flux state of square lattice. As a result, it is natural to expect that the SO(5) WZW theory appears from the Dirac semi-metal state at low energies when SU(2) holon excitations become gapped, increasing the Hubbard interaction above a critical value. Notice that the quantum phase transition from this deconfined critical phase to the Dirac semi-metallic state can be identified with the Higgs transition described by the condensation of the SU(2) holon-matrix field. When interactions are increased more, pairing correlations between nearest neighbor sites become enhanced in the singlet channel, destabilizing the critical spin-liquid state described by the SO(5) WZW theory, where charge fluctuations become more suppressed. As a result, spinon excitations are gapped due to their pairing orders. Actually, the appearance of the gapped spin-liquid state turns out to be a solution near the critical spin-liquid state in the saddle-point approximation \cite{Kim_Tien_SU2SR}. An interesting point is that the nature of this gapped spin-liquid state is characterized by local time reversal symmetry breaking not globally, given by an unconventional pairing symmetry, where a detailed configuration is not relevant for the present discussion \footnote{There exist two Weyl points in the graphene structure, where one lies at $\bm{K}$ and the other at $-\bm{K}$ in the Brillouin zone. Although the symmetry of the spinon-pairing order parameter is given by $d_{x^{2} - y^{2}} + i d_{xy}$ at $\bm{K}$, thus breaking time reversal symmetry, where $d_{x^{2} - y^{2}}$ and $d_{xy}$ denote a two-dimensional irreducible representation of the honeycomb lattice, the other valley of $-\bm{K}$ allows $d_{x^{2} - y^{2}} - i d_{xy}$ for the symmetry of the pairing order parameter, which turns out to preserve the time reversal symmetry as a total system \cite{Kim_Tien_SU2SR}. This is somewhat analogous to cancelation of parity anomaly.}. In terms of the SO(5) WZW theory, a valence-bond liquid state appears instead of the valence-bond solid phase discussed in the extended Heisenberg model on square lattice. The existence of such a gapped spin-liquid state seems to be consistent with an interesting simulation result recently performed \cite{Graphene_Z2SL_Simulation}. An essential question is how we can describe such quantum phase transitions beyond the saddle-point analysis, taking into account effects of gauge fluctuations. See Fig. \ref{Phase_Diagram_Honeycomb_Lattice} for a phase diagram of this QCD$_{3}$ plus nonabelian Higgs model, based on the mean-field approximation.

%%%%%%%%%%%%%%%%%%%%%%%%%%%%%%%%%%%%%%%%%%%%%%%%%%%%%%%%%%%%%%%%%%%%%%%%
%%%%%%%%%%%%%%%%%%%%%%%%%%%%%%%%%%%%%%%%%%%%%%%%%%%%%%%%%%%%%%%%%%%%%%%%

\begin{figure}[htp]
\centerline{\includegraphics[scale=0.6]{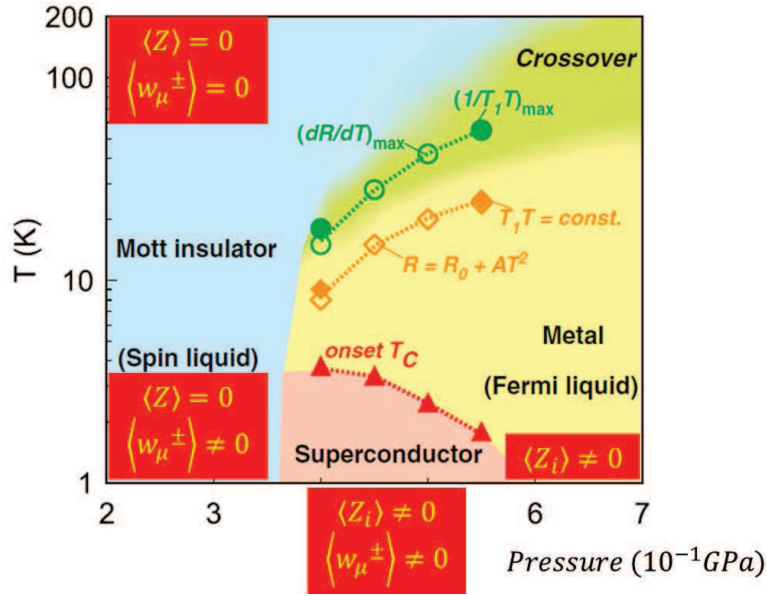}}
\caption{ (Color online) Phase diagram of an organic material [$\kappa$-(BEDT-TTF)$_{2}$Cu$_{2}$(CN)$_{3}$]. Based on the effective SU(2) gauge theory [Eq. (\ref{SU2GT_Higgs})], one may propose that the spin-liquid to superconducting transition is identified with the Higgs transition in the presence of gluon condensation, where the gluon condensation drives pairing correlations of spinons (SU(2) charges). See the text for more details. (Figure from K. Kurosaki et al. [55])} \label{Phase_Diagram_Triangular_Lattice}
\end{figure}

If we apply the SU(2) slave-rotor effective theory to the case of triangular lattice and perform the continuum approximation, we obtain an effective SU(2) gauge theory at half filling \begin{eqnarray} Z_{eff}(m^{2};v^{2}) &=& \int D F D Z D a_{\mu}^{k} e^{- \int_{0}^{\beta} d \tau \int d^{2} \bm{r} {\cal L}_{eff}(m^{2};v^{2})} , \nn {\cal L}_{eff}(m^{2};v^{2}) &=& F^{\dagger} (\partial_{\tau} - \mu_{s}\tau^{3} - ia_{\tau}^{k} \tau_{k} ) F + \frac{1}{2m_{s}} |(\partial_{i} - i a_{i}^{k} \tau_{k} )F|^{2} - \frac{1}{4 e^{2}} f_{\mu\nu}^{k} f_{\mu\nu}^{k} \nn &+& \bm{tr} \Bigl(- i Z \partial_{\tau} Z^{\dagger} + a_{\tau}^{k} \tau_{k}\Bigr)^{2} + v^{2} \bm{tr} \Bigl(- i Z \bm{\nabla} Z^{\dagger} + \bm{a}^{k} \tau_{k}\Bigr)^{2} + m^{2} \Bigl( \bm{tr} Z Z^{\dagger} - 1 \Bigr) , \label{SU2GT_Higgs} \nn \end{eqnarray} where nonrelativistic spinons near a Fermi surface interact with SU(2) matrix holons through SU(2) gauge fluctuations. An interesting question is how we can deduce the phase diagram of an organic material which shows a spin-liquid state at low temperatures and ambient pressure, a metal-insulator transition increasing pressure, and an unconventional superconducting phase in the vicinity of the metal-insulator transition, based on this effective field theory beyond the mean-field approximation. See Fig. \ref{Phase_Diagram_Triangular_Lattice}. An idea toward unconventional superconductivity near the Mott transition is to introduce gluon condensation in an irreducible representation of the corresponding point group of a lattice structure. If we decompose SU(2) gauge fields as follows \bqa && a_{\mu}^{3} \equiv a_{\mu} , ~~~~~ w_{\mu}^{\pm} = \frac{1}{\sqrt{2} e} (a_{\mu}^{1} \pm i a_{\mu}^{2}) , \eqa the Yang-Mills Lagrangian reads \bqa {\cal L}_{YM} = - \frac{1}{4 e^{2}} f_{\mu\nu}^{k} f_{\mu\nu}^{k} &=& - \frac{1}{4 e^{2}} f_{\mu\nu} f_{\mu\nu} - (D_{\nu} w_{\mu}^{+}) (D_{\nu} w_{\mu}^{-}) + (D_{\mu} w_{\mu}^{+}) (D_{\nu} w_{\nu}^{-}) \nn &-& 2 i w_{\mu}^{+} f_{\mu\nu} w_{\nu}^{-} + \frac{e^{2}}{4} (w_{\mu}^{+} w_{\nu}^{-} - w_{\nu}^{+} w_{\mu}^{-})^{2} , \eqa where $D_{\mu} w_{\nu}^{\pm} = (\partial_{\mu} \pm i a_{\mu}) w_{\nu}^{\pm}$ and $f_{\mu\nu} = \partial_{\mu} a_{\nu} - \partial_{\nu} a_{\mu}$. When $\langle w_{\mu}^{\pm} \rangle \not= 0$, the gluon condensation drives spinon pair condensation, where it serves an effective pairing potential as the case of the BCS theory. This mechanism will lead U(1) gauge gauge fluctuations gapped, where Z$_{2}$ gauge fields emerge \footnote{Since the order parameter is not gauge invariant as the BCS Cooper pair, it needs much care to define such an order. Recall the footnote d to discuss the projective symmetry group briefly.}. This gapped spin-liquid state is expected to evolve into the superconducting phase when holons become condensed, decreasing the Hubbard interaction strength through increasing pressure. We also speculate that the gluon condensation mechanism may explain why the valence-bond liquid state appears from the critical spin-liquid state of the SO(5) WZW theory on honeycomb lattice instead of the valence-bond solid phase.

%%%%%%%%%%%%%%%%%%%%%%%%%%%%%%%%%%%%%%%%%%%%%%%%%%%%%%%%%%%%%%%%%%%%%%%%
%%%%%%%%%%%%%%%%%%%%%%%%%%%%%%%%%%%%%%%%%%%%%%%%%%%%%%%%%%%%%%%%%%%%%%%%

\subsection{Discussion: Quantization of topological excitations and emergent gauge theory}

We have discussed that strongly coupled field theories appear as both abelian and nonabelian gauge theories with bosonic and fermionic matter fields in antiferromagnetic doped Mott insulators and metal-insulator transitions quite commonly, rather unexpected. In particular, we speculated how to simulate nonperturbative physics of topological excitations within the perturbative framework, where the emergent enlarged global symmetry and the phenomenological introduction of confinement have been suggested. Interestingly, the appearance of the issue on nonperturbative physics of topological excitations is not limited in such strongly interacting lattice models as either the large$-U$ limit of the Hubbard model or an intermediate regime involved with metal-insulator transitions. In this section we discuss that this issue also arises in the ``weakly" interacting regime of the Landau's Fermi-liquid state.

\begin{figure}[htp]
\centerline{\includegraphics[scale=0.6]{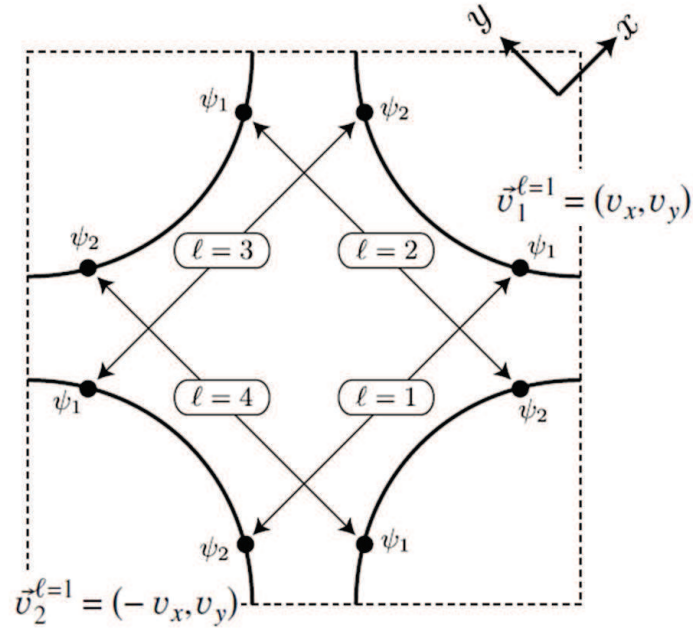}}
\caption{ (Color online) A Fermi surface of the tight-binding model with nearest-neighbor, next-nearest-neighbor, ... hopping parameters near half-filling. There exist four pairs ($l = 1, ..., 4$) connected by the antiferromagnetic wave-vector $\bm{Q} = (\pi, \pi)$, where each pair is given by $\psi_{1}$ and $\psi_{2}$. (Figure from Max A. Metlitski and S. Sachdev [23])} \label{Fermi_Surface_Square_Lattice}
\end{figure}

Quantum phase transitions involved with Fermi-surface instabilities in the Landau's Fermi-liquid state are described by condensation of local order parameters, breaking associated global symmetries \cite{Landau_Damping_QCP_Review}. Here, we consider antiferromagnetic quantum criticality as one example, given by the following effective field theory \cite{Chubukov_Spin_Fermion_Model} \bqa && Z_{AFQCP} = \int D \psi_{n \sigma}^{l} D \phi^{k} \exp\Bigl[ - \int_{0}^{\beta} d \tau \int d^{2} \bm{r} \Bigl\{ \psi_{n\sigma}^{l \dagger} (\partial_{\tau} - i \bm{v}_{n}^{l} \cdot \bm{\nabla}) \psi_{n\sigma}^{l} \nn && + \frac{\lambda}{\sqrt{N_{\sigma}}} \phi^{k} \psi_{n\sigma}^{l\dagger} \bm{\sigma}^{k}_{\sigma\sigma'} \psi_{-n \sigma'}^{l} + \phi^{k} (- \partial_{\tau}^{2} - v_{\phi}^{2} \bm{\nabla}^{2} + m^{2}) \phi^{k} \Bigr\} \Bigr] . \eqa $\psi_{n\sigma}^{l}$ represent low-energy electron excitations near several hot points ($n = 1, 2$ and $l = 1, ..., 4$) of the Fermi surface, given in Fig. \ref{Fermi_Surface_Square_Lattice}, where electrons at $n = 1$ with a fixed $l$ scatter into those at $n = 2$ with the same $l$, involving spin rotations through antiferromagnetic fluctuations with their transfer momentum $\bm{Q} = (\pi, \pi)$ and described by the interaction vertex $\frac{\lambda}{\sqrt{N_{\sigma}}} \phi^{k} \psi_{n\sigma}^{l\dagger} \bm{\sigma}^{k}_{\sigma\sigma'} \psi_{-n \sigma'}^{l}$. The number of such hot spots is $l = 1, ..., 4$. An important point is that the Fermi-velocity $\bm{v}_{1}^{l}$ is not parallel (antiparallel) to $\bm{v}_{2}^{l}$, i.e., $|\bm{v}_{1}^{l} \times \bm{v}_{2}^{l}| \not= 0$, which plays an important role in renormalization for dynamics of antiferromagnetic fluctuations. Dynamics of antiferromagnetic fluctuations is assumed to enjoy their relativistic spectrum at UV. This effective field theory may be regarded to be a minimal model for antiferromagnetic quantum criticality.

%%%%%%%%%%%%%%%%%%%%%%%%%%%%%%%%%%%%%%%%%%%%%%%%%%%%%%%%%%%%%%%%%%%%%%%%
%%%%%%%%%%%%%%%%%%%%%%%%%%%%%%%%%%%%%%%%%%%%%%%%%%%%%%%%%%%%%%%%%%%%%%%%

Increasing the coupling constant $\lambda$ above a certain critical value $\lambda_{c}$, the sign of the mass of antiferromagnetic fluctuations becomes negative, giving rise to condensation of such an order parameter. This state is identified with an antiferromagnetic phase, allowing two types of antiferromagnons with the well-known relativistic dispersion, nothing but Goldstone bosons. Tuning $\lambda$ at $\lambda_{c}$, the mass gap vanishes and antiferromagnetic fluctuations become critical. Resorting to essentially the same strategy as the previous section to introduce self-energy corrections self-consistently through the Luttinger-Ward functional approach, we obtain \bqa && Z_{AFQCP} = \int D \psi_{n \sigma}^{l} D \phi^{k} \exp\Bigl[ - \int_{0}^{\beta} d \tau \int d^{2} \bm{r} \Bigl\{ \psi_{n\sigma}^{l \dagger} \Bigl(- i \frac{c_{\psi}}{N_{\sigma}} (- \partial_{\tau}^{2})^{\frac{1}{4}} - i \bm{v}_{n}^{l} \cdot \bm{\nabla}\Bigr) \psi_{n\sigma}^{l} \nn && + \frac{\lambda}{\sqrt{N}_{\sigma}} \phi^{k} \psi_{n\sigma}^{l\dagger} \bm{\sigma}^{k}_{\sigma\sigma'} \psi_{-n \sigma'}^{l} + \phi^{k} (\gamma_{\phi} \sqrt{- \partial_{\tau}^{2}} - v_{\phi}^{2} \bm{\nabla}^{2}) \phi^{k} \Bigr\} \Bigr] , \label{AFQCP_CFT} \eqa where nonanalytic derivative expressions in time for self-energy corrections are well defined in the frequency space. It is straightforward to see that this effective theory is a critical field theory, where the spin-fermion coupling constant $\lambda$ is marginal under the scale transformation \bqa && \tau = b \tau' , ~~~~~ \bm{r} = b^{1/2} \bm{r}' \longrightarrow \psi_{n\sigma}^{l}(\bm{r},\tau) = b^{- 3/4} {\psi_{n\sigma}^{l}}'(\bm{r}',\tau') , ~~~~~ \phi^{k}(\bm{r},\tau) = b^{- 1/2} {\phi^{k}}'(\bm{r}',\tau') \nn \eqa with a scale parameter $b$. This marginality leads one to introduce the spin degeneracy $N_{\sigma}$, making the fixed point lie in the weak coupling region, which allows him to perform a controlled expansion, referred to as $1/N_{\sigma}$. Actually, it has been argued that vertex corrections give rise to higher order quantum corrections in $1/N_{\sigma}$, where the $z = 2$ critical spin dynamics is governed by the Gaussian fixed point since local interactions between such spin fluctuations are irrelevant at low energies \cite{Chubukov_Spin_Fermion_Model}. More precisely speaking, such local interactions turn out to be dangerously irrelevant, which break the hyperscaling relation and do not allow the $\omega/T$ scaling physics for the susceptibility of order parameter fluctuations, generally speaking \cite{Landau_Damping_QCP_Review}.

Recently, this problem has been revisited. First of all, vertex corrections turn out to be not subleading in the $N_{\sigma} \rightarrow \infty$ limit \cite{SSL_Large_N_Failure}, sometimes more singular in the $1/N_{\sigma}$ expansion \cite{Sachdev_Large_N_Failure}. This means that the critical field theory is strongly coupled even in the $N_{\sigma} \rightarrow \infty$ limit. Quite recently, two ways have been proposed for the controlled expansion: One is to introduce a parameter $x$ into the kinetic energy of order parameter fluctuations in addition to $N_{\sigma}$, where $x$ gives rise to a nonanalytic interaction potential between renormalized electrons which deviates from a standard Coulomb one \cite{Double_Expansion}, and the other is to consider a dimensional regularization in the scheme of the fermion renormalization group analysis \cite{SSL_Dimensional_Regularization}, both keeping the fixed point within the weak coupling regime. Although the existence of a perturbative fixed point has been demonstrated quite nicely, it is still not clear how these perturbative fixed points reflect nonperturbative physics originating from self-consistent vertex corrections, which may allow the $\omega/T$ scaling physics for the susceptibility of order parameter fluctuations.

An idea starts from the fact that some-types of instanton excitations may play an important role in such a strongly coupled regime, where vertex corrections are expected to encode the nonperturbative physics. Introducing instanton excitations into the effective field theory for antiferromagnetic quantum criticality [Eq. (\ref{AFQCP_CFT})], we would reach the following expression \bqa && Z_{eff} = \sum_{N_{t} = N + \bar{N} \in even} \frac{N ! \bar{N} !}{N_{t} !} y^{N+\bar{N}}\int D \mathcal{M} D f_{n \sigma}^{l} D \pi^{k} \nn && \exp\Bigl[ - \int_{0}^{\beta} d \tau \int d^{2} \bm{r} \Bigl\{ f_{n\sigma}^{l \dagger} \Bigl(- i \frac{c_{\psi}}{N_{\sigma}} (- \partial_{\tau}^{2})^{\frac{1}{4}} - i \bm{v}_{n}^{l} \cdot \bm{\nabla}\Bigr) f_{n\sigma}^{l} + \frac{\lambda}{\sqrt{N}_{\sigma}} \pi^{k} f_{n\sigma}^{l\dagger} \bm{\sigma}^{k}_{\sigma\sigma'} f_{-n \sigma'}^{l} \nn && + \pi^{k} (\gamma_{\phi} \sqrt{- \partial_{\tau}^{2}} - v_{\phi}^{2} \bm{\nabla}^{2}) \pi^{k} \Bigr\} - \mathcal{S}_{I}[\mathcal{M}] - \mathcal{S}_{eff}^{f-I}[f_{n\sigma}, \mathcal{M}] - \mathcal{S}_{eff}^{\pi-I}[\pi^{k}, \mathcal{M}] \Bigr] . \label{AFQCP_Instanton} \eqa $N$ ($\bar{N}$) represents the number of instantons (anti-instantons), set to be equal $N = \bar{N}$ in the respect of energy cost and thus, their total number is $N_{t} = 2 N$. $y = e^{- S_{I}}$ is fugacity of single instanton excitations, where $S_{I}$ is the corresponding instanton action. $\mathcal{M}$ means the moduli space of instantons such as their sizes, center-of-mass coordinates, and so on, referred to as collective coordinates and utilized for the first quantization. $f_{n \sigma}^{l}$ and $\pi^{k}$ represent scattering states of itinerant electrons and smooth spin fluctuations given by instanton and anti-instanton fluctuations. $\mathcal{S}_{I}[\mathcal{M}]$ describes dynamics of instantons. $\mathcal{S}_{eff}^{f-I}[f_{n\sigma}, \mathcal{M}]$ and $\mathcal{S}_{eff}^{\pi-I}[\pi^{k}, \mathcal{M}]$ keep scattering physics between itinerant electrons and instantons and between smooth spin fluctuations and instanton excitations, respectively. Unfortunately, the procedure from the critical field theory of Eq. (\ref{AFQCP_CFT}) to this effective field theory has not been clarified at all. This expression should be regarded to be formal, where a reliable derivation itself is quite a big business.

Suppose that this effective field theory has been constructed in a certain way. Then, the next question is how to integrate over instanton and anti-instanton fluctuations consistently. In the context of nonabelian gauge theories (QCD$_{4}$) it has been suggested that instanton fluctuations of SU(2) gauge fields give rise to effective interactions between light quarks, referred to as 't Hooft effective interactions, where topologically protected fermion zero modes play an essential role \cite{tHooft_Interaction}. One may expect similar effective interactions between itinerant electrons and antiferromagnetic spin fluctuations, based on Eq. (\ref{AFQCP_Instanton}). Unfortunately, it is difficult to find the robustness of fermion zero modes with instantons even if such states exist because the Lorentz symmetry encoded into the Dirac operator is explicitly broken and such zero modes are not protected topologically \footnote{In section 2 the origin of the valence-bond order in the instanton core is that there exists a fermion zero mode, where the symmetry of the instanton field is determined by that of the fermion zero mode, responsible for deconfined quantum criticality.}. On the other hand, one may expect that zero-mode-type states can exist as quasi-bound states, which assign a nontrivial quantum number to the hedgehog configuration at least approximately, where the skyrmion current can be conserved asymptotically. In particular, this second scenario is to benchmark the deconfined quantum critical physics of the SO(5) WZW theory. Unfortunately, the conditions for the second scenario are not clarified at all, associated with the nature of the conformal invariant fixed point in metallic antiferromagnetic quantum phase transitions.

In order to realize this scenario, we decompose the collective field $\bm{\phi}$ as follows \bqa && \bm{\phi} = \phi \bm{n} = \frac{\phi}{2} z_{\sigma}^{\dagger} \bm{\sigma}_{\sigma\sigma'} z_{\sigma'} , \eqa where $\phi$ is an amplitude of an antiferromagnetic order parameter, frozen and determined self-consistently in the last stage, and $z_{\sigma}$ denotes a spin direction of $\bm{\phi}$, carrying the half spin quantum number of $\phi$ and referred to as a spinon field. Then, the kinetic-energy term of antiferromagnetic spin fluctuations can be reformulated as follows \bqa && \int_{0}^{\beta} d \tau \int d^{2} \bm{r} \phi^{k} (- \partial_{\tau}^{2} - v_{\phi}^{2} \bm{\nabla}^{2} + m^{2}) \phi^{k} \nn && \longrightarrow J \sum_{\mu\nu} \phi_{\mu}^{k} \phi_{\nu}^{k} + m^{2} \sum_{\mu} \phi_{\mu}^{k 2} = \frac{J \phi^{2}}{4} \sum_{\mu\nu} z_{\mu\alpha}^{\dagger} \bm{\sigma}_{\alpha\alpha'} z_{\mu\alpha'} \cdot z_{\mu\beta}^{\dagger} \bm{\sigma}_{\beta\beta'} z_{\mu\beta'} + m^{2} \sum_{\mu} \phi^{2} \nn && = - \frac{J \phi^{2}}{4} \sum_{\mu\nu} z_{\mu\sigma}^{\dagger} z_{\nu\sigma} z_{\nu\sigma'}^{\dagger} z_{\mu\sigma'} + m^{2} \sum_{\mu} \phi^{2} \nn && \longrightarrow - \frac{J \phi^{2}}{4} \sum_{\mu\nu} ( z_{\mu\sigma}^{\dagger} \chi_{\mu\nu} z_{\nu\sigma} + H.c. ) + \frac{J \phi^{2}}{4} \sum_{\mu\nu} |\chi_{\mu\nu}|^{2} + m^{2} \sum_{\mu} \phi^{2} \nn && \approx - \frac{J \chi \phi^{2}}{4} \sum_{\mu\nu} ( z_{\mu\sigma}^{\dagger} e^{i a_{\mu\nu}} z_{\nu\sigma} + H.c. ) + \frac{J \phi^{2}}{4} \sum_{\mu\nu} \chi^{2} + m^{2} \sum_{\mu} \phi^{2} \nn && \longrightarrow \int_{0}^{\beta} d \tau \int d^{2} \bm{r} \Bigl\{ \frac{J \chi \phi^{2}}{4} |(\partial_{\mu} - i a_{\mu}) z_{\sigma}|^{2} + \frac{z J \phi^{2}}{4} \chi^{2} + m^{2} \phi^{2} \Bigr\} , \eqa where $\chi$ is a Hubbard-Stratonovich field to renormalize the velocity of spinons and $a_{\mu}$ is an emergent U(1) gauge field to count low-lying transverse fluctuations of the hopping field $\chi$. The interaction vertex can be decomposed as follows \bqa && \frac{\lambda}{\sqrt{N_{\sigma}}} \phi^{k} \psi_{n\sigma}^{l\dagger} \bm{\sigma}^{k}_{\sigma\sigma'} \psi_{-n \sigma'}^{l} = \frac{\lambda \phi}{\sqrt{N_{\sigma}}} z_{\alpha}^{\dagger} \bm{\sigma}_{\alpha\alpha'} z_{\alpha'} \cdot \psi_{n\sigma}^{l\dagger} \bm{\sigma}_{\sigma\sigma'} \psi_{-n \sigma'}^{l} \longrightarrow  - \frac{\lambda \phi}{\sqrt{N_{\sigma}}} \psi_{n\sigma}^{l\dagger} z_{\sigma} z_{\sigma'}^{\dagger} \psi_{-n \sigma'}^{l} \nn && \longrightarrow - \frac{\lambda \phi}{\sqrt{N_{\sigma}}} \psi_{n\sigma}^{l\dagger} z_{\sigma} f_{-n}^{l} - \frac{\lambda \phi}{\sqrt{N_{\sigma}}} f_{n}^{l \dagger} z_{\sigma}^{\dagger} \psi_{-n \sigma}^{l} + \frac{\lambda \phi}{\sqrt{N_{\sigma}}} f_{n}^{l \dagger} f_{-n}^{l} , \eqa where $f_{n}^{l}$ is an emergent fermion field, the saddle-point value of which is $\langle f_{n}^{l \dagger} \rangle = \langle \psi_{n\sigma}^{l\dagger} z_{\sigma} \rangle$, expected to originate from scattering between a monopole configuration and an itinerant electron near the Fermi surface. As a result, we reach the following expression for an effective field theory \bqa && Z_{eff} = \int D \psi_{n \sigma}^{l} D z_{\sigma} D f_{n}^{l} D a_{\mu} \exp\Bigl[ - \int_{0}^{\beta} d \tau \int d^{2} \bm{r} \Bigl\{ \psi_{n\sigma}^{l \dagger} (\partial_{\tau} - i \bm{v}_{n}^{l} \cdot \bm{\nabla}) \psi_{n\sigma}^{l} \nn && - \frac{\lambda \phi}{\sqrt{N_{\sigma}}} \psi_{n\sigma}^{l\dagger} z_{\sigma} f_{-n}^{l} - \frac{\lambda \phi}{\sqrt{N_{\sigma}}} f_{n}^{l \dagger} z_{\sigma}^{\dagger} \psi_{-n \sigma}^{l} + \frac{\lambda \phi}{\sqrt{N_{\sigma}}} f_{n}^{l \dagger} f_{-n}^{l} + \frac{J \chi \phi^{2}}{4} |(\partial_{\mu} - i a_{\mu}) z_{\sigma}|^{2} \nn && + m_{z}^{2} (|z_{\sigma}|^{2} - 1) + \frac{1}{2 g^{2}} (\epsilon_{\mu\nu\lambda} \partial_{\nu} a_{\lambda})^{2} + \frac{z J \phi^{2}}{4} \chi^{2} + m^{2} \phi^{2} + \frac{u}{N} \phi^{4} \Bigr\} \Bigr] . \eqa An emergent U(1) gauge theory results, taking into account nonperturbative physics of topological excitations. We emphasize that this scenario is to benchmark the physics of the SO(5) WZW theory, where the skyrmion current is preserved at the quantum critical point, supposing emergent U(1) gauge fluctuations defined on the noncompact U(1) space.

\section{Symmetry-protected-topological states and sigma models with topological terms}

\subsection{Motivation}

In the preceeding sections a recurrent theme was the way in which
skyrmions and other topological
excitations of an antiferromagnet
can affect the low energy effective theory of strongly correlated electrons through
their coupling to the fermionic degree of freedom.
The spin sector of the theory typically took the form of a nonlinear sigma model
with a Wess-Zumino term.
With the recent advent on physics of topological phases of matter-where prominent examples
include topological insulators and superconductors, an interesting twist
was added lately to the list of condensed matter physics problems which can be addressed by such
topological terms (and the closely related $\theta$-terms).
This new addition is the subject of symmetry protected topological (SPT) states.
Here we will switch gears and attempt
to provide a flavor of the basic ideas involved in this ongoing development,
following and building on the results of Ref. \refcite{Takayoshi}.
It is our hope to convey to the reader, through the treatments of the earlier sections
combined with the brief account that follows, the breadth of the physics which can arise
when these topological terms are present.

SPT states are the conceptual generalization of topological insulators (whose existence was predicted within the framework
of noninteracting band electron theory) to strongly interacting electrons as well as non-fermionic many-body systems, such as
bosonic cold atoms and magnetic systems. While topological insulators have been categorized within a symmetry-based
scheme into several distinct classes,
much of the current experimental effort focuses on insulators in three spatial dimensions
with a strong spin-orbit coupling and time reversal invariance. With the term ^^ ^^ topological insulator"
we will hereafter always be referring to this particular class of insulators,
which constitutes an example of an SPT state protected
by time-reversal symmetry: Kramers doublets
emerge at special points in momentum space, and their presence is robust against
%a moderate degree of
disorder and/or perturbation as long as time reversal symmetry is respected. When the mapping
from momentum space (the Brillouin zone) to the band structure is endowed with a topologically nontrivial
invariant, this feature gives rise to a robust surface Dirac cone, which in turn leads to numerous
exotic quantum effects. It is not possible to continuously deform, using time reversal invariant perturbations,
a topological insulator into a trivial one (defined as the state for which the above mentioned topological invariant is zero)
without closing the bulk energy gap, i.e. without encountering a quantum phase transition.
It is thus appropriate to view the topological insulator as a phase of matter
which is distinct from a trivial insulator. Once time reversal symmetry is violated, however, the Kramers theorem
no longer applies, and the topological distinction between the two insulating states is destroyed.

The notion of a general SPT state can be inferred from the topological insulator story just summarized.
In the second subsection we will illustrate this concept through a simple example of
an SPT state realized in quantum spin chains under an applied magnetic field.
As the counterpart of a charge insulator, we will
exclusively consider gapped spin states.
We take a topologically trivial state
to be that which is described as a direct product of local states,
in other words a state lacking a global quantum entanglement\footnote{In
SPT states, in contrast, the system is typically percolated with a network of
short-ranged entanglements.}.
A crucial ingredient for the  construction of an SPT state is a symmetry
which is to be imposed onto the system
(there can in general be several of such symmetries),
which prevents the systems from being adiabatically deformed
into to a trivial state.
After discussing these basic features of an SPT states through our physical example,
we will put the field theoretical construction of our sample SPT state
into the wider context of nonlinear sigma models with topological terms.
This will allow us to extend our approach to 3d SPT states, where effective
theories intimately related to those we have encountered in earlier sections
will play a central role.

\subsection{Physical example of an SPT state in one spatial dimension}

To carry out the program just mentioned in as simple a setting as possible which at the same time is physically motivated,
we consider an antiferromagnetic spin chain in an external magnetic field.
A minimal Hamiltonian describing this circumstance consists of a
Heisenberg exchange interaction and a Zeeman coupling of the spins to the magnetic field,
\begin{equation}
 {\cal H}=J\sum_{j}\bol{S}_{j}\cdot\bol{S}_{j+1}
%   +D\sum_{j}(S_{j}^{z})^{2}
-H\sum_{j}S_{j}^{z}.
 \label{eq:Hamil}
\end{equation}
The classical picture would imply that by turning on the magnetic field and increasing its strength,
a magnetization would eventually appear, and continue to increase monotonically up to its
saturation value. In reality though, there can appear, as observed in experiments and numerical studies,
finite intervals within the magnetization curve where the magnetization stays constant.
This magnetization plateau has been a subject of considerable interest for researchers working on
quantum spin systems. Here we wish to illustrate that an antiferromagnet in the plateau regime can
under appropriate conditions be a typical example of an SPT state.

\begin{figure}[bt]
%\centerline{\psfig{file=ijmpbf1.eps,width=3.65in}}
\centerline{\psfig{file=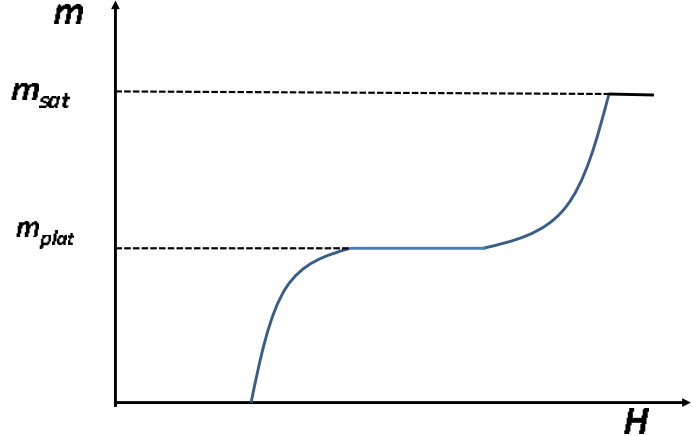,width=3.65in}}
\vspace*{8pt}
\caption{Schematic illustration of a magnetization plateau appearing within an M-H curve.}
\end{figure}

Let us try to  deduce the form of the effective field theory which captures the low-energy physics
of our spin chain. (A more careful derivation can be found in Ref. \refcite{Tanaka09}.) 
Since the magnetic field optimizes the magnetization and thereby effectively
suppresses the fluctuation of the spins in the z-direction,
 we expect that the relevant degree of freedom is the fluctuation within the xy plane.
Thus our theory is basically a quantum XY model. But in analogy to superfluids that are also
described as an XY model, the action should generally contain a topological (Berry phase) term.
For the superfluid case, the topological term is known to have several important implications:
for example it will influence vortex dynamics \cite{Fisher},
and govern the superfluid-Mott insulator
quantum phase transition \cite{Wen,Herbut}.
In a similar vein it is essential to incorporate this term in order to gain the correct understanding of our spin system's quantum
mechanical behavior.
Bearing in mind that the canonical conjugate of a planar angular variable is the (spin) angular momentum, from which in this case
a portion $m$, the magnetization per site, has been segregated off due to the magnetic field, it is not difficult to reason that the
action written in imaginary time and in the continuum limit takes the form
\begin{equation}
{\cal S}[\phi(\tau, x)]=\int d\tau dx\left[
i\frac{S-m}{a}\partial_{\tau}\phi +\frac{1}{2g}(\partial_\mu \phi)^2
\right],
\label{continuum action}
\end{equation}
where $\phi$ is the angular variable representing the orientation of the in-plane
staggered spin moment,
$S$ is the spin quantum number, $a$ is the lattice constance and $g$ a model-dependent coupling constant.

The first term ${\cal S}^{\rm B}\equiv i\frac{S-m}{a}\int d\tau dx \partial_{\tau}\phi$ of
Eq. (\ref{continuum action}) can be be derived systematically by simply noting that the Berry phase associated with
the spin at each site, in terms of spherical angular variables $(\theta, \phi)$,
is $\int S(1-\cos\theta)d\phi$, and by further observing that in the present situation
$S \cos\theta\equiv m$.
If not for this Berry phase term,
the Feynman weight $e^{-{\cal S}}$ for each configuration $\phi(\tau, x)$
resembles a Boltzmann weight for a 2d classical XY model. The system should then
undergo a Kosterlitz-Thouless transition at some value of $g$, giving rize to a plasma phase characterized by vortex condensation.
However, the Berry phase will generally
assign an individual phase factor to each Feynman weight,
and in particular will lead to a destructive interference among $\phi$-vortices,
which will prevent the system from entering the plasma phase regardless of the value of $g$
(see the discussion below on how to count the Berry phase associated with vortex configurations). Exceptions will occur if special conditions
rendering the Berry phase to be ineffective are realized.
This can happen under the condition
\begin{equation}S-m\in \mathbb{Z}.\label{condition}\end{equation}
There are several ways to see this\cite{Tanaka09}. For instance, by putting the system back on a lattice, for which case ${\cal S}^{\rm B}\rightarrow \sum_{j}{\cal S}_{j}^{\rm  B}=i\sum_{j}(S-m)\partial_{\tau}\phi_j$, one sees that a $2\pi\times W_j$-winding
of the phase $\phi_j$ along the imaginary-time direction at a given site, where $W_j\in \mathbb{Z}$, will contribute
a trivial phase factor : $e^{-i2\pi (S-m)W_j}\equiv1$ to the path integral
$Z=\int D\phi e^{-{\cal S}[\phi]}$.
Since the total contribution from any space-time configuration,
including those with vorticities,
reduces to a product of these on-site phase factors with some spatial distribution of the winding number $W_j$,
it follows that that will also have no effect on the physics, i.e. the system does not suffer a quantum phase interference and can undergo a transition into the plasma phase. This disordered phase with a short-ranged correlation and a spectral gap above the ground state
is none other than the magnetization plateau. A similar argument for the irrelevance of vortices
can be made for rational values
of the quantity $(S-m)$. That, however, can be treated by a suitable generalization of the integer-valued case where the unit cell and hence
the effective lattice constant spans several sites. We shall therefore focus hereafter
on the plateau states satisfying Eq. (\ref{condition}).

In order to see how the above relates to SPT states, and at the same time establish a
connection with the topological $\theta$ term of a nonlinear sigma model,
we rewrite the Berry phase term into a slightly different form. For this purpose
we once again place the system on a lattice (it proves convenient to work on a space-time grid)
and apply the easily verified identity
$
 {\cal S}^{\rm B}_{j}[\phi]
   =2i(S-m)\int d\tau\partial_{\tau}\phi_{j}
    -{\cal S}^{\rm B}_{j}[\phi_j]
$
to all even sites, which allows us to extract a portion written as a staggered sum
\begin{align}
 {\cal S}^{\rm B}
   =&\sum_{j}(-1)^{j}{\cal S}_{j}^{\rm B}[\phi]+\sum_{j}
     i(S-m)\int d\tau\partial_{\tau}\phi_{j}.
 \label{eq:TotalBerryPhase}
\end{align}
We now take the continuum limit.
First  observe that the second term of (\ref{eq:TotalBerryPhase}) becomes
\begin{equation}
 \sum_{j}i(S-m)\int d\tau\partial_{\tau}\phi_{j}\;\to\;
   i\int dxd\tau\frac{S-m}{a}\partial_{\tau}\phi,
 \label{eq:UniformBP}
\end{equation}
which recovers the Berry phase term we have discussed before. As already mentioned,
this term can be discarded under the assumed condition $S-m\in\mathbb{Z}$.
This leaves us to deal with just the staggered summation. It is easy to check that
this term counts up the vorticity associated with plaquettes
belonging to the
odd-even rows (but not the even-odd rows)
of the space-time grid \cite{Sachdev}.  The continuum form reads
 \begin{align}
 {\cal S}^{\rm B}
   \to& i\frac{S-m}{2}\int d\tau dx
     (\partial_{\tau}\partial_{x}-\partial_{x}\partial_{\tau})
     \phi(\tau,x)\nonumber\\
\equiv&i\pi(S-m)Q_{\rm v},
 \label{eq:TotalBPSphereCoord}
\end{align}
where $Q_{\rm v}$ is the
net vorticity throughout space-time (see Ref. \refcite{Sachdev} for a pictorial derivation
of this term within a different context). Note that this is a total derivative;
in essence, therefore, what we have done in this seemingly trivial rewriting is to
%pick up
salvage a surface-term contribution to the Berry phase, in addition to the previously
known \cite{Tanaka09} bulk Berry phase term which for the case at hand is irrelevant.

To make sense of the expression (\ref{eq:TotalBPSphereCoord}),
it turns out to be useful to compare with another system
-a spin-$S$ 1d planar Heisenberg antiferromagnet {\it not} subjected to a magnetic field.
A widely used low energy effective field theory for an antiferromagnetic
spin chain is the O(3) nonlinear sigma model with a topological
$\theta$-term ${\cal S_{\theta}}$ \cite{Auerbach}.
The vacuum angle (i.e. the coefficient of ${\cal S}_{\theta}$) which famously
governs the behavior of the ground state is $\theta=2\pi S$.
%The latter term takes the form
%\begin{equation}
 %{\cal S}_{\theta}=i\frac{\theta}{2\pi}\int d\tau dx
   %(\partial_{\tau}a_{x}-\partial_{x}a_{\tau}),\nonumber
%\end{equation}
%where the coefficient $\theta$, the vacuum angle, is $\theta=2\pi S$, and we
%have employed the CP$^{1}$ representation \cite{Auerbach},
%which relates the unit vector $\bol{n}$
%of the O(3) model to a unit-norm CP$^{1}$ spinor $\bol{z}$ through the relation
%$\bol{z}^{\dagger}\bol{\sigma}\bol{z}=\bol{n}$, which in turn induces
%the U(1) gauge connection $a_{\mu}\equiv i\bol{z}^{\dagger}\partial_{\mu}\bol{z}$.
%In the planar limit $\bol{n}\equiv{}^{t}(\cos\phi, \sin\phi, 0)$,
%we can choose the gauge  $\bol{z}\equiv{}^{t}(1/\sqrt{2},
%e^{i\phi(\tau,x)}/\sqrt{2})$,
%which leads to $a_{\mu}=\frac{1}{2}\partial_{\mu}\phi$.
%Substituting this into ${\cal S}_{\theta}$ results in
In the next subsection we will show that for an antiferromagnetic spin chain
at the planar limit, this term can be written as
 \begin{equation}
 {\cal S}_{\theta}=i\frac{\theta}{2}Q_{\rm v}=i\pi SQ_{\rm v},
   \label{eq:PlanarThetaTerm}
\end{equation}
where, as before, $Q_{\rm v}$ is the
vorticity associated with the space-time configuration of the angular field $\phi(\tau, x)$,
which in this case gives the orientation of the unit-modulus
antiferromagnetic order parameter.
(A derivation of the lattice version of Eq. (\ref{eq:PlanarThetaTerm})
can be found in Ref. \refcite{Sachdev}.)
We thus find that Eq. (\ref{eq:TotalBPSphereCoord}) has a form identical
to the $\theta$ term of Eq. (\ref{eq:PlanarThetaTerm}) but with a
vacuum angle of $\theta=2\pi(S-m)$,
i.e. to the case where the spin quantum number of the planar antiferromagnet is $S-m$.
In fact, since the kinetic term of our theory also
take the form of an O(3) nonlinear sigma model action in which the planar limit
has been taken, we see that our effective theory for the magnetic plateau
coincides precisely with that for a spin $S-m$ planar antiferromagnetic spin chain
(again recall that $S-m\in \mathbb{Z}$) without a magnetic field.\footnote{The difference in the coupling constant of the kinetic terms is irrelevant
since they are nonuniversal and in any case flows under a renormalization group procedure.}

This fact can be understood naturally in the following way. (Our argument is in part a repetition of
the reasoning leading to Eq. (\ref{continuum action}), but will also motivate us to incorporate
the insightful valence-bond-solid picture \cite{Auerbach}.) The physics of the
Haldane gap state, which is the ground state of an integer-$S$ Heisenberg spin chain (corresponding to the vacuum angle $\theta=2\pi S$),
can be represented qualitatively as a valence bond solid (VBS) state. To construct
this state, one begins by imagining that the spin-$S$ sitting at each site is a composite
object made of $2S$ spin-1/2 degrees of freedom. The VBS state is then built by letting
each of these spin-halves form a singlet bond with a second $S=1/2$ degree of freedom
belonging to an adjacent site. (Subsequently the spin-halves on each site are symmetrized so that the
state is projected onto the correct Hilbert space for spin-$S$ chains.) The translationally
invariant VBS state with $S$ singlet bonds forming on every link on the chain
is known to represent well the basic features of the spin-$S$ Haldane-gap state.
It is also known that this Haldane gap phase persists down to the planar limit.
Returning to the magnetization plateau state, we have a situation where on each site,
the spin fluctuation in the field direction is pinned down, and as mentioned before,
a portion $m$ out of the total $S$ spin moment therefore effectively drops out
from the spin dynamics. The remaining moment $S-m$ fluctuating within the plane is basically
free to form a subsystem which is in essence a spin-$(S-m)$
planar antiferromagnetic spin chain.  In VBS language, we can pictorially depict this by
letting $2m$ of the spin-halves polarize parallel to the field, while
the residual $2(S-m)$ spin-halves
participate in forming a translationally invariant VBS state (we encourage the reader to
consider the simplest example of $(S, m)=(3/2, 1/2$)). It is therefore reasonable that the
low energy physics of the magnetization plateau for  $S-m\in \mathbb{Z}$ should
reduce to that of the planar limit of the spin $S-m$ Haldane-gap state.

\begin{figure}[bt]
\centerline{\psfig{file=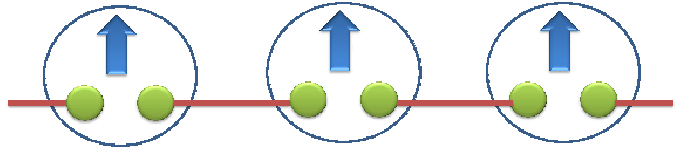,width=3.65in}}
\vspace*{8pt}
\caption{VBS-like depiction of a magnetization plateau with $S=\frac{3}{2}$ and $m=\frac{1}{2}$.}
\end{figure}

We now discuss what can be inferred about the ground state wavefunction given the
topological term of Eq. (\ref{eq:TotalBPSphereCoord}), following
the methods of Ref. \refcite{XuSenthil}.
For the sake of simplicity we employ the strong coupling limit,
where the action reduces to that of a topological
nonlinear sigma model consisting only of the $\theta$ term,
although the global structure which we will find
remains unchanged when we move away from this limit. Let us consider
expanding the state vector $|\Psi\rangle$ for the ground state in terms of the
snapshot configuration $\phi(x)$, i.e.
$ |\Psi\rangle=\int {\cal D}\phi(x)
   \Psi[\phi(x)]|\phi(x)\rangle$. The coefficient $\Psi[\phi(x)]$ is the
ground state wavefunctional, which is the probability amplitude for the
configuration $\phi(x)$ to be realized in the ground state.
This main idea here is to observe that this quantity can formally be expressed
as the following Euclidean path integral,
\begin{equation}
\Psi[\phi(x)]\propto \int_{\phi(\tau_{\rm i}, x)=\phi_{\rm i}(x)}
^{\phi(\tau_{\rm f}, x)\equiv\phi(x)}
{\cal D}\phi(\tau, x)e^{-{\cal S}_{\theta}[\phi(\tau, x)]}.
\label{eq:path integral}
\end{equation}
which describes the amplitude for starting at the initial time $\tau_{\rm i}$ with
the configuration
$\phi_{\rm i}(x)$, and evolving at the final time $\tau_{\rm f}$ into the configuration of interest
$\phi(x)$. Since the evolution proceeds in imaginary time,
the system is projected onto the groundstate. (To be precise this step
should be followed by a summation over
the initial configuration $\phi_{\rm i}$.)
To focus on bulk properties we will assume a periodic boundary condition
in the spatial direction. Plugging the action
${\cal S}_{\theta}=i(S-m)\int d\tau dx (\partial_{\tau}a_{x}-\partial_{x}a_{\tau})$ with
$a_{\mu}=\partial_{\mu}\phi/2$
into Eq. (\ref{eq:path integral}), we readily find that
\begin{eqnarray}
\Psi[\phi(x)]&\propto& e^{-i\frac{S-m}{2}\oint dx \partial_{x}\phi(\tau_{\rm f}, x)}\equiv
e^{-i\pi(S-m)W}\\ \nonumber
&=&(-1)^{(S-m)W},
\end{eqnarray}
where $W=\frac{1}{2\pi}\oint dx \partial_x \phi(x) \in \mathbb{Z}$
is the winding number of the angular variable $\phi(x)$
along the spatial extent of the system. Clearly the structure of the wavefunctional
is sharply distinguished by the parity of the integer valued quantity $S-m$.
For even $S-m$, the wavefunctional reduces to that in the absence
of the topological term while for odd $S-m$ it has a
nontrivial global structure $\Psi[\phi(x)]\propto (-1)^W$, suggesting that the ground state for the two cases belong to different phases. It is interesting to note that this distinction arose
from a temporal surface term of ${\cal S}_{\theta}$, in a manner completely analogous
to the way a spatial boundary contribution from the $\theta$ term
gives rise to the fractional spin moment
emerging at the end of an antiferromagnetic spin chain \cite{Ng}, which is a hallmark
of the Haldane-gap phase.

This parity dependence is also manifested in a dual vortex field theory description, which we sketch for later use.  For this purpose, we begin again with the effective lagrangian density
${\cal L}=\frac{1}{2g}(\partial_{\mu}\phi)^{2}+i\pi(S-m)q_{\rm v}$, where $q_{\rm v}$
is the vortex density, i.e. $\int d\tau dx q_{\rm v}=Q_{\rm v}$, and submit it to a
Hubbard-Stratonovich transformation $\frac{1}{2g}(\partial_{\mu}\phi)^{2}\rightarrow
\frac{g}{2}{\rm J}_{\mu}^{2}+i{\rm J}_{\mu}\partial _{\mu}\phi$. Decomposing $\phi(\tau, x)$ into
fields with and without vorticity and integrating over the latter
leads to the constraint $\partial_{\mu}{\rm J}_{\mu}=0$, which is explicitly solved by
introducing a dual field $\varphi(\tau, x)$ (which is vortex-free) defined through the relation
${\rm J}_{\mu}=\frac{1}{2\pi}\epsilon_{\mu\nu}\partial_{\nu}\varphi$.
Combining these, we obtain
${\cal L}=\frac{g}{8\pi^2}(\partial_{\mu}\varphi)^{2}+i(\pi(S-m)-\varphi)q_{\rm v}$,
where the coupling constant $g$ has inverted ($\propto 1/g$) as is characteristic of a dual theory.
 Integrating out $\varphi$ gives us the first-quantized vortex theory
${\cal L}=\frac{\pi^{2}}{g}q_{\rm v}\frac{1}{-\partial^2}q_{\rm v}+i\pi(S-m)q_{\rm v}$.
To obtain a second-quantized vortex-field theory, we perform a standard small-fugacity expansion,
restricting the vorticity entering into the grand-partition function of the vortex gas
to $\pm 1$.  Taking into account the Berry phase factors $e^{\pm i\pi(S-m)}$
which accompany the Feynman weight for vortex/antivortex events, we arrive at
the sine-Gordon action
\begin{equation}
 {\cal L}_{\rm dual}=\frac{g}{8\pi^{2}}(\partial_{\mu}\varphi)^{2}
   +2z\cos\big(\pi(S-m)\big)\cos\varphi,
 \label{eq:SineGordon}
\end{equation}
where $z$ is the fugacity of the vortices/antivortices.
As we are interested in the disordered (magnetization plateau) phase, we will assume that
vortex proliferation has occurred, i.e. that the cosine term is relevant.
Since the sign of the cosine term is dependent on the parity of $S-m$, the optimal
value of the field $\varphi$ differs for even and odd $S-m$, which is another indication
that the two cases lie in distinct phases. In fact, from the discussion below, it is
reasonable to assert that if an inversion symmetry about the link center in the spin chain is imposed, an intervening mass-vanishing point would be necessary
to go from one state to another.

In order to consider the possibility of an SPT state/phase, it is pertinent to
sort out the symmetry properties of the system in question.
In the present setup, time-reversal symmetry is broken due to the presence of
the magnetic field, while the spatial-inversion symmetry with respect
to the center of a link connecting a pair of adjacent spins is present.
Let us thus consider the effect of applying a staggered magnetic field to our spin chain,
which explicitly breaks this symmetry.
Noting that this perturbation adds a sign-alternating component to the
magnetization: $m\rightarrow m+(-1)^{j} \delta m$, we can repeat the
derivation leading to Eq. (\ref{eq:TotalBerryPhase}),
to find that the only part of this action which receives a modification is
the staggered summation over the onsite Berry phase terms,
$\sum_{j}(-1)^j {\cal S}^{\rm B}$, and the net change can be summarized as a
shift in the value of the effective vacuum angle, $\theta=2\pi(S-m)\rightarrow
2\pi(S-m-\delta m)$\footnote{This statement should be take with some caution, though. Since the external field dictates how a spin residing at the vortex core should polarize, the structure of the vortex
is different from that in, e.g. easy plane antiferomagnets which can have
meron and antimeron-like vortex structures (see the next subsection). 
Hence the above does not imply that a massless point will be encountered, as in the usual
Haldane gap problem,
when the system traverses the point $\theta=\pi$ as $\delta m$ is varied.
This is most easily verified from the dual vortex field theory described in the text.}.
(It is not difficult to show how a generic perturbation breaking this
same symmetry affects the Berry phase term in a similar manner,  solely through a modified
value of $\theta$.) This in turn implies that the wave functional is also modulated
into the form $\Psi[\phi(\tau, x)]\propto e^{-i\pi(S-m-\delta m)}$. Hence by sweeping
the value of $\delta m$, we can now interpolate between the two previously distinct
structures exhibited by the wavefunctional dependent on the parity of $S-m$. The crucial
question now is whether this interpolation can be achieved without closing an
energy gap. To see that this is indeed possible, we merely need to note that the
dual vortex field theory, in accordance with the above change in Berry phases,
is now given by
\begin{equation}
 {\cal L}_{\rm dual}=\frac{g}{8\pi^{2}}(\partial_{\mu}\varphi)^{2}
   +2z\cos\big(\pi(S-m-\delta m)-\varphi\big).
\end{equation}
It is clear from this action that the optimal value of $\varphi$ will
change continuously, following the variation of $\delta m$, all the while
keeping the cosine potential at the fixed value of $-2z$.
With all of the above combined, we can conclude that the magnetization plateau
state characterized by the topological term of Eq. (\ref{eq:TotalBPSphereCoord}) or equivalently the effective theory
of Eq. (\ref{eq:SineGordon}), belongs to an SPT phase protected by link-centered
inversion symmetry if $S-m$ is an odd integer.  This has been confirmed by a
numerical study of magnetization plateaus preformed for the cases 
$(S, m)=(3/2, 1/2), (3, 1)$, and $(3, 2)$\cite{Takayoshi}.
There it was found that the entire ^^ ^^ entanglement spectrum", a quantity generally
believed  to provide the fingerprints of topological order,
was two-fold degenerate only for odd $S-m$. A rigourous projective symmetry group
analysis carried out for explicit VBS-type wavefunctions
infers that this degeneracy is indeed the signature of an SPT state\cite{Takayoshi}.

\subsection{SPT states and the interplay of $\theta$- and Wess-Zumino terms}

The discussions of the previous subsection
suggest a natural mathematical extension to 3+1d
which we will outline below.
As a preparation for discussing the
3+1d problem, let us begin by recollecting the
basic steps relevant to the 1d case within the framework of the
O(3) nonlinear sigma model. For our purpose it is convenient to employ the
CP${}^{1}$ representation.
The three component
unit vector appearing in the O(3) theory $\bol{n}={}^{t}(\sin\theta\cos\phi,
\sin\theta\sin\phi, \cos\theta)$ can be expressed
in terms of the CP$^{1}$ spinor $\bol{z}={}^{t}(\cos\frac{\theta}{2}, \sin\frac{\theta}{2}e^{i\phi})\in\mathbb{C}^2$
via $\bol{z}^{\dagger}{\bf \sigma}\bol{z}=\bol{n}$.
The U(1) connection induced by this first
Hopf map $S^3 \rightarrow S^2$ is $a_{\mu}=i\bol{z}^{\dagger}\partial_{\mu}\bol{z}
=\frac{1}{2}(1-\cos\theta)\partial_{\mu}\phi$.  Finally, the $\theta$ term of the
O(3) model in 1+1d can be written in CP$^{1}$ language in terms of the first Chern number,
\begin{eqnarray}
{\cal S}_{\theta}^{1+1d}&=&i\frac{\theta}{4\pi}\int d\tau dx \bol{n}\cdot\partial_{\tau}\bol{n}\times\partial_{x}\bol{n}
\nonumber \\
&=&i\frac{\theta}{2\pi}\int d\tau dx (\partial_{\tau}a_{x}-\partial_{x}a_{\tau}).
\label{O3 theta}
\end{eqnarray}
%In the magnetization plateau problem, we found that for $S-m\in\mathbb{Z}$ the problem %effectively reduces to the O(3) model (without an external field)
%with the  subsidiary condition that $S\cos\theta\equiv m$ so that the fluctuation of the %order parameter occurs solely within the xy-plane of the target manifold. This anisotropy
%makes the task of correctly expressing the Berry phase term in the continuum a delicate
%problem, and it was shown that for this purpose we should invoke
%the second representation of Eq. (\ref{O3 theta}).

Since the $\theta$ term is a total divergence,
surface terms will arise if the integration of Eq. (\ref{O3 theta}) is performed
on a base-manifold with surfaces. It follows then that there should also be a correspondence
which equates the respective surface terms arising
from the two expressions in Eq. (\ref{O3 theta}). That correspondence takes the form
\begin{equation}
{\cal S}_{\rm WZ}^{0+1d}[\bol{n}(\tau)]={\cal S}_{\rm CS}^{0+1d}[a_{\tau}(\tau)],
\label{0+1 WZ-CS}
\end{equation}
where the Wess-Zumino term
\begin{equation}
{\cal S}_{\rm WZ}^{0+1d}\equiv ik\frac{2\pi}{area(S^{2})}\int_{0}^{1}du\int d\tau
\tilde{\bol{n}}\cdot\partial_{u}\tilde{\bol{n}}\times\partial_{\tau}\tilde{\bol{n}}
\end{equation}
 is just the
Berry phase term for a single spin with spin quantum number $S=k/2$,
where $\tilde{\bol{n}}$ is the extension of $\bol{n}(\tau, x)$ such that
$\tilde{\bol{n}}(u=0, \tau, x)={}^{t}(0,0,1)$ and
$\tilde{\bol{n}}(u=1, \tau, x)=\bol{n}(\tau, x)$.
The CP${}^1$ counterpart
\begin{equation}
{\cal S}_{\rm CS}^{0+1d}\equiv ik\int d\tau a_{\tau}
\end{equation}
is the level $k$ 0+1d abelian  Chern-Simons term
(we remind the reader that the general Chern-Simons term
takes the form $\propto{\rm Tr}\bol{A}\wedge \bol{F}^{n}$).
It is easy to check that the explicit expression for $a_{\mu}$ in spherical coordinates
reproduces the correct spin Berry phase.

%For the later purpose of extending to
%higher dimensions,
It is instructive to briefly reiterate in terms of the CP${}^{1}$ framework
how we had treated the planar fluctuations in the previous subsection,
making remarks on the way on the more topological aspects of the problem.
Instead of repeating the whole procedure for the magnetization plateau situation,
we will mainly concern ourselves here with 1d antiferromagnets (in the absence of
external fields) in the Haldane gap state.
The basic structure of our argument consists of carrying out the {\it reverse process}
of extracting the surface Wess-Zumino/Chern-Simons term of Eq. (\ref{0+1 WZ-CS})
from the bulk $\theta$ term of Eq. (\ref{O3 theta}).
We start by adding up spin Berry phase terms in a staggered fashion along a
spatial direction. One can check that this result in:
\begin{eqnarray}
{\cal S}_{\rm B}^{\rm tot}&=&\sum_{j}(-1)^{j}{\cal S}_{\rm CS}^{0+1d}[a_{\tau}(\tau, j)]
\nonumber \\
&\rightarrow&\frac{ik}{2}\int d\tau dx \epsilon_{\mu\nu}\partial_{\mu}a_{\nu}.
\label{CP1 theta}
\end{eqnarray}
where on taking the continuum limit in the second line,
we converted finite differences to derivatives and
added terms as appropriate to guarantee gauge invariance
(An alternative procedure
would be to work on a space-time lattice and use Stoke's theorem before moving on to the
continuum limit.) This is the CP${}^{1}$ representation for the $\theta$ term
(the second expression in Eq. (\ref{O3 theta})) with $\theta=\pi k (=2\pi S)$.
At this point we restrict the dynamics to {\it planar fluctuations}
\footnote{See remarks at the end of this subsection regarding the necessity
of making this somewhat artificial restriction.}.
The simplest
example is to take the planar limit where $\cos\theta\equiv 0$.
(As we are focused on spin-gapped systems, $S$ needs to be an integer for this
case on account of the Haldane conjecture.)
As the CP${}^{1}$ spinor corresponding to the planar unit vector
$\bol{n}\equiv{}^{t}(\cos\phi, \sin\phi, 0)$,
we can choose  $\bol{z}\equiv\frac{1}{\sqrt{2}}^{t}(1,
e^{i\phi(\tau,x)})$, for which
the U(1) connection is $a_{\mu}=\frac{1}{2}\partial_{\mu}\phi$.
Note that despite appearances,
this is not a pure gauge, owing to the factor 1/2.
Inserting this into the action of
Eq. (\ref{CP1 theta}) we arrive at Eq. (\ref{eq:PlanarThetaTerm}),
${\cal S}_{\rm B}^{\rm tot}=i\pi S Q_{\rm v}$.
We next turn to the ground state wavefunctional. It is easy to verify that this is
given by
\begin{equation}
\Psi[a_{\mu}]\propto e^{-ik\oint dx a_{x}}=e^{-i\pi S W}
\end{equation} where
\begin{equation}
W=\frac{1}{2\pi}\oint dx \partial_{x}\phi\in\mathbb{Z}
\label{snapshot winding 1+1}
\end{equation}
is a winding number
associated with the snapshot configuration. The wavefunctional thus
differs fundamentally in its topological structure depending on whether $S$ is odd or even.
This is consistent with the recent findings \cite{Pollman} that revealed that the Haldane gap state can be characterized as an SPT phase only for odd $S$. As already mentioned in the
previous subsection, this nontrivial topological structure has common routes with
the emergence of fractional spin moments at the spatial ends of spin chains in the Haldane
gap state. It is important to note that since the $\theta$ term is a total divergence,
a surface term arises at whatever boundary happens to be avaliable
on the domain of the space-time integration.
For a spin chain with open ends (a spatial boundary), a surface Berry phase term arises,
where the fractionalization of the spin quantum number comes from the factor of 1/2
which is present in the second line of Eq. (\ref{CP1 theta}) \cite{Ng}. When considering
the evolution of the state vector along a finite extent of the imaginary time axis,
a similar fractionalized surface term appears at the initial and final times. Here again the
factor of 1/2 is essential for the topological discrimination between odd and even $S$
that we saw above. We will
observe later that the exact same phenomena will arise in 3d when the physics is
governed by a $\theta$ term.
To complete the analysis, we turn on a staggered magnetic field in the z-direction,
breaking the inversion symmetry with respect to a link center.
(Physically the planar limit can be interpreted as having its origin in a strong easy-plane
anisotropy. The coupling to an external field can induce a z-component.) Precisely as in the
previous subsection this will induce a staggered magnetization $\delta m$, which in turn
will shift the vacuum angle to $\theta=2\pi(S-\delta m)$\footnote{As noted before, this
should not be taken literary as regards the dependence of the energy gap on $\delta m$: 
$\theta=\pi$ as a function of $\delta m$ does not in this case imply a massless point. 
This owes to the fact that due to the imposed staggered field, the vortex core structure is different from an easy-plane antiferromagnet in the absence of external fields. 
In the latter, for a given vortex configuration, a spin residing at the core can  escape out of plane and point in either the up or down direction. Both (meron) configurations must be 
incorporated to arrive at the correct dual  sine-Gordon theory. Once an external field is 
switched on, the orientation of the core spin is preassigned by the field, leading to a 
slightly different form of sine-Gordon theory.  
One should therefore  
map the system into the dual sine Gordon theory to derive the correct information on the energy gap.}.
Thus the discrimination of the
wavefunctional form between odd and even $S$ is destroyed. Furthermore, we can perform
a duality transformation and map the system into a gas of vortices. Tracing the
arguments of the previous subsection we can show that the interpolation between
the two forms of wavefunctionals can be done without encountering a gap-closing, and
we conclude\footnote{See Ref. \refcite{Pollman} for the other symmetries that can protect the odd-$S$ Haldane gap phase.} that the odd-$S$ Haldane gap phase (at least for the planar case) is an
SPT phase protected by link-centered
inversion symmetry.
%\footnote{A subtle difference arises here, however.
%The spins at the vortex core can escape into the z-direction, turning the vortices
%into merons and antimerons. The two types of defects need to be incorporated to arrive
%at the proper dual (sine-Gordon) theory for arbitrary $S$.}.

\begin{figure}[bt]
%\centerline{\psfig{file=ijmpbf1.eps,width=3.65in}}
\centerline{\psfig{file=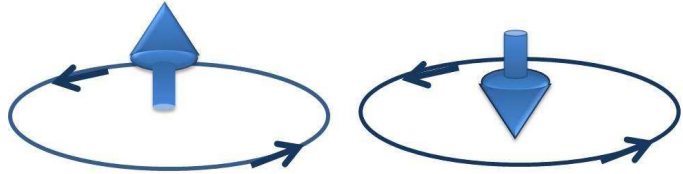,width=3.65in}}
\vspace*{8pt}
\caption{Schematics of a meron, where the spin at the core of a given 
vortex-like configuration can escape out of plane into one of two orientations, up or down. 
Application of an external field destroys this option. The case with and without external 
fields thus map into dual sine-Gordon-type field theories with subtle but crucial differences.}
\end{figure}

 In the previous subsection we considered a state with a finite magnetization $m$,
i.e. $S\cos\theta=m$. By imposing the plateau condition
$S-m\in\mathbb{Z}$, we were able to rewrite the initially unstaggered
sum over the spin Berry phase terms into a staggered one,
and
everything goes through as before with the only change
being that the gauge field now becomes $a_{\mu}=\frac{1}{2}(1-\frac{m}{S})\partial_{\mu}\phi$.
The action and the wavefunctional each become, as discussed in
the previous subsection, ${\cal S}=i\pi(S-m)Q_{\rm v}$
and $\Psi[a_{\mu}]\propto e^{-i\pi(S-m)W}$.

%; to see the difference with the
%pure gauge situation, consider making the gauge transformation
%$a\rightarrow a+u^{-1}du$ where $u\in$U(1). This induces the following trivial
%change in the Chern-Simons term (which is the pure gauge contribution): $ik\oint  %u^{-1}du=2\pi k\times$integer. This merely demonstrates that the Chern-Simons term is
%gauge invariant.

%Meanwhile when $a=\frac{1}{2}d\phi$ we have  ${\cal S}_{\rm CS}^{0+1d}=\frac{ik}{2}\int d\tau
%\partial_{\tau }\phi(\tau)$, which for generic $k$ {\it can} give rise to nontrivial effects.
%Stacking up these terms in a staggered fashion along a spatial
%direction the Berry phase action becomes ${\cal S}_{\rm BP}^{\rm tot}
%=i\sum_{j}(-1)^{j}S\int d\tau\partial_{\tau}\phi(\tau, j)$ which can be related to the
%net space-time vorticity $Q_{\rm v}$ as we have already discussed,
%i.e. ${\cal S}_{\rm BP}^{\rm tot}={\cal S}_{\theta}^{1+1d}
%[a=\frac{1}{2}d\phi]\vert_{\theta=2\pi S}=i\pi S Q_{\rm v}$. Similarly,
%for the plateau states of the previous section we insert
%$Sa=\frac{1}{2}(S-m)d\phi$ into the Chern-Simons term, and the staggered summation
%over this quantity
%leads to ${\cal S}_{\rm BP}^{\rm tot}={\cal S}_{\theta}^{1+1d}
%[a]\vert_{\theta=2\pi (S-m)}=i\pi (S-m) Q_{\rm v}$.

With this preparation we turn to 3+1d. A procedure completely parallel
to the 1+1d case can be carried out by use of the
second Hopf map $S^{7}\rightarrow S^{4}$, or more precisely, the
quaternionic projective (HP${}_{1}$) representation of the O(5) nonlinear sigma model.
The building blocks of this construction are thus five component unit vectors, each living in
its own  2d (xy) spatial plane, which will later be stacked up along a third spatial direction.
This can for instance represent, as in the earlier sections
of this article, the competition between antiferromagnetic and VBS orders.
Let us then denote this vector as $\bol{N}={}^{t}(\sin\theta\cos\phi, \sin\theta\sin\phi\bol{n},
\cos\theta)\in S^{4}$ where $\bol{n}\in S^{2}$ is a three component unit vector.
This can be recasted into a two component unit quaternionic spinor $\bol{q}\in\mathbb{H}^2$ in the
form
\begin{equation}
 \bol{q}\equiv
 \begin{pmatrix}
   \cos\frac{\theta}{2}\\ \sin\frac{\theta}{2}g
 \end{pmatrix},\nonumber
\end{equation}
where an explicit expression for $g\in\mathbb{H}$ can be obtained
by first writing it as an element of SU(2), i.e. $g=e^{i\phi\bol{n}\cdot\bol{\sigma}}$
and subsequently making the replacements $i\sigma_{x}\rightarrow\bol{i},
i\sigma_{y}\rightarrow\bol{j}, i\sigma_{z}\rightarrow -\bol{k}$, where
we have taken advantage of the fact that the algebra spanned by the three quaternionic
imaginary units $\bol{i}, \bol{j}, \bol{k}$ form the same algebra as the
Pauli matrices. The second Hopf map can be constructed explicitly via \cite{Demler}
$\bol{q}^{\dagger}\gamma_{\alpha}\bol{q}=N_{\alpha}, 1 \leq \alpha \leq 5$, where
\begin{align}
 &\gamma_{1}=
   \begin{pmatrix} {} & 1 \\ 1 & {} \end{pmatrix},\quad
  \gamma_{2}=
   \begin{pmatrix}
{}&-\bol{i}\\ \bol{i}&{}\end{pmatrix}\quad
 \gamma_{3}=
   \begin{pmatrix} {} & -\bol{j} \\ \bol{j} & {} \end{pmatrix},\nonumber\\
 &\gamma_{4}=
   \begin{pmatrix} {} & -\bol{k} \\  \bol{k} & {} \end{pmatrix},\quad
  \gamma_{5}=\begin{pmatrix} 1 &{} \\ {} & -1 \end{pmatrix}.\nonumber
\end{align}
Note that in correspondence to the presence of three imaginary units,
there are three matrices here bearing the same structure as
the Pauli matrix $\sigma_y$.
Below we will treat $g$ for the most part as an SU(2) valued matrix
with the understanding that a suitable conversion into
the quaternionic language can always be made. (The subtleties related to this conversion
can be checked for each equation along the lines of Refs. \refcite{Nash Sen} and
\refcite{Naber}.)
The gauge field which arises in this representation can thus be
treated as an SU(2) gauge field,
$b_{\mu}=-\bol{q}^{\dagger}\partial_{\mu}\bol{q}=\frac{1}{2}(1-\cos\theta)
g^{-1}\partial_{\mu}g$.

We shall start the construction of an SPT state in 3+1d
by writing down the 2+1d counterpart of Eq. (\ref{0+1 WZ-CS}),
\begin{equation}
{\cal S}_{\rm WZ}^{2+1d}[\bol{N}(\tau, x, y)]={\cal S}_{\rm CS}^{2+1d}[b_{\mu}(\tau, x, y)].
\label{2+1 WZ-CS}
\end{equation}
The left hand side of this equation represents the Wess-Zumino term of the O(5) nonlinear sigma model,
\begin{eqnarray}
{\cal S}_{\rm WZ}^{2+1d}&=&ik\int_{B\times[0,1]}
\tilde{\bol{N}}^{*}
\omega_{S^{4}}
\nonumber \\
&=&ik\frac{2\pi}{area(S^{4})}\int_{0}^{1}du\int d\tau dxdy
\epsilon_{abcde}\tilde{N}^{a}
\partial_{u}\tilde{N}^{b}\partial_{\tau}\tilde{N}^{c}
\partial_{x}\tilde{N}^{d}\partial_{y}\tilde{N}^{e}.
\label{2+1 O5 WZ}
\end{eqnarray}
In the first line of the above, $B$ denotes the base manifold of the O(5) nonlinear sigma model,
(i.e. $\bol{N}: (\tau, x, y)\in B\rightarrow S^{4}$), and $B\times [0, 1]$ is its
extension which is necessary for constructing the Wess-Zumino term,
$\tilde{\bol{N}}: (u, \tau, x, y)\in B\times[0, 1]\rightarrow S^4$ is the corresponding
extension of the map $\bol{N}$. Furthermore, $\omega_{S^4}$ stands for the
volume form of the target manifold $S^4$ and $\tilde{\bol{N}}^{*}$ is its pullback
to $B\times [0,1]$  \cite{Tanaka11}. The second line expresses this explicitly in coordinates.
Meanwhile the action appearing on the right hand side of Eq. (\ref{2+1 WZ-CS})
is the Chern-Simons term
\begin{equation}
{\cal S}_{\rm CS}^{2+1d}=i2\pi k\cdot \frac{1}{8\pi^2}
{\rm Tr}\int_{B} \epsilon_{\mu\nu\rho}\left(
b_{\mu}\partial_{\nu}b_{\rho}+\frac{2}{3}b_{\mu}b_{\nu}b_{\rho}
\right).
\label{2+1 HP1 CS}
\end{equation}
We note that upon making the large gauge tranformation
$b\rightarrow u^{-1}bu+u^{-1}du$ where $u\in$SU(2) and we are
here resorting to differential form notations, this action changes by
$-i2\pi k W$, where $W=\frac{1}{24\pi^{2}}{\rm Tr}\int_{B}
(u^{-1}\partial_{\mu}u)(u^{-1}\partial_{\nu}u)(u^{-1}\partial_{\rho}u)
\in\mathbb{Z}$, which guarantees its gauge invariance.
The validity of Eq. (\ref{2+1 WZ-CS}) will become evident shortly by comparing the
the 3+1d actions which is obtained by the ^^ ^^ stacking layer" construction
carried out for both the Wess-Zumino and Chern-Simons terms. We will now turn to
this step.

As in the previous 0+1d$\rightarrow$1+1d construction,
we now place the 2+1d systems regularly along a third (z) spatial axis
and add up their topological terms in a sign-alternating
fashion, which is followed by taking the continuum limit.
This will be the natural thing to do, for instance, if the five components of
$\bol{N}$ stood for an antiferromagnetic-VBS competition as mentioned
above, since an anti-parallel alignment of the three components in the antiferromagnetic
sector (let us call them $N_x , N_y , N_z $) would clearly induce a sign change in ${\cal S}_{\rm WZ}^{2+1d}$.
(In analogy to the magnetization plateau in 1+1d,
we can also consider a situation where one of these three
components (say the $N_z$) has a fixed {\it uniform}
value due to the coupling to an external field,
e.g. a magnetic field applied in the z-direction. Precisely as in the treatment of the
1d case, we can still staggerize the sum, i.e. make the conversion
$\sum_{j}{\cal S}_{\rm WZ/CS}^{2+1d}\vert_{j}
\rightarrow\sum_{j}(-1)^{j}{\cal S}_{\rm WZ/CS}^{2+1d}\vert_{j}$,
for special fixed values of $N_z$ which renders
the residual terms that arise with this rewriting trivial. We will come back to this point later.)
Carrying out the sign-alternating summation for the Wess-Zumino term, Eq. (\ref{2+1 O5 WZ}), is quite straightforward
 \cite{Tanaka06}.
As before, a crucial factor of 1/2 comes out when
taking the continuum limit in the z-direction,
and we obtain
\begin{eqnarray}
{\cal S}_{\rm B}^{\rm tot}
&\equiv&
\sum_{j}(-1)^{j}{\cal S}_{WZ}^{2+1d}\vert_j
%&=&ik\int_{B\times[0,1]}
%\tilde{\bol{N}}^{*}
%\omega_{S^{4}}
\nonumber \\
&=&i\frac{2\pi S}{area(S^{4})}\int d\tau d^3 \bol{r}
\epsilon_{abcde}{N}^{a}
\partial_{\tau}{N}^{b}\partial_{x}{N}^{c}
\partial_{y}{N}^{d}\partial_{z}{N}^{e},
\label{3+1 O5 theta}
\end{eqnarray}
where $S=k/2$. This is the $\theta$-term of the $3+1d$ O(5) nonlinear sigma model
with a vacuum angle of $\theta=2\pi S$.
 As for the Chern-Simons term, we replace like before
finite differences with a z-derivative and supplement terms to maintain gauge invariance
(or apply Stoke's theorem) to find
\begin{eqnarray}
{\cal S}_{\rm B}^{\rm tot}&=&i2\pi S \cdot
\frac{1}{32\pi^2}{\rm Tr}
\int d^4 x \epsilon_{\mu\nu\rho\sigma}F_{\mu\nu}F_{\rho\sigma}
\nonumber \\
&=&i2\pi S\cdot \frac{1}{8\pi^2}
{\rm Tr}\int \bol{F}\wedge \bol{F}
\label{3+1 HP1 theta}
\end{eqnarray}
where $F_{\mu\nu}$ is the gauge curvature %$D_{\mu}$ is the covariant derivative
and $\bol{F}=\frac{1}{2!}F_{\mu\nu}dx^{\mu}
\wedge x^{\nu}$. This is the $\theta$ term of the SU(2) Yang-Mills theory (or of the
HP${}^{1}$ nonlinear sigma model \cite{Demler}), also
with vacuum angle $\theta=2\pi S$. The mathematical proof that
the two expressions, Eq. (\ref{3+1 O5 theta}) and
Eq. (\ref{3+1 HP1 theta}) are equivalent can be found in the literature, 
e.g. ref \refcite{Hatsugai}.
This is an indication that our point of departure, Eq. (\ref{2+1 WZ-CS}) is valid.

Up to this point the argument has been general, i.e. holds for generic configurations
of the field $\bol{N}$. Following the procedure of the
1+1d problem we now evaluate Eq. (\ref{3+1 HP1 theta})
for the ^^ ^^ planar configuration"
$\cos\theta=0$ (in the competing-orders context, this corresponds
to setting the z-component of the antiferromagnetic sector to zero),
for which case the gauge field becomes $b_{\mu}=\frac{1}{2}g^{-1}\partial_{\mu}g$,
which is, as in the 1+1d analog, not a pure gauge. As before, in this limit we require that
$S$ (which is just defined through the relation with the level $k=2S$ and
in 3+1d is not directly related to the spin quantum number) be an integer.
This results in
\begin{equation}
{\cal S}_{\theta}[b=\frac{1}{2}g^{-1}dg]\vert_{\theta=2\pi S}=i\pi S Q_{\rm m}
\label{monopole BP term}
\end{equation}
where
\begin{equation}
Q_{\rm m}=\frac{1}{32\pi^2}{\rm Tr}\int d^4 x \epsilon_{\mu\nu\rho\sigma}
\partial_{\mu}\left(
(g^{-1}\partial_{\nu}g)(g^{-1}\partial_{\rho}g)(g^{-1}\partial_{\sigma}g)
\right)\in\mathbb{Z}
\label{monopole charge}
\end{equation}
is the monopole charge throughout space-time. Note that Eq. (\ref{monopole BP term})
is completely analogous to the vortex Berry phase of Eq. (\ref{eq:PlanarThetaTerm}).
The becomes even more apparent by noting that the total space-time vorticity
can be written, using the notation $g\equiv e^{i\phi}\in$U(1) as
$Q_{\rm v}=\frac{1}{2\pi i}\int d^2 x \epsilon_{\mu\nu}\partial_{\mu}
(g^{-1}\partial_{\nu}g)\in\mathbb{Z}$, where the analogy with Eq. (\ref{monopole charge})
is clear. The ground state wavefunctional can also be evaluated easily, using the fact
that a temporal surface term arises owing to the well known relation
$d{\rm Tr}(bdb+\frac{2}{3}b^3)\propto{\rm Tr}F\wedge F$. This leads to the result
\begin{eqnarray}
\Psi [b] &\propto&
e^{-\frac{1}{2}{\cal S}_{\rm CS}^{3+0d} \left[b=\frac{1}{2}g^{-1}dg\right]}
\nonumber \\
&=&
e^{
i\frac{k}{8\pi}{\rm Tr}\int dxdydz \left[\epsilon_{\alpha\beta\gamma}
(b_{\alpha}\partial_{\beta}b_{\gamma}+\frac{2}{3}b_{\alpha}b_{\beta}b_{\gamma})
\right]_{b=\frac{1}{2}g^{-1}dg}
}
\nonumber \\
&=& e^{-i\pi S W},
\end{eqnarray}
where $W$ is the winding number associated with the snapshot configuration,
\begin{equation}
W=\frac{1}{24\pi^2}{\rm Tr}\int dxdydz\left[\epsilon_{\alpha\beta\gamma}
(g^{-1}\partial_{\alpha}g)
(g^{-1}\partial_{\beta}g)
(g^{-1}\partial_{\gamma}g)\right]\in\mathbb{Z}.
\label{snapshot winding 3+1}
\end{equation}
There is thus a clear distinction between the topological structure of the
wavefunctional for odd and even $S$, which is the 3d analog of the topological
disctinction between the wavefunctionals of Haldane gap states
for odd and even spin quantum number. By switching on a perturbation
which induces a staggered component $\delta N_{z}$to $N_z$, i.e. the analog of the
staggered magnetic field employed in the 1d case, one readily sees that the
value of $\theta$ changes continuously with $\delta N_{z}$. The causes the
wavefunctional to interpolate between the forms corresponding
to odd and even $S$\footnote{By carrying out a fugacity expansion into the
dual action of a monopole (instanton) gas ensemble, one sees that this interpolation
occurs without closing the energy gap.}. Thus the odd $S$ case represents an
SPT phase protected by the link-centered inversion symmetry along the z-direction.
Clearly we can extend the above to the case
where instead of the ^^ ^^ planar limit" $N_{z}=0$,  the value of
$N_{z}$
is set to a finite and constant value.
%i.e.  $\cos\theta\equiv \alpha$=const.,
%and hence $b=\frac{1}{2}(1-\alpha)g^{-1}dg$,
%the total Berry phase action is a
%uniform (unstaggered) sum over the individual 2+1d Chern-Simons terms, since
%only two of the five components within the antiferromagnetic sectors tend to align
%in a staggered way and will not give rise to a sign change in the action.
%However, as in the magnetization plateau case, one can verify that
%there are special values of $\alpha$ which allow the sum to be
%rewritten as a staggered summation.  For those cases, it is straightforward to
%show that the behavior of the
%ground state wavefunctional  modifies to the form
%$\Psi\propto e^{-i\pi \frac{S(3\alpha-\alpha^3 )}{2}W}$, where the above requirement of
%recasting into a staggered sum implies that the quantity $\frac{S(3\alpha-\alpha^3 )}{2}$ %appearing in the exponent be an integer.
%Depending on whether that integer is odd or even, the system belong to an SPT phase  or a
%topologically trivial phase. From our construction it is clear that
%the symmetry protecting the SPT state is the link-inversion symmetry along the z-axis.
%To show that this is so, one can mimick the analysis of the previous subsection by deriving an
%effective theory of the monopole gas ensemble, which can be used to show that a
%perturbation which breaks this symmetry will destroy the topological distinction
%of the wavefunctional and connect the two phases without encountering a gap-closing.
There are special values of $N_{z}$ for which the sum over the Berry phases can be
turned into a staggered sum as in the planar limit, and one is lead to the 3d counterpart to
the findings of the previous subsection.

Having outlined how the topological terms of nonlinear sigma models can be turned into a
tool to discriminate SPT and non-SPT states,
we wish to reflect here on the somewhat curious fact that in the above,
the reduction of the number of ^^ ^^ active" components of the
unit vectors $\bol{n}$ (for the 1+1d case) and $\bol{N}$ (for 3+1d) by one was necessary
in order to build in a nontrivial topological structure into the wavefunctional \cite{XuSenthil}.
The mathematical reason for this necessity can be understood by noting that the winding numbers of Eqs. (\ref{snapshot winding 1+1})
and (\ref{snapshot winding 3+1}) relevant to this purpose each involve unit vectors
with two and four components, each with one component less than their initial sigma model actions. In other words, the reduction was required so that the {\it surface} action derived 
out of the bulk $\theta$ term 
can be endowed with a nontrivial topology.
While this is a powerful scheme, one may opt for an
analysis which is less restrictive, since, for the 1d case for instance, the Haldane gap state
does not require that the system be in the planar limit. The resolution to this 
apparently important aspect of the problem
is beyond the scope of this short review and will be taken up elsewhere.

\subsection{Summary}

To summarize this section,
we have described the stepwise construction of SPT state in 1+1d and 3+1d,
which are both described in terms of nonlinear sigma models with topological terms.
Here a rich interplay between topological terms differing by
one space-time dimension-$\theta$ terms
and Wess-Zumino/Chern Simons terms came into play. The 1+1d case was applied to
Haldane gap states at the planar limit, as well as to
magnetization plateau in a 1d spin chain. Although at present it is not clear
as to exactly what physical system our 3d effective theory describes, it is intriguing that
the building blocks are the same actions which appear in the theory of
competing orders which were taken up in the earlier part of this article.
Interesting topological states of matter may thus be in store in systems related to
those discussed there.
It is also suggestive that effective theories of SPT phases, e.g. those concerning
certain quantum Hall states, also featuring nonlinear sigma models
and their topological terms have appeared in work based on other  approaches \cite{AshwinSenthil,XuSenthil}. A complete classification of SPT states is
an important subject which is still under development. A powerful framework
for this purpose utilizing group cohomology was put forth in Ref. \refcite{Chen12}.
There the authors start from
special 1d ground state wavefunctions (the so-called matrix product states) which are
simple enough so that their symmetry properties can be put to detailed analysis, which enables one to explicitly construct
a wide class of SPT states.
These wavefunctions were further recasted into
a path integral over a topological nonlinear sigma model (where only the $\theta$ term is present) taking values on abstract symmetric spaces. This field-theoretical interpretation
of wavefunctions of SPT states has the virtue of being generalizable to higher
dimensions. While there are some apparent similarities between that approach and
the semiclassical theories taken up in the above, much remains to be done to
understand the whole picture of this rich subject.

\section{Conclusion}

%
%Emergence of enhanced global symmetries, resulting from competing physics between almost degenerate two orders, has been proposed to be responsible for deconfinement of fractionalized excitations and a scenario for the pseudogap phase in high T$_{c}$ cuprates. On the other hand, disappearance of such competing physics has been suggested to confine spinons and holons forming electron excitations, expected to describe non-Fermi liquid physics near optimal doping in high T$_{c}$ cuprates. An interesting point is that this nonperturbative physics seems to be found in the perturbative framework, where topological excitations for such nonperturbative physics look introduced into an effective field theory through the enhanced emergent symmetry.
%

Topological excitations and their nonperturbative effects are two key words in the present review article. The first part covered their nonperturbative effects on strongly coupled field theories in the presence of itinerant fermions and the second part included their roles in classifying interacting topological phases in the absence of itinerant fermions. In both parts topological terms played their essential roles in the nonperturbative physics of topological excitations, where the first subject was on the role of topological excitations in conformal invariant fixed points in the presence of topological terms while the second subject was on that in gapped phases characterized by the interplay between topological excitations and topological terms.

For ``metallic" systems, we have two kinds of well defined quantum phases as our starting points: One is the 
%Landau's 
Fermi-liquid state described by %the 
Landau's Fermi-liquid theory and the other is the Luttinger liquid state described by the Luttinger liquid theory. In the first part we made a conjecture that one dimensional physics can be generalized into two dimensions when electron correlations are strong enough, referred to as universal one dimensional physics. Our reference was the SO(3) nonlinear $\sigma-$model with the Berry phase term, regarded to be a well-known example for the Luttinger-liquid physics. An essential point was that this UV effective field theory will flow into the SO(4) WZW theory at IR, well known to be as an exact solution for the Heisenberg spin chain with spin $1/2$, although the underlying renormalization group path from the UV to the IR has not been understood, where nonperturbative effects of skyrmions (instantons) in the nonlinear $\sigma-$model %are not known in the presence of the Berry phase term. 
in the presence of the Berry phase term are unknown. 
However, it seems to be clear that the nonperturbative physics of the UV effective field theory can be introduced within the perturbative analysis of the IR conformal field theory, where the perturbative renormalization group analysis for the SO(4) WZW theory results in the deconfined critical physics of the spin chain. The underlying mechanism for this surprising result is that the Berry phase term assigns an additional quantum number involved with valence bond ordering to the skyrmion core and such valence bond fluctuations become symmetry equivalent with antiferromagnetic fluctuations. The interplay between topological excitations and topological terms in critical systems may allow enhancement of global symmetry from UV to IR, making the nonperturbative physics of topological excitations at UV visible in the perturbation framework with enhanced global symmetries at IR.

This universal one dimensional physics has been generalized into two dimensions, referred to as deconfined quantum criticality and described by the SO(5) WZW theory. The first part of this review article focused on how to generalize the SO(5) WZW theory in the presence of doped holes. In this respect Eq. (\ref{Mu_QED3_SO5_WZW}) contains the main message of the first part. First of all, scattering between valence bond fluctuations and doped holes will be quite difficult to describe in the present technology if we start from the UV nonlinear $\sigma-$model field theory, where valence bond fluctuations should be described by monopole-type excitations. However, we have well defined vertices for the scattering problem within the SO(5) WZW formulation, 
expected to allow us %dealing with 
to deal with 
such effective interactions perturbatively. In other words, the nonperturbative problem involved with scattering between itinerant electrons and magnetic monopoles is translated into the perturbative one associated with that between doped holes and valence bond excitations. Performing the perturbative renormalization group analysis, 
%we suspect 
we are lead to suspect 
that the high T$_{c}$ superconducting state may be identified with the two-dimensional generalization of the Emery-Luther phase, which valence bond fluctuations are responsible for. We would like to mention that this is a meaningful progress on the 
issue %how to 
on how to 
introduce the nonperturbative physics into strongly coupled conformal field theories.

We believe that the universal one dimensional physics is not limited %on 
to the strong-correlation limit, i.e., the case when the ratio between the interaction energy and kinetic one is infinite. Indeed, we observed that essentially the same situation can appear near the metal-insulator transition particularly %in 
on a 
honeycomb lattice. We applied this scenario to the case of ``weak" coupling metallic antiferromagnetic quantum criticality, where itinerant electrons still remain strongly coupled with antiferromagnetic fluctuations, implying that vertex corrections should be incorporated consistently beyond the Hertz-Moriya-Millis framework. Since the nature of the conformal fixed point is not clarified yet, this application should be addressed more carefully.

%The universal one dimensional physics has also been generalized toward three dimensions in 
The generalization of %universal 
one dimensional physics towards higher dimensions 
also plays an important role when   
classifying gapped topological phases, which is the subject covered in the second part. 
A central mathematical apparatus employed here 
is the fact that the $(1+1)d$ $\theta-$term in the SO(3) nonlinear $\sigma-$model 
can be expressed as the staggered summation over 
a collection of  $(0+1)d$ WZW terms (single spin Berry phases) 
aligned along a one dimensional spatial extent, as is well known from the work of Haldane.  
It proves convenient for our purpose to observe that 
this scheme can also be carried out in the complex projective (CP$^{1}$) representation of the SO(3) nonlinear $\sigma-$model, 
% with the $\theta-$term, 
where the staggered summation for the $(0+1)d$ Chern-Simons term results in the CP$^{1}$ representation of the $(1+1)d$ $\theta-$term. 
After discussing the $(1+1)d$ setup,  
this mathematical parallelism was generalized to three spatial dimensions, where the starting point corresponding to the $(0+1)d$ WZW term in the SO(3) nonlinear $\sigma-$model is now the $(2+1)d$ WZW term in the SO(5) nonlinear $\sigma-$model. Meanwhile when we resort to the quaternionic projective (HP$^{1}$) representation, the gauge-field term corresponding to the $(0+1)d$ Chern-Simons term is the $(2+1)d$ nonabelian Chern-Simons term which appears in the HP$^{1}$ representation of the SO(5) WZW theory. The idea was to perform a stacking layer construction along the $z-$direction which amounts to a staggered summation over these  $(2+1)d$ topological terms. It is verified easily that starting with the WZW term, we arrive at the $(3+1)d$ $\theta-$term of the $(3+1)d$ SO(5) nonlinear $\sigma-$model. In the same way, the staggered summation 
over the $(2+1)d$ nonabelian Chern-Simons terms 
gives us the $\theta-$term of the SU(2) Yang-Mills theory, i.e. we obtain the HP$^{1}$ representation of the SO(5) nonlinear $\sigma-$model which contains an $F\wedge F$ type  $\theta-$term. This set of interrelated effective field theories allows us to discuss symmetry protected topological phases in three dimensions based on a solid platform. 

Classification of gapped topological phases can be achieved by investigating the 
dynamics of topological excitations, where instanton excitations %are supposed to 
carry nontrivial quantum numbers %assigned 
inherited from the topological $\theta-$term, 
which in turn can modify the dynamics of instantons in a significant way.  
%pretty much. 
In this respect it is a natural strategy to attempt to construct an effective dual field theory in terms of topological excitations, %where 
into which the role of the topological term is encoded. 
In $(1+1)d$, the sine-Gordon action for vortices in the planar limit of the Heisenberg spin chain turns out to have 
two distinct vortex phases, depending on the even-odd parity of the coefficient of the $\theta-$term 
(which, physically, is just the parity of the spin quantum number), which classifies insulating states into topologically nontrivial and trivial ones. This classification  
can also be reached through a study the ground-state wave function. 
It turns out that the ground-state wave function 
%shows %the 
%a sign ambiguity, determined by the even-odd parity of the topological charge, 
for the odd spin case exhibits a characteristic sign dependence on the even-odd parity of 
the topological charge,  
while their amplitude (norm) remains %topologically identical. 
the same for all topological sectors. Meanwhile, the wave function 
for the even spin case does not have this sign dependence 
and can thus be regarded as being topologically trivial. 
The crucial point is that the former ground state cannot be connected adiabatically 
to the latter as long as inversion symmetry is preserved.  Such states are referred to as symmetry protected topological phases. It proves straightforward to extend 
this classification scheme for one dimensional physics is to three dimensions, 
where a similar sine-Gordon-type action can be constructed to describe 
the dynamics of SU(2) instantons in the presence of the topological $\theta-$term. 
However, various types of topological excitations can arise 
based on the SO(5) nonlinear $\sigma-$model with the $(3+1)d$ $\theta-$term or its HP$^{1}$ representation, depending on the symmetries of the considered system. 
In order to fully classify three dimensional gapped phases, it is thus necessary to investigate the dynamics of various topological excitations in more depth, and to see how they depend on the nature of topological terms and global symmetries.

\section*{Acknowledgements}

This study was supported by the Ministry of Education, Science, and Technology (No. 2012R1A1B3000550 and No. 2011-0030785) of the National Research Foundation of Korea (NRF) and by TJ Park Science Fellowship of the POSCO TJ Park Foundation. KS would like to express sincere thanks to his collaborators, Hyun-Chul Kim (PNJL study in Sec. 2.3), Minh-Tien Tran (Spin-liquid study in Sec. 2.4), Mun Dae Kim (SU(2) slave-rotor theory in Sec. 2.4), C. P\'epin (Eliashberg theory), and A. Benlagra (Luttinger-Ward functional approach). The work of AT on SPT states was supported in part by KAKENHI through grant no. (C) 23540461. He thanks S. Takayoshi , K. Totsuka and T. Morimoto for collaborations on related subjects.

%\appendix{Heading of Appendix}
%
%\subappendix{This is the subappendix}
%
%\subsubappendix{Sub-subappendix}
%
%\appendix{Another Appendix}


\begin{thebibliography}{0}
\bibitem{Shankar_RG} R. Shankar, Rev. Mod. Phys. {\bf 66}, 129 (1994).
\bibitem{BCS_Textbook} J. R. Schrieffer, \textit{Theory of Superconductivity}, (Westview Press, 1999).
\bibitem{Kondo_Textbook} A. C. Hewson, \textit{The Kondo Problem to Heavy Fermions}, (Cambridge University Press, New York, 1993).
\bibitem{Luttinger_Liquid_Textbook} A. O. Gogolin, A. A. Nersesyan, and A. M. Tsvelik, \textit{Bosonization and Strongly Correlated Systems} (Cambridge University Press, New York, 2004).
\bibitem{SSL_Large_N_Failure} S.-S. Lee, Phys. Rev. B {\bf 80}, 165102 (2009).
\bibitem{Mahan_Textbook} G. D. Mahan, \textit{Many-Particle Physics 3th ed.}, (Kluwer Academic/Plenum Publishers, New York, 2000).
\bibitem{Boson_Vortex_Duality_Textbook} I. Herbut, \textit{A Modern Approach to Critical Phenomena}, (Cambridge University Press, New York, 2007).
\bibitem{Maslov_Review_LL} D. L. Maslov, arXiv:cond-mat/0506035 (unpublished).
\bibitem{Solitons_Instantons_Textbook} R. Rajaraman, \textit{Solitons and Instantons} (Elsevier Science, New York, 2003).
\bibitem{Kim_SO5_WZW_mu_QED3} Ki-Seok Kim, Phys. Rev. B {\bf 78}, 195113 (2008).
\bibitem{Haldane} F. D. M. Haldane, Phys. Rev. Lett. {\bf 61}, 1029 (1988).
\bibitem{Read_Sachdev} N. Read and S. Sachdev, Phys. Rev. Lett. {\bf 62}, 1694 (1989); N. Read and S. Sachdev, Phys. Rev. B {\bf 42}, 4568 (1990); S. Sachdev and N. Read, Int. J. Mod. Phys. B {\bf 5}, 219 (1991).
\bibitem{Tanaka_SO5} A. Tanaka and X. Hu, Phys. Rev. Lett. {\bf 95}, 036402 (2005).
\bibitem{DQCP} T. Senthil, A. Vishwanath, L. Balents, S. Sachdev, and M. P. A. Fisher, Science {\bf 303}, 1490 (2004); T. Senthil, L. Balents, S. Sachdev, A. Vishwanath, and M. P. A. Fisher, Phys. Rev. B {\bf 70}, 144407 (2004).
\bibitem{Fukushima_PNJL} K. Fukushima, Phys. Lett. B {\bf 591}, 277 (2004).
\bibitem{PNJL_Review} C. Ratti, S. Roessner, and W. Weise, Phys. Lett. B {\bf 649}, 57 (2007); S. Roessner, T. Hell, C. Ratti, and W. Weise, Nucl. Phys. A {\bf 814}, 118 (2008);
K. Fukushima, Phys. Rev. D {\bf 79}, 074015 (2009); W.-J. Fu, Z. Zhang, and Y.-X. Liu, Phys. Rev. D {\bf 77}, 014006 (2008).
\bibitem{Kim_Kim_PNJL} Ki-Seok Kim and Hyun-Chul Kim, J. Phys.: Condens. Matter {\bf 23}, 495701 (2011).
\bibitem{Kim_SU2SR} Ki-Seok Kim, Phys. Rev. Lett. {\bf 97}, 136402 (2006); Ki-Seok Kim, Phys. Rev. B {\bf 75}, 245105 (2007); Ki-Seok Kim and Mun Dae Kim, Phys. Rev. B {\bf 81}, 075121 (2010).
\bibitem{Kim_Tien_SU2SR} Minh-Tien Tran and Ki-Seok Kim, Phys. Rev. B {\bf 83}, 125416 (2011).
\bibitem{Kim_Spin_Fermion_Model_AFQCP} Ki-Seok Kim, arXiv:1403.1136, to be published in Phys. Rev. B.
\bibitem{Kim_Spin_Fermion_Model_FMQCP} Ki-Seok Kim, arXiv:1408.2993 (unpublished).
\bibitem{Chubukov_Spin_Fermion_Model} Ar. Abanov, A. V. Chubukov, and J. Schmalian, Adv. Phys. {\bf 52}, 119 (2003).
\bibitem{Sachdev_Large_N_Failure} Max A. Metlitski and S. Sachdev, Phys. Rev. B {\bf 82}, 075128 (2010).
\bibitem{Sandvik_DQCP} A. W. Sandvik, Phys. Rev. Lett. {\bf 98}, 227202 (2007); A. W. Sandvik, Phys. Rev. Lett. {\bf 104}, 177201 (2010).
\bibitem{Tanaka_SO4} A. Tanaka and X. Hu, Phys. Rev. Lett. {\bf 88}, 127004 (2002).
\bibitem{Quantum_Spins_Textbook} A. Auerbach, \textit{Interacting Electrons and Quantum magnetism} (Springer-Verlag, New York, 1994).
\bibitem{SU2SBGT} I. Affleck, Z. Zou, T. Hsu, and P. W. Anderson, Phys. Rev. B {\bf 38}, 745 (1988); E. Dagotto, E. Fradkin, and A. Moreo, Phys. Rev. B {\bf 38}, 2926 (1988).
\bibitem{Affleck_Marston_pi_Flux} I. Affleck and J. B. Marston, Phys. Rev. B {\bf 37}, 3774 (1988).
\bibitem{Kotliar_pi_Flux} G. Kotliar and J. Liu, Phys. Rev. B {\bf 38}, 5142 (1988).
\bibitem{Wen_Symmetry} Y. Ran and X.-G. Wen, arXiv:cond-mat/0609620v3 (unpublished).
\bibitem{Wen_Textbook} X.-G. Wen, \textit{Quantum Field Theory of Many-Body Systems} (Oxford University Press, New York, 2008).
\bibitem{Abanov_WZW} A. G. Abanov and P. B. Wiegmann, Nucl. Phys. B {\bf 570}, 685 (2000).
\bibitem{Anomaly_Cancelation} F. D. M. Haldane, Phys. Rev. Lett. {\bf 61}, 2015 (1988).
\bibitem{Z4_Vortex_Spinon} M. Levin and T. Senthil, Phys. Rev. B {\bf 70}, 220403 (2004).
\bibitem{Lee_Nagaosa_Wen_SL_Review} P. A. Lee, N. Nagaosa, and X.-G. Wen, Rev. Mod. Phys. {\bf 78}, 17 (2006).
\bibitem{KS_MD_SC} K.-S. Kim and M. D. Kim, Phys. Rev. B {\bf 77}, 125103 (2008); K.-S. Kim and Mun Dae Kim, Phys. Rev. B {\bf 75}, 035117 (2007).
\bibitem{ARPES_Review} A. Damascelli, Z.-X. Shen, Z. Hussain, Rev. Mod. Phys. {\bf 75}, 473 (2003).
\bibitem{ASL_Mother} M. Hermele, T. Senthil, and M. P. A. Fisher, Phys. Rev. B {\bf 72}, 104404 (2005).
\bibitem{Sachdev_SO5} C. Xu and S. Sachdev, Phys. Rev. Lett. {\bf 100}, 137201 (2008).
\bibitem{Witten_QED2_NLsM} E. Witten, Nucl. phys. B {\bf 149}, 285 (1979).
\bibitem{Shankar_SF_tJ} R. Shankar, Phys. Rev. Lett. {\bf 63}, 203 (1989); R. Shankar, Nucl. phys. B {\bf 330}, 433 (1990).
\bibitem{Luttinger_Ward_Functional} J. M. Luttinger and J. C. Ward, Phys. Rev. {\bf 118}, 1417 (1960); G. Baym and L. P. Kadanoff, Phys. Rev. {\bf 124}, 287 (1961).
\bibitem{Luttinger_Ward_Functional_Kim_Pepin} A. Benlagra, K.-S. Kim, and C. P\'epin, J. Phys.: Condens. Matter {\bf 23}, 145601 (2011).
\bibitem{Landau_Damping_QCP_Review} H. V. Lohneysen, A. Rosch, M. Vojta, and P. Wolfle, Rev. Mod. Phys. {\bf 79}, 1015 (2007).
\bibitem{Z3_Gauge_Fields_Perturbation} A. M. Tsvelik, \textit{Quantum Field Theory in Condensed Matter Physics}, (Cambridge University Press, Cambridge, 1995).
\bibitem{Nagaosa_Lee_SM} N. Nagaosa and P. A. Lee, Phys. Rev. Lett. {\bf 64}, 2450 (1990); P. A. Lee and N. Nagaosa, Phys. Rev. B {\bf 46}, 5621 (1992).
\bibitem{YB_Wen_Lee} Y.-B. Kim, A. Furusaki, X.-G. Wen, and P. A. Lee, Phys. Rev. B {\bf 50}, 17917 (1994).
\bibitem{Ployakov_Loop} A. M. Polyakov, \textit{Gauge Fields and Strings} (Harwood Academic Publishers, New York, 1987).
\bibitem{Weiss_PNJL} N. Weiss, Phys. Rev. D {\bf 24}, 475 (1981).
\bibitem{IoffeLarkin} L. B. Ioffe and A. I. Larkin, Phys. Rev. B {\bf 39}, 8988 (1989).
\bibitem{Kim_TR_Boltzmann} K.-S. Kim and C. P\'epin, Phys. Rev. Lett. {\bf 102}, 156404 (2009); K.-S. Kim and C. P\'epin, J. Phys.: Condens. Matter {\bf 22}, 025601 (2010); Ki-Seok Kim, Phys. Rev. B {\bf 84}, 085117 (2011).
\bibitem{Data} H. Takagi, B. Batlogg, H. L. Kao, J. Kwo, R. J. Cava, J. J. Krajewski, and W. F. Peck, Jr, Phys. Rev. Lett. {\bf 69}, 2975 (1992).
\bibitem{Florens_Georges} S. Florens and A. Georges, Phys. Rev. B {\bf 70}, 035114 (2004).
\bibitem{Graphene_Z2SL_Simulation} Z. Y. Meng, T. C. Lang, S. Wessel, F. F. Assaad, and A. Muramatsu, Nature {\bf 464}, 847 (2010).
\bibitem{Kanoda_Phase_Diagram} K. Kurosaki, Y. Shimizu, K. Miyagawa, K. Kanoda, and G. Saito, Phys. Rev. Lett. {\bf 95}, 177001 (2005).
\bibitem{Double_Expansion} D. F. Mross, J. McGreevy, H. Liu, and T. Senthil, Phys. Rev. B {\bf 82}, 045121 (2010).
\bibitem{SSL_Dimensional_Regularization} D. Dalidovich and S.-S. Lee, Phys. Rev. B {\bf 88}, 245106 (2013).
\bibitem{tHooft_Interaction} G. 't Hooft, Phys. Rev. Lett. {\bf 37}, 8 (1976); G. 't Hooft, Phys. Rev. D {\bf 14}, 3432 (1976).



\bibitem{Takayoshi}
S. Takayoshi, K. Totsuka and A. Tanaka, arXiv:1412.4029.
\bibitem{Tanaka09}A. Tanaka, K. Totsuka and X. Hu,
Phys. Rev. B {\bf 79}, 064412 (2009).
\bibitem{Fisher}M. P. A. Fisher, Physica {\bf 177}, 553 (1991).
\bibitem{Wen}X.-G. Wen, {\it Quantum Field Theory of Many-Body Systems}
(Oxford Univ. press, U.K. 2004).
\bibitem{Herbut}I. Herbut, {\it A Modern Approach to Critical Phenomenna}
(Cambridge Univ. Press, Cambridge, U.K. 2007).
\bibitem{Sachdev}S. Sachdev, Physica A {\bf 313}, 252 (2002).
\bibitem{Auerbach}A. Auerbach, {\it Interacting Electrons and Quantum Magnetism}
(Springer-Verlag, New York, U.S.A. 1994).
\bibitem{XuSenthil}C. Xu and T. Senthil. Phys. Rev. B {\bf 87}, 174412 (2013).
\bibitem{Ng}T.-K. Ng, Phys. Rev. B {\bf 50}, 555 (1994).
\bibitem{Demler}E. Demler and S.-C. Zhang, Ann. Phys. {\bf 271}, 83 (1999).
\bibitem{Nash Sen}C. Nash and S. Sen,  {\it Topology and Geometry for Physicists}
(Dover Publications, New York, U. S. A. 2011).
\bibitem{Naber}G. L. Naber, {\it Topology, Geometry and Gauge Fields: Interactions}
(Springer, New York, 2011.)
\bibitem{Pollman}F. Pollman, A. M. Turner, E. Berg and M. Oshikawa,
Phys. Rev. B {\bf 81}, 064439 (2010); F. Pollman, E. Berg, A. M. Turner and M. Oshikawa,
Phys. Rev. B {\bf 85}, 075125 (2012).
\bibitem{Tanaka11}A. Tanaka, J. Phys. Conf. Ser. {\bf 320}, 012020 (2011).
\bibitem{Tanaka06}A. Tanaka and X. Hu, Phys. Rev. B {\bf 74}, 140407(R) (2006).
\bibitem{Hatsugai}Y. Hatsugai, New J. Phys. {\bf 12}, 065004 (2010).
\bibitem{AshwinSenthil}A. Vishwanath and T. Senthil. Phys. Rev. X {\bf 3}, 011016 (2013).
\bibitem{Chen12}
X. Chen, Z.-C. Gu, Z.-X. Liu, X.-G. Wen, Science {\bf 338},
%\href{http://dx.doi.org/10.1126/science.1227224}
{1604} (2012).
\end{thebibliography}
\end{document}